\documentclass[a4paper,cite,11pt]{article}

\usepackage[T1]{fontenc}
\usepackage[latin1]{inputenc}

\usepackage{graphicx,color}

\usepackage{latexsym ,amsmath,amsfonts,amssymb,amstext,mathrsfs,upgreek,dsfont}

\newcommand{\Res}[1]{\mathsf{Res}\left[ #1\right]}
\newcommand{\hyper}[5]{\;_{#1}{\rm F}_{#2} \left(\left.\begin{matrix} {#3}
\\ {#4} \end{matrix}\right| #5\right) }

\allowdisplaybreaks[1]

\begin{document}

\begin{titlepage}

\title{\bf On convergent series representations of Mellin-Barnes integrals} 

\vspace{2cm}

\author{Samuel FRIOT~$^\dag$ and David GREYNAT~$^\ddag$}

\vspace{1cm}

\date{}

\maketitle

\begin{center}
$^\dag$~Institut de Physique Nucl\'eaire d'Orsay\\ Universit\'e Paris-Sud 11, 91405 Orsay Cedex, France

\vspace{0.5cm}

$^\ddag$~Departamento de F\'isica Te\'orica\\ 
Universidad de Zaragoza, Cl Pedro Cerbuna 12, E-50009 Zaragoza, Spain
\end{center}

\thispagestyle{empty}

\vspace{1cm}

\begin{abstract}
Multiple Mellin-Barnes integrals are often used for perturbative calculations in particle physics. In this context, the evaluation of such objects may be performed through residues calculations which lead to their expression as multiple series in powers and logarithms of the parameters involved in the problem under consideration. However, in most of the cases, several series representations exist for a given integral. They converge in different regions of values of the parameters, and it is not obvious to obtain them.
For twofold integrals we present a method which allows to derive straightforwardly and systematically:
\begin{enumerate}
\item[(a)] different sets of poles which correspond to different convergent double series representations
of a given integral,
\item[(b)] the regions of convergence of all these series (without an \textit{a priori} full knowledge
of their general term),
\item[(c)] the general term of each series (this may be performed, if necessary, once the
relevant domain of convergence has been found).
\end{enumerate}

This systematic procedure is illustrated with some integrals which appear, among others, in the
calculation of the two-loop hexagon Wilson loop in $\mathcal{N} = 4$ SYM theory.

 Mellin-Barnes integrals of higher dimension are also considered.
\end{abstract}

\end{titlepage}

\section{Introduction}

The Mellin-Barnes (MB) representation is an important tool of asymptotic analysis and it is also very useful in high energy physics. In the context of the perturbative calculations of particle physics, for instance, the evaluation of a multiple MB integral, after resolving the $\epsilon$ singularities, may often be obtained as a decomposition in terms of some MB integrals of lower dimension (typically onefold, twofold and threefold integrals) that have to be expressed as (multiple) series. Each of these MB integrals of lower dimension, which in general depend on a number of parameters equal to its dimension, do have several convergent series representations.  These different series converge in different regions and are, as a rule, analytic continuations of each other. 

In this paper, we mainly focus on the case of twofold MB integrals (see however section \ref{MBofHD}) and, after presenting a general method allowing to derive these different series representations from a given twofold integral without any use of analytic continuation properties, we show that for a large class of MB integrals met in the perturbative calculations of particle physics (if not all), one may derive the regions of convergence of such series without the full knowledge of their general term\footnote{We will see that this property is in fact not restricted to the case of double series.}. Therefore, it enables one, if necessary, to choose the relevant region of convergence before a possibly extensive calculation. 

Although the method allows to get the complete series (and thus exact results) in the computation of a given quantity, it may also be interesting if one only needs the first few terms in the expansions of this quantity for different values of the parameters (which of course belong to different regions of convergence). Indeed, one does not need to perform the resummation of the series corresponding to one region, followed by an analytic continuation, in order to obtain the result in another region.

We illustrate the presentation of our method with some examples of integrals which appear, among others, in the evaluation of the two-loop hexagon Wilson loop in $\mathcal{N}=4$ SYM.

\section{Onefold Mellin-Barnes integrals\label{Onefold}}

In this introductory section we would like to briefly summarize the main different situations that may be encountered when one is interested in the series representations of onefold Mellin-Barnes integrals. To fix ideas, we begin with two simple illustrative examples.

The first one is the toy integral $Z(0)$ corresponding to the zero-dimensional version of the vacuum-to-vacuum generating functional of $\lambda\,\phi^4$ theory, used in \cite{Friot:2009fw} for the exposition of a method allowing to obtain non-perturbative asymptotic improvements directly from divergent perturbative expansions:
\begin{equation}
\label{vacuum}
Z(0)=\frac{1}{\sqrt{2\pi}}\int_{-\infty}^{+\infty}d\phi\ e^{-\frac{1}{2}\phi^2-\frac{\lambda}{4!}\phi^4}.
\end{equation}
It was shown in \cite{Friot:2009fw} that the following Mellin-Barnes representation for $Z(0)$ may be obtained:
\begin{equation}
\label{MBZ}
Z(0)=\frac{1}{\sqrt{\pi}}\int\limits_{c-i\infty}^{c+i\infty}\frac{ds}{2i\pi}\ \left(\frac{\lambda}{3!}\right)^{-s}\Gamma(s)\Gamma\left(\frac{1}{2}-2s\right),
\end{equation}
where $c={\rm Re\,} s\in]0,\frac{1}{4}[$, and that by closing the contour of integration to the left, by Cauchy theorem, one obtains the asymptotic expansion
\begin{equation}\label{asymp2}
Z(0)\underset{\lambda\rightarrow 0}{\sim} \frac{1}{\sqrt{\pi}}\sum_{k=0}^{+\infty}\frac{(-1)^k}{k!}\Gamma\Big(\frac{1}{2}+2k\Big)\left(\frac{\lambda}{3!}\right)^k,
\end{equation}
which, from d'Alembert's ratio test, diverges for \textit{any} value of $\lambda$.

In \cite{Friot:2009fw}, the quantity of interest (and starting point of the non-perturbative asymptotic analysis) was the divergent expansion (\ref{asymp2}), therefore it was not mentioned that one may also close the contour of (\ref{MBZ}) to the right to get the series
\begin{equation}\label{ConvZ0}
Z(0)= \frac{1}{2\sqrt{\pi}}\sum_{k=0}^{+\infty}\frac{(-1)^k}{k!}\Gamma\Big(\frac{k}{2}+\frac{1}{4}\Big)\left(\frac{3!}{\lambda}\right)^{\frac{k}{2}+\frac{1}{4}},
\end{equation}
which, this time, converges for \textit{any} value of $\lambda$.

Thus, (\ref{asymp2}) and (\ref{ConvZ0}) are two very different series representations of $Z(0)$.

\begin{figure}[h]
\begin{center}
\includegraphics[height=2.5cm]{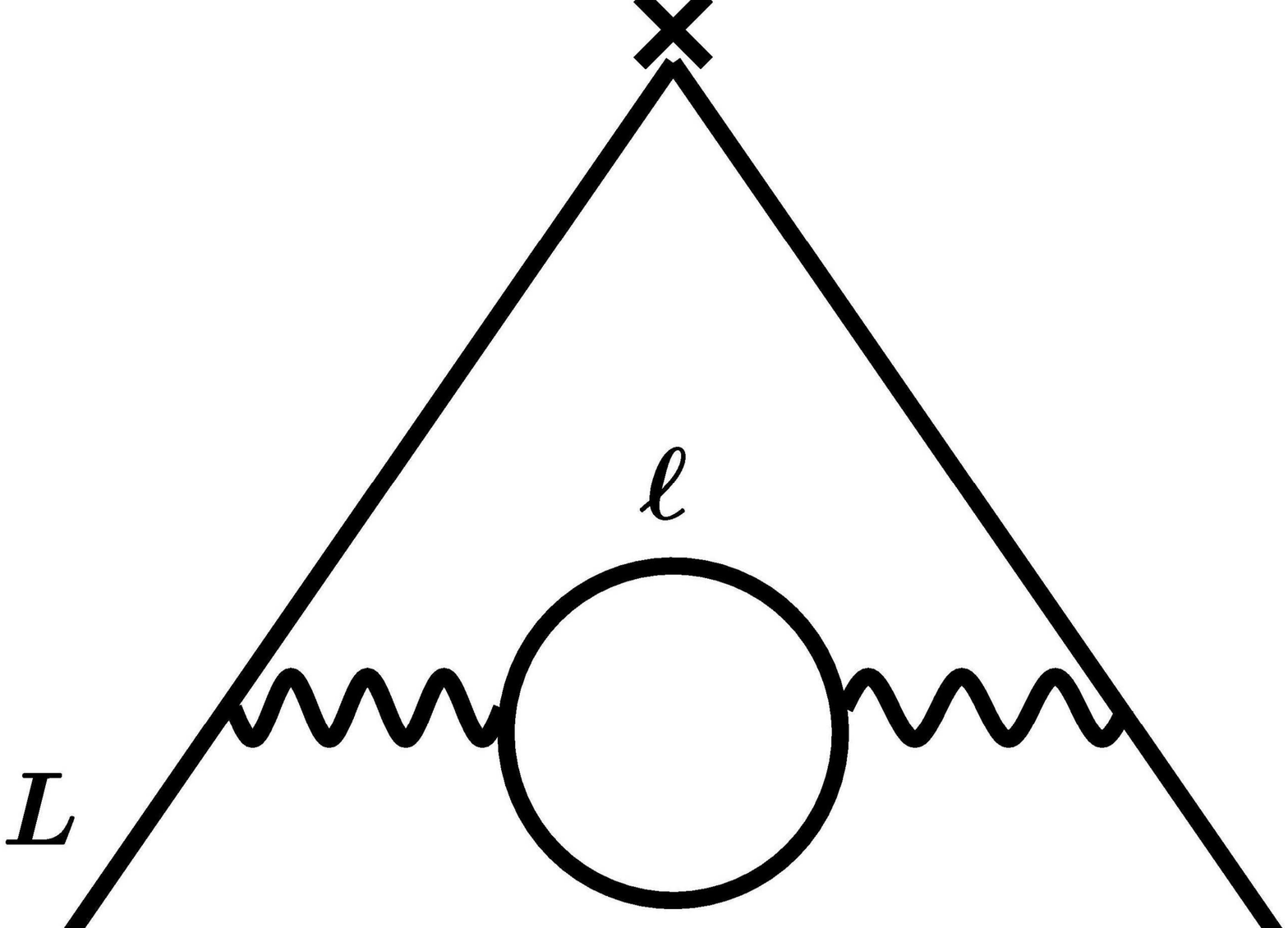} 
\caption{2-loop QED vacuum polarisation contribution to $g-2$ of a lepton $L$.\label{g-21looplepton}} 
\end{center}
\end{figure}

Let us come to our second example, the QED contribution to the anomalous magnetic moment of a lepton $a_L^{\ vac.\ pol.}$ of the diagram of Figure \ref{g-21looplepton}.  

Following \cite{Friot:2005cu} or \cite{Aguilar:2008qj} it is easy to obtain the corresponding Mellin-Barnes representation
\begin{equation}\label{vacpolMB}
 a_L^{\ vac.\ pol.}
=\left( \frac{\alpha}{\pi} \right)^2 \int\limits_{c-i\infty}^{c+i\infty}\!\! \frac{ds}{2i\pi} \left(\frac{m_\ell^2}{m_L^2} \right)^{-s}\left(\frac{\pi}{\sin \pi s }\right)^2\frac{1-s}{(2+s)(1+2s)(3+2s)}\ , 
\end{equation}
where $c\in]0,1[$. By closing the contour of integration to the right, one gets the series\footnote{We correct a sign misprint of \cite{Friot:2005cu}.}
\begin{equation}\label{amu01}
a_L^{\ vac.\ pol.}\underset{\frac{m_\ell^2}{m_L^2}\geq1}=\left(\frac{\alpha}{\pi}\right)^2\sum_{n=0}^\infty	\left(\frac{m_L^2}{m_\ell^2}\right)^{\! n+1}\!\!\left[\frac{45\!-\!28n^2\!-\!8n^3}{[(3+n)(3\!+\!2n)(5\!+\!2n)]^2}-\!\frac{n}{(3\!+\!n)(3\!+\!2n)(5\!+\!2n)}\ln\frac{m_\ell^2}{m_L^2}\!\right],
\end{equation}
which, from d'Alembert's and Raabe-Duhamel's ratio tests, converges in the region $\frac{m_\ell^2}{m_L^2}\geq 1$.

Alternatively, closing the contour to the left, one gets the series 
\begin{equation}\label{amu02}
a_L^{\ vac.\ pol.} \underset{\frac{m_\ell^2}{m_L^2}\leq 1}=\left(\frac{\alpha}{\pi}\right)^2\left\{\frac{1}{6}\ln\frac{m_L^2}{m_\ell^2}-\frac{25}{36}+\frac{\pi^2}{4} \frac{m_\ell}{m_L}+\left(\frac{m_\ell^2}{m_L^2}\right)\left[-2\ln\frac{m_L^2}{m_\ell^2}+3 \right]\right. \nonumber
\end{equation}
\begin{equation} 
\hspace{1.5cm} -\frac{5\pi^2}{4}\left(\frac{m_\ell}{m_L}\right)^3 + \left(\frac{m_\ell^2}{m_L^2}\right)^2\left[\frac{1}{2}\ln^2\frac{m_\ell^2}{m_L^2}- \frac{7}{3}\ln\frac{m_\ell^2}{m_L^2}+ \frac{44}{9} + \frac{\pi^2}{3}\right]\nonumber
\end{equation}
\begin{equation}
 \hspace{1.5cm}\left.\!+\!\!\sum_{n=3}^{\infty}\!\left(\frac{m_\ell^2}{m_L^2}\right)^{\!\!n}\!\!\left[ \frac{1+n}{(2-n)(1-2n)(3-2n)}\ln\!\frac{m_L^2}{m_\ell^2}\! +\! \frac{-25+32n+4n^2-8n^3}{(2-n)^2(1-2n)^2(3-2n)^2}\!\right]\!\right\},
\end{equation}

which converges in the region $\frac{m_\ell^2}{m_L^2}\leq 1$.

In fact, the two simple examples (\ref{MBZ}) and (\ref{vacpolMB}) reflect the particular situations that may be encountered in the computation of the following more general onefold Mellin-Barnes integral:
\begin{equation}\label{onefold}
I(x) = \int\limits_{c-i\infty}^{c+i\infty}{d s\over 2\pi i}\,
x^{-s}\frac{\prod_{j=1}^m\Gamma(a_js+b_j)}{\prod_{k=1}^n\Gamma(c_ks+d_k)}\,,
\end{equation}
where $c$ is chosen such that the straight line of the integration contour does not intersect the poles of the integrand and $a_j, b_j, c_k$ and $d_k$ are real numbers.  

Let us define the quantities 
\begin{equation}
\Delta\doteq\sum_{j=1}^ma_j-\sum_{k=1}^nc_k \hspace{1cm}\text{and}\hspace{1cm}\alpha\doteq\sum_{j=1}^m|a_j|-\sum_{k=1}^n|c_k|.
\end{equation}

The region where the integral (\ref{onefold}) converges is $|\arg x|<\frac{\pi}{2}\alpha$, see for instance \cite{Paris} (one may sometimes extend the region of convergence on its boundary as for instance when $\alpha=0$).

Three cases are in order \cite{Tsikh:1994}, \cite{Passare:1996db}:
\\

$\bullet$ If $\Delta>0$, closing the integration contour to the left leads to a series representation of the integral (\ref{onefold}) which converges for any value of $x$, but closing the contour to the right\footnote{In some cases there might be only a finite number of poles or no pole at all in this part of the complex $s$-plane, so that Cauchy theorem does not lead to a useful asymptotic information. This is the signal of an exponentially suppressed asymptotic expansion and other mathematical techniques have to be used.} gives a divergent asymptotic expansion.
\\

$\bullet$ If $\Delta<0$, closing the contour to the right leads to a series representation which converges for any value of $x$, but closing the contour to the left gives a divergent asymptotic expansion.
\\

$\bullet$ If $\Delta=0$, closing the contour to the left and to the right gives two convergent series (if the $a_j$ are all of the same sign, one series is terminating), the first one converging in some disk $|x|<R$ whereas the other converges outside this disk. Moreover, if $\alpha>0$, each of the two series is an analytic continuation of the other.
\\

It is easy to see that (\ref{MBZ}) has $\Delta=-1$ and $\alpha=3$, and that (\ref{vacpolMB}) may be reduced to the form (\ref{onefold}) with $\Delta=0$ and $\alpha=4$.

\begin{figure}[h]
\begin{center}
\hspace{-1cm}
\includegraphics[width=0.7\textwidth]{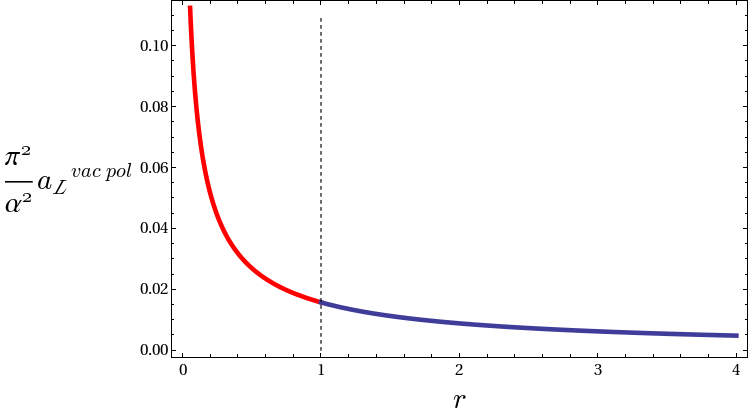}  
\caption{$\left(\frac{\pi}{\alpha}\right)^2a_L^{\ vac.\ pol.}$ as a function of $r$: in blue, eq. (\ref{amu1}) and in red, eq. (\ref{amu2}). \label{ac}} 
\end{center}
\end{figure}

Therefore, the two expressions (\ref{amu01}) and (\ref{amu02}), which may be expressed in terms of special functions as follows\footnote{Other equivalent representations may be found in the literature.} (with $r=m_{\ell}^2/m_L^2$):
\begin{align}\label{amu1}
 a_L^{\ vac.\ pol.}&\underset{r\geq 1} = \left( \frac{\alpha}{\pi} \right)^2 \left[ -\frac {1}{4} - r + \frac{\Upphi\left(\frac {1} {r},2, \frac{3}{2}\right)}{4r} - \frac {5 \Upphi\left(\frac {1} {r}, 2, \frac {5}{2} \right)}{4r}  + \frac {1}{2} \sqrt{r} \text{ArcCoth}\left(\sqrt {r} \right) \ln r - \frac {\ln r} {6}\right.\nonumber \\
&\hspace{2cm}  \left. + \frac {3}{2} r \ln(r)- \frac {5}{2} r^{3/2}\text {ArcCoth}\left(\sqrt{r}\right) \ln r - r^2 \ln \left(1 - \frac {1}{r}\right) \ln r + r^2 \text{Li}_2 \left(\frac{1}{r}\right)\right] 
\end{align}
and 
\begin{align}
a_L^{\ vac.\ pol.} &\underset{r<1} =\left( \frac{\alpha}{\pi} \right)^2 \left[- \frac{25}{36} - \frac{\pi^2}{4}r^{1/2} + 3 r - \frac{5\pi^2}{4}r^{3/2} + \left(\frac{44}{9}+ \frac{\pi^2}{3}\right)r^2 + \frac{5}{4} r^3 \Upphi\left(r,2,\frac{3}{2}\right) \nonumber\right.\\ 
&\hspace{1.8cm}- \frac{1}{4} r^3 \Upphi\left(r,2,\frac{5}{2}\right) - \frac{1}{6} \ln r + \frac{3}{2} r \ln r + \frac{1}{2}\sqrt{r}\text{ArcTanh}\left(\sqrt {r} \right)\ln r\nonumber \\ 
&\hspace{1.8cm}\left.- \frac{5}{2}r^{3/2}\text{ArcTanh}\left(\sqrt {r} \right) \ln r - r^2 \ln(1-r)\ln r + \frac{1}{2} r^2\ln ^2 r - r^2 \text{Li}_2 \left(r\right) \right],\label{amu2}
\end{align}
where $\Phi(z,s,a)$ denotes the Lerch's function  $\Phi (z,s,a)\doteq\sum_{n=0}^{\infty}\frac{z^n}{(a+n)^s}$ for $|z|<1$ and $a\neq 0, -1,-2,...$, are analytic continuations of one another (see Figure \ref{ac}, where the point $r=1$ corresponds to the case of identical leptons $\ell=L$).

In this paper, we are interested in the generalization of the results of this section to Mellin-Barnes integrals of higher dimension. We will be more particularly concerned with the extension of the $\Delta=0$ case which seems to be the main situation met in the perturbative calculations of particle physics and we will see that for an integral which fulfills this property, several convergent multiple series representations may be obtained. As a rule, these series are analytic continuations of each other and we will show that it is possible to get the regions of convergence of these multiple series before computing their full expressions.

\section{Twofold Mellin-Barnes integrals\label{twofoldMBint}}

In this section we consider twofold Mellin-Barnes integrals depending on two (complex) parameters and we present a method which allows to derive straightforwardly and systematically:\\

$\bullet$ different sets of poles which correspond to different convergent (double) series representations of a given integral,\\

$\bullet$ the regions of convergence of all these series (without an \textit{a priori} complete knowledge of their general term),\\

$\bullet$ the general term of each series (this last step may therefore be performed, if necessary, once the domain of convergence of interest has been found).\\

This systematic procedure will be illustrated with some integrals which appear in the intermediate calculations of the two-loop hexagon Wilson loop in $\mathcal{N}=4$ SYM theory (see \cite{DelDuca:2010zg}), and on the example of the scalar box integral with one external mass (in the latter case we will also show how one can resolve the $\epsilon$ singularities from a graphical point of view with this method). 

We saw, in the case of the two onefold integrals explicitly calculated in section \ref{Onefold}, that the integration contours were closed to the left or to the right, the poles where one had to compute residues in the $s$-plane being either on the negative or on the positive $\textrm{Re}\ s$ axis. We also saw that when $\Delta=0$, by closing the contour to the left or to the right, one obtains two convergent series. The generalization at the twofold level of this particular case does not lead to only two series but in fact to several. These convergent (double) series are, as a rule, analytic continuation of each other and the poles where one has to take the residues which correspond to each series belong to some specific sectors in the $(\textrm{Re}\ z_1,\textrm{Re}\ z_2)$ plane, where $z_1$ and $z_2$ are the two integration variables \cite{Tsikh:1998}. Of course in general these sectors (the so-called cones, in the following) are not accessible by naively closing the two contours of the twofold integral to the left or to the right, but we will show that it is anyhow easy to find the cones by simple intuitive rules or with the help of a geometrical procedure.

To expose the method in all details, we will first consider the simple twofold Mellin-Barnes integral $R_{-1}(u_1,u_2,u_3)$ (first line of eq. (3.6) in \cite{DelDuca:2010zg}). We will show that for this integral five different cones may be obtained and, after recalling the way that residues which correspond to a given cone are calculated, that each of the five associated double series converges in a particular region of the ($|u_1|$, $|u_2|$) plane\footnote{Indeed, in can be seen in \cite{DelDuca:2010zg} that $R_{-1}(u_1,u_2,u_3)$ does not depend on $u_3$, therefore in the following we do not keep the notation of these authors and, instead, we will write $R_{-1}(u_1,u_2)$.} that we will give explicitely.  Then, we will detail how to get two of these five series from their respective cone.

We insist to say that one of the important points of this paper is that the region of convergence of a given series may be obtained without the full knowledge of the general term of this series. Therefore, in the cases where one knows the numerical value of the expansion parameters (like it is, for instance, in the QED calculations done in \cite{Aguilar:2008qj},  
where the parameters are lepton mass ratios) this allows to select the relevant cone right from the beginning without having to compute everything. However, we will also see in the example of $R_{-1}(u_1,u_2)$ that there exists a (semi) infinite region in the ($u_1$, $u_2$) plane that is not accessible to any of the five series representations that our method allows to find. Similar "out of reach" regions will also be met in the case of the two other integrals under study in this section. 
The problem of the direct obtaining of a series representation which converges in these (at first sight) inaccessible regions (without using the traditional approach, \textit{i.e} without performing analytic continuation of the resummation of the known result corresponding to one of the accessible regions) is presently under investigation with a complementary original method but will not be treated in this paper.

After the calculation of $R_{-1}(u_1,u_2)$ we will consider in this section the slightly more complicated twofold Mellin-Barnes integral $R_{-1-z_1}(u_1,u_2,u_3)$ (second line of eq. (3.6) in \cite{DelDuca:2010zg}) and the scalar box with one external mass. This will allow us to explain, among other things, a technical point of the method which may be important to underline for the reader and which concerns "spurious" sets of poles which, if misinterpreted, could be thought as being part of the set of relevant poles inside a given cone. These undesired singularities should however not be taken into consideration in the calculations and we will see that it is easy to distinguish them. Moreover we should add that if one were anyway trying to compute them by mistake, one would fall on undecidable ill-defined results which would therefore signal their presence. Apart from this fact, the $R_{-1-z_1}(u_1,u_2,u_3)$ and scalar box integrals are not harder to compute than $R_{-1}(u_1,u_2)$ since it is worth recalling that our method is a systematic one. In particular, we will not need any introduction of Euler integral representation or other integral manipulations as was performed in \cite{DelDuca:2010zg}: keeping right from the beginning the original twofold Mellin-Barnes representations and applying our method directly on them allows to obtain the different convergent double series straightforwardly.

\subsection{The $R_{-1}(u_1,u_2)$ integral \label{R-1}}

In the following we consider $u_1$ and $u_2$ as complex numbers.
Let us start by giving the definition of $R_{-1}(u_1,u_2)$ (first line of eq. (3.6) in \cite{DelDuca:2010zg}):
\begin{align}
\hspace{-2cm}
&R_{-1}(u_1,u_2) \nonumber \\
&= -\int\limits_{c-i\infty}^{c+i\infty}{d z_1\over 2\pi i} \int\limits_{d-i\infty}^{d+i\infty} {d z_2\over 2\pi i}\,
u_1^{z_1}\, u_2^{z_2}\,\Gamma^2 \left(-z_1\right)\, \Gamma\left(z_1\right)\, \Gamma^2 \left(-z_2\right)\, \Gamma \left(z_2\right)\Gamma \left(z_1+z_2+1\right)\;,
\label{Rmoins1}
\end{align}
where $c=-\frac{1}{3}$ and  $d=-\frac{1}{4}$. Note that there is a symmetry of this integral by exchanging $u_1$ and $u_2$ since we could take $c=d=-\frac{1}{3}$ or $c=d=-\frac{1}{4}$ without changing (\ref{Rmoins1}).

It will of course allow us to obtain some of the series representations directly from others.

\begin{figure}[h]
\begin{center}
\includegraphics[width=0.5\textwidth]{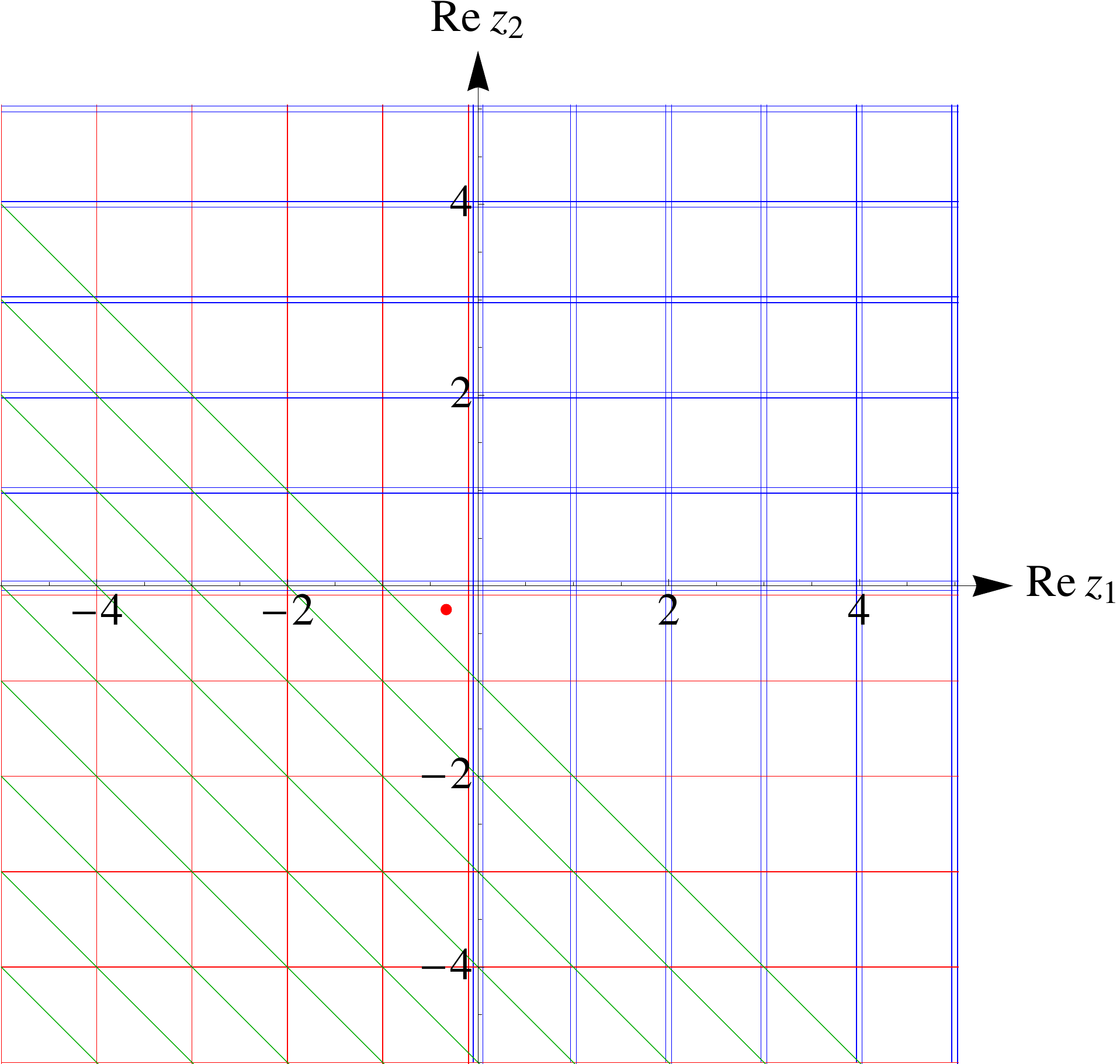}
\end{center}
\caption{Singular structure of the integrand of $R_{-1}(u_1,u_2)$. Coinciding singular lines have been slightly separated for convenience. The red dot is the point $(c,d)$. \label{plane}}
\end{figure}

The singular structure of the integrand of (\ref{Rmoins1}), in the $(\textrm{Re}\ z_1,\textrm{Re}\ z_2)$ plane, is represented in Figure \ref{plane}. It is easy to see that, with $n\in\mathds{N}$, it corresponds to a drawing of the (singular) lines $-z_1=-n$ (the vertical blue lines on figure \ref{plane}) and $-z_2=-n$ (horizontal blue lines) which both appear twice, of the lines $z_1=-n$ (vertical red lines), $z_2=-n$ (horizontal red lines) and of the lines $z_1+z_2+1=-n$ (oblique green lines).

At the intersections of these singular lines are the singular points where one will have to compute residues of the integrand, the resummation of these in the different cones\footnote{As we said above, there are cases where some of the intersections are irrelevant, but this does not occur in the computation of $R_{-1}$.
} leading to various convergent (double) series representations of (\ref{Rmoins1}). 

It is clear from Figure \ref{plane} that there is only a finite number of intersection types (and a finite number of domains in the $(\textrm{Re}\ z_1,\textrm{Re}\ z_2)$ plane where one will always find the same intersection type), which makes the resummations possible. Moeover, apart from a few points as for instance the point $(0,0)$, which is by itself its own type, to a given type of intersection correspond one series of residues. For instance, as we will see, in the highest right-corner of Figure \ref{plane}, for $(z_1,z_2)=(1+m,1+n)$ where $m\geq 0$ and $n\geq 0$ are integers, the intersection type consists in a crossing of two horizontal and two vertical (blue) lines and corresponds to a double series. Still in this corner, for $(z_1,z_2)=(0,1+m)$ (resp. $(z_1,z_2)=(1+m,0)$) where $m\geq 0$ is an integer, the intersection type consists in a crossing of two horizontal blue lines (resp. two vertical blue lines) and three vertical lines, one red and two blue (resp. three horizontal lines, same colors). The latter correspond obviously to single series, since the parametrization of these singular points do only depend on the $m$ summation indice. For $(z_1,z_2)=(0,0)$, the intersection type consists in a crossing of three horizontal and three vertical lines (one red and two blue in both cases). In fact, if one takes the residues at the intersections which belong to this complete quarter of plane, one matches exactly equation (C.2) of \cite{DelDuca:2010zg} which has been obtained by closing both contours of (\ref{Rmoins1}) to the right. 

We will show now that one can obtain this cone as a particular result of the application of a simple rule which allows more generally to exhibit various other sets of singular points from a given twofold Mellin-Barnes integral (\textit{i.e.} other compatible cones), corresponding to other convergent series representations of the considered integral.

\subsubsection{Determination of the cones\label{det_cone}}

\begin{figure}[h]
\hspace{1.2cm}
\begin{tabular}{cc}
\includegraphics[width=0.4\textwidth]{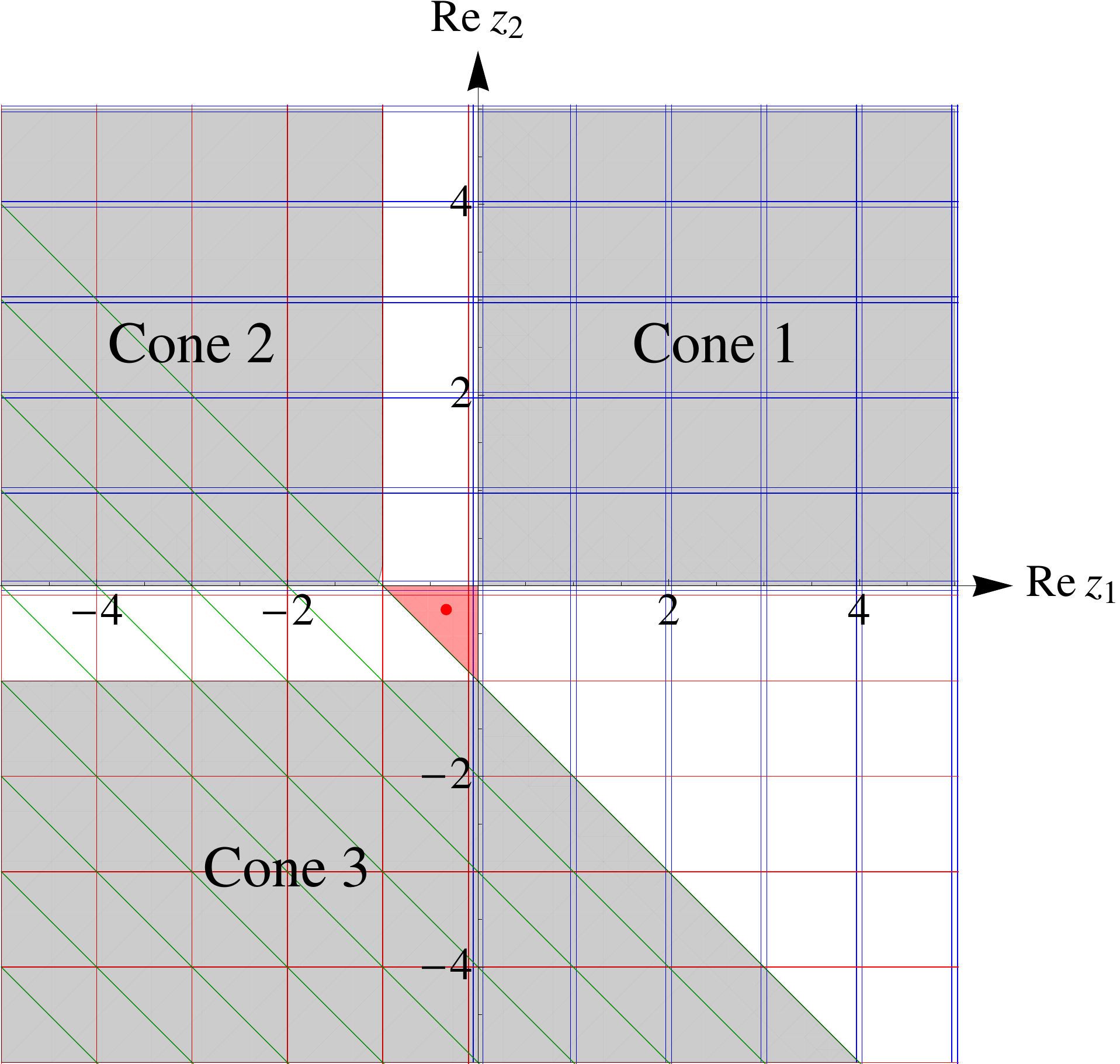} &
\includegraphics[width=0.4\textwidth]{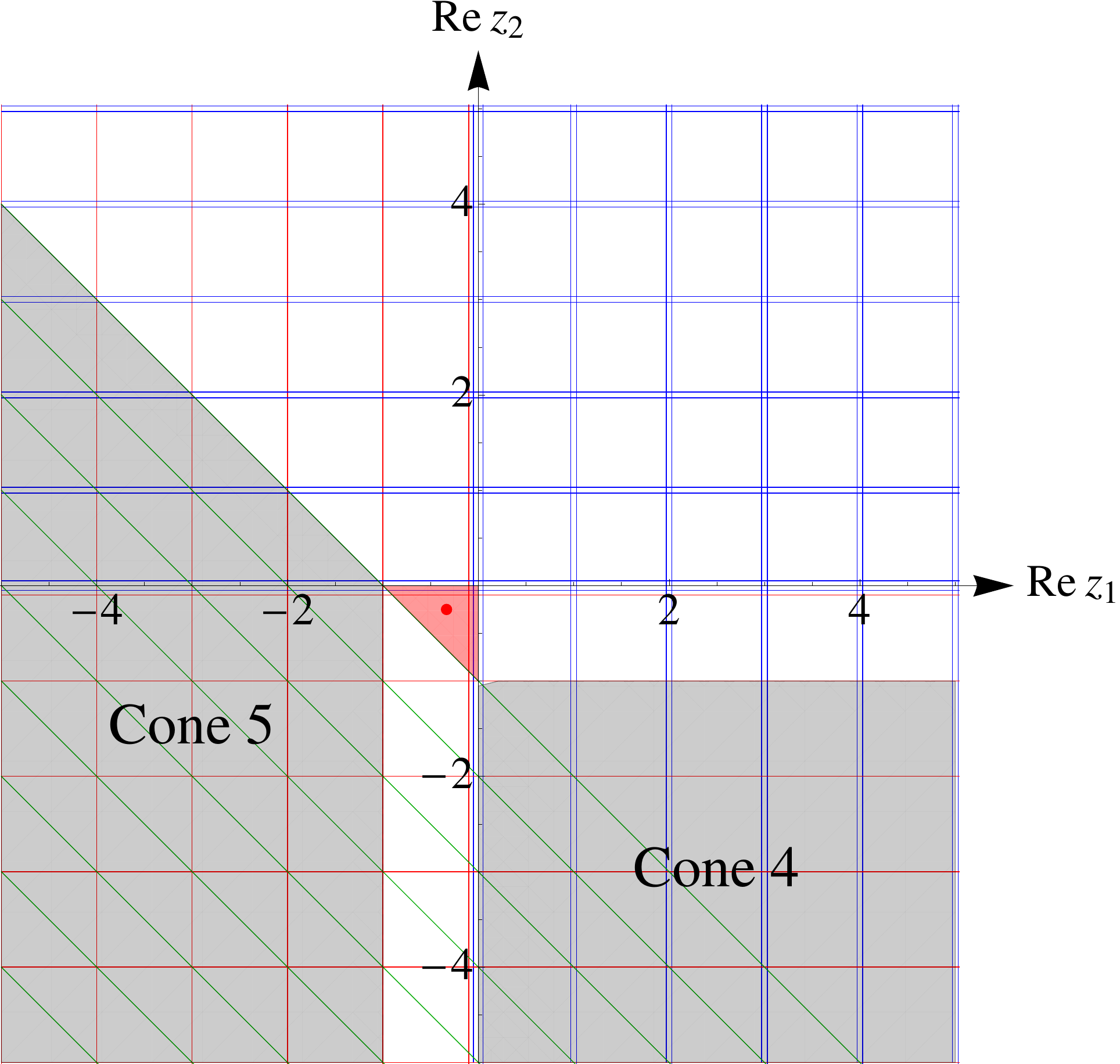} \\
\end{tabular}
\caption{In grey, the five different cones  of $R_{-1}(u_1,u_2)$. \label{conesR1}}
\end{figure}

First, we have to find the fundamental polygone of (\ref{Rmoins1}), \textit{i.e} the region of convergence of the integral in the $(\textrm{Re}\ z_1,\textrm{Re}\ z_2)$ plane where a continuous change of the point $(c,d)$ from its initial value $\left(-\frac{1}{3},-\frac{1}{4}\right)$ may be performed without modifying the value of the integral. It is dictated by the simultaneous constraints $\textrm{Re}\ z_1<0$, $\textrm{Re}\ z_2<0$ and $\textrm{Re}(1+z_1+z_2)>0$ (leading to the red triangle in Figure \ref{conesR1}).

Then, by a simple inspection of the constraints implied by the different couples\footnote{A given couple of gamma functions should of course not depend on only one of the variables $z_1$ or $z_2$.} of gamma functions of the integrand of (\ref{Rmoins1}), whose singular lines intersect each other in some regions of the $(\textrm{Re}\ z_1,\textrm{Re}\ z_2)$ plane, it is straightforward to get the different cones. 
Indeed, from the eight different couples of gamma functions which may be obtained, only five lead to non-trivial regions which are compatible with the fundamental triangle, in the following way:

\vspace{0.3cm}

$\bullet$ $(\Gamma \left(-z_1\right),\Gamma \left(-z_2\right))$ leads to Cone 1: $\textrm{Re}\ z_1>0$, $\textrm{Re}\ z_2>0$.

\vspace{0.3cm}

$\bullet$ $(\Gamma \left(z_1\right),\Gamma \left(-z_2\right))$ leads to Cone 2: $\textrm{Re}\ z_1<-1$, $\textrm{Re}\ z_2>0$.

\vspace{0.3cm}

$\bullet$ $(\Gamma \left(z_2\right),\Gamma \left(z_1+z_2+1\right))$ leads to Cone 3: $\textrm{Re}\ z_2<-1$, $\textrm{Re}(1+z_1+z_2)<0$.

\vspace{0.3cm}

$\bullet$ $(\Gamma \left(-z_1\right),\Gamma \left(z_2\right))$ leads to Cone 4: $\textrm{Re}\ z_1>0$, $\textrm{Re}\ z_2<-1$.

\vspace{0.3cm}

$\bullet$ $(\Gamma \left(z_1\right),\Gamma\left(z_1+z_2+1\right))$ leads to Cone 5: $\textrm{Re}\ z_1<-1$, $\textrm{Re}(1+z_1+z_2)<0$.

\vspace{0.3cm}

The first three cones are depicted on the first picture of Figure \ref{conesR1}, the two others on the second picture.

However, it may be more convenient for the reader to know that the cones may also be obtained following a simple general geometrical procedure \cite{Tsikh:1998}. As described in this reference, one has to consider the value of the vector 
\begin{equation}
{\bf\Delta}=
\begin{pmatrix}
\sum_{j=1}^{m}a_j-\sum_{k=1}^{p}c_k\\ 
\\
\sum_{j=1}^{m}b_j-\sum_{k=1}^{p}d_k\,
\end{pmatrix},
\end{equation}
which generalizes the quantity $\Delta$ of section \ref{Onefold}, and where the real coefficients $a_j$, $b_j$, $c_k$ and $d_k$ are defined in the general twofold Mellin-Barnes integral as follows
\begin{equation}
I(x,y) = \int\limits_{c-i\infty}^{c+i\infty} {d z_1\over 2\pi i}\int\limits_{d-i\infty}^{d+i\infty}{d z_2\over 2\pi i}\,
\; x^{-z_1}\; y^{-z_2}\; \frac{\prod_{j=1}^m\Gamma(a_jz_1+b_jz_2+e_j)}{\prod_{k=1}^p\Gamma(c_kz_1+d_kz_2+f_k)}\,,
\end{equation}
where $c$ and $d$ are chosen such that the integration contours do not intersect the poles of the integrand. 

Now the statements at the end of section \ref{Onefold} may be generalized as follows \cite{Tsikh:1998}:
\\

$\bullet$ If ${\bf\Delta}=(0,0)$, we are in a so-called degenerate case where several relevant cones coexist, corresponding to different convergence domains. Therefore, several convergent double series expansions may be obtained which, as a rule, are analytic continuations of each other.
\\

$\bullet$ If ${\bf\Delta}\ne(0,0)$ and none of the vectors $(a_j,b_j)$ is proportional\footnote{The case where ${\bf\Delta}\ne (0,0)$ and at least one of the $(a_j,b_j)$ is proportional to ${\bf\Delta}$ is also a degenerate case (in \cite{Dorokhov:2008cd} a quantum field theory calculation with this particular type of degeneracy has been considered).} to ${\bf\Delta}$, then we are in a simpler non-degenerate situation where there is only one double series representation which converges absolutely for all non-zero complex values of $x$ and $y$. 
\\

Looking at (\ref{Rmoins1}), one concludes that $R_{-1}(u_1,u_2)$ has ${\bf\Delta}=(0,0)$. In fact, it seems that in the high energy physics perturbative computations, one often faces twofold Mellin-Barnes integrals with ${\bf\Delta}=(0,0)$ (in this paper we only consider this case).

Let us now come back to the geometrical procedure that we mentioned above for the determination of the different cones, in order to describe it in detail.
For this one has to draw on Figure \ref{plane} an arbitrary straight line $l$ that goes through the point $(c,d)$, in our case the point $(-\frac{1}{3},-\frac{1}{4})$. Obviously there is an infinity of such lines, but only a finite number of compatible cones can be deduced from this infinite set. 

Once the straight line $l$ is drawn, the extraction of the corresponding compatible cone(s)\footnote{In some cases one may find two such cones for a given $l$: one in each of the two half-planes that are separated by $l$.} follows from the following rule: the singular lines that cross $l$ on one of its "sides" (the point $(c,d)$ defines the separation point between each of the two "sides" of $l$) intersect the singular lines that cross $l$ on its other side at some points which are inside the compatible cone \cite{Tsikh:1998}.

\begin{figure}[h]
\begin{center}
\includegraphics[width=0.5\textwidth]{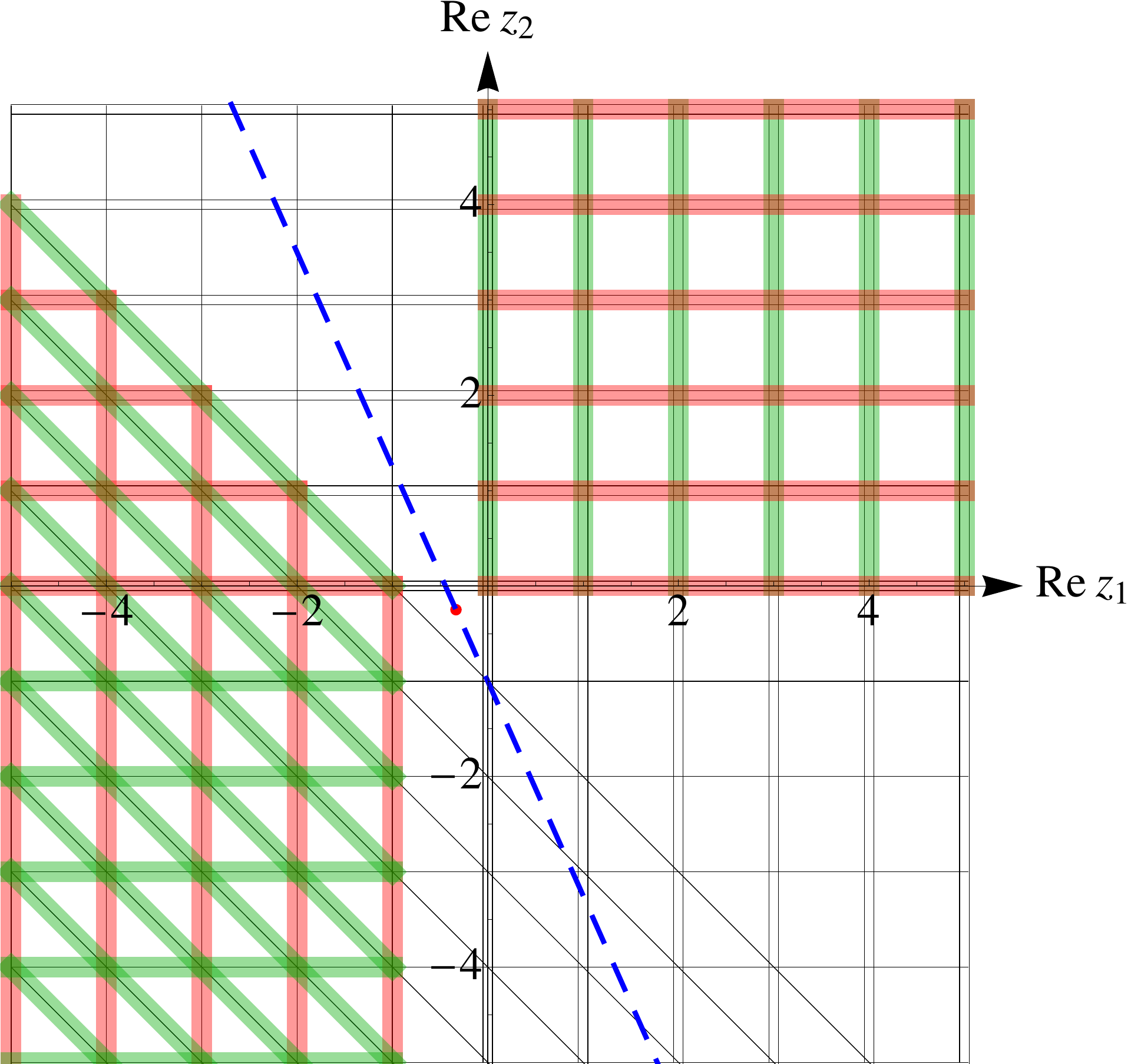}
\end{center}
\caption{Cone 1 and Cone 5.\label{cones1}}
\end{figure}

In Figure \ref{cones1} we show one possible straight line from which two compatible cones emerge (the colored regions), the right 90 degree cone corresponding, as we already said, to the results given in eqs (C.2), (C.3), (C.4) and (C.5) in Appendix C of \cite{DelDuca:2010zg}.

Straightforwardly, the three other cones of Figure \ref{conesR1} may be obtained by a rotation of the $l$-line around $(c,d)$.

All cones being found, the integral $R_{-1}(u_1,u_2)$ may then be computed as follows (we use some of the notations of \cite{Tsikh:1998} for later purpose):
\begin{align}
R_{-1}(u_1,u_2&)=\frac{1}{(2\pi i)^2}\int\limits_{\gamma+i\mathds{R}^2}\,\omega\\
&\underset{(u_1,u_2)\in \mathcal{R}_i}=\sum_{\text{Cone}\ i}
\Res{\omega}\;,
\label{Rmoins1_Tsikh}
\end{align}
where $\omega$ is the meromorphic 2-form
\begin{equation}
\label{omega_meromorf}
\omega=u_1^{z_1}\, u_2^{z_2}\,\Gamma^2 \left(-z_1\right)\, \Gamma\left(z_1\right)\, \Gamma^2 \left(-z_2\right)\, \Gamma \left(z_2\right)\Gamma \left(z_1+z_2+1\right)\ dz_1\wedge dz_2\,,
\end{equation}
$\gamma=(c,d)$, and where $\Res{.}$ is meant for residue.

The sum in (\ref{Rmoins1_Tsikh}) has to be understood as a sum over the residues taken at the intersections inside the $i^{\text{th}}$ cone and $\mathcal{R}_i$ is the region of convergence of such a sum.

Let us now explain how to get the different regions of convergence $\mathcal{R}_i$ associated to each of the five cones. 

\subsubsection{Regions of convergence\label{reg_conv_section}}

The different cones corresponding to (\ref{Rmoins1}) have been found and an obvious need now is to know in which region of the $(|u_1|,|u_2|)$ plane the corresponding double series converge.
Although it is possible to get the region of convergence of a given series once one has its general term, this order of doing things is not always interesting since in many instances one knows right from the beginning the numerical values of the expansion parameters, and therefore one would prefer to know to which region of convergence these values belong, in order not to compute all the cones one by one but only the "right-one" directly. We will show that it is possible to find the different regions of convergence without a full computation of the general terms of the corresponding double series, thanks to a nice property of the integrand of the Mellin-Barnes integrals met in high energy physics perturbative calculations. To explain this in full details, it is now necessary to make a digression in order to define several quantities that will be needed. 

\paragraph{Residues computational method} 

Inside a given cone, the first important thing to do is to consider \textit{exclusively} the intersections where at least one singular line crosses one side of the line $l$ and one singular line crosses the other side of $l$ (the sides of $l$ have been defined in section \ref{det_cone}). Indeed, it may happen that, for a given cone, part of the intersections of a singular structure are formed only by singular lines which cross the same side of $l$. These are the spurious singularities that we talked about in the introduction of section \ref{twofoldMBint}. The reason why these intersections should not be considered is simply that they do not comply with the definition that we gave in section \ref{det_cone} for a singularity to belong to a cone, and we prefer to insist on this point in order to prevent the reader from an incorrect use of the method.  In the calculation of $R_{-1}(u_1,u_2)$ no such spurious singularities will be met, but as we already said this will happen for some of the cones of $R_{-1-z_1}(u_1,u_2,u_3)$ (the small black circles in Figure \ref{conesR2} distinguish these spurious singularities) and in many other instances as for example the scalar box integral discussed later in this section (see the small black filled circles in Figure \ref{conesBox}) or the QED calculations that have been performed in \cite{Aguilar:2008qj}.

Once the possible spurious singularities have been discarded from the calculation, one has to classify the different subsets of relevant singularities (indeed, we have seen that inside a given cone, there exists in general several types of intersections). This way, one may deal with infinite subsets of residues at the same time, which may be convenient if the aim is to get an exact result. If only the first few terms of the expansion are needed, they come from the contributions of the intersections which are the closest to the fundamental polygon and, of course, they may be computed one by one (this was the strategy adopted in \cite{Aguilar:2008qj}). 

The coordinates of a particular subset are parametrized by affine functions $f$ and $g$ as
\begin{equation}
(z_1, z_2) = (f(m,n), g(m,n)),
\end{equation}
where $m$ and $n$ are integers. $f$ and $g$ are trivially obtained either by looking at the picture representing the singular structure of the integrand (Figure \ref{plane} in our present case) or by solving the system of the singular lines equations coming from two gamma functions chosen between all the gamma functions which are involved in the considered type of intersections.

Then, for the computation of the corresponding residues, the following four steps have to be performed \cite{Aguilar:2008qj, Aguilar}:
\\

$\bullet$ One first performs on (\ref{omega_meromorf}) the change of variables  $z_1\mapsto z_1+f(m,n)$ and $z_2\mapsto z_2+g(m,n)$ so as to bring the singularity to the origin $(0,0)$.
\\

$\bullet$ One then uses the generalized reflexion formula\footnote{One may also use the identity $\Gamma(-n+ z)=\frac{\Gamma(1+ z)}{ z\,\Pi_{i=1}^{n}(-i+ z)}$ but it is often more convenient to use (\ref{reflexion}).}
\begin{equation}
\Gamma(-k + z)=(-1)^k\; \frac{\pi}{\sin(\pi z)} \; \frac{1}{\Gamma(k+1-z)} = (-1)^k \; \frac{\Gamma(1+ z)\Gamma(1- z)}{ z\ \Gamma( k+1 - z)}
\label{reflexion}
\end{equation} 
(where $k\geq 0$ is an integer) in order to explicitly extract, in the denominator, the singular structure of the singular gamma functions.
\\

$\bullet$ Two different cases may then appear, depending on the type of intersection considered (or, in other words, the type of arguments of the singular gamma functions which contribute to this kind of singularity):
\\

\underline{First case:} if the singular lines crossing each other at the intersections are only horizontal and vertical lines (\textit{i.e} if the arguments of all the singular gamma functions do not depend on $z_1$ and $z_2$ simultaneously), one finds for the 2-form an expression of the type
\begin{equation}
\omega=\frac{h(z_1,z_2)}{z_1^n\ z_2^m}\ dz_1\wedge dz_2\,,\label{Cauchy_form}
\end{equation}
where $n$ and $m$ are positive integers and the function $h$ is analytic at the origin.

\underline{Second case:} if at least one of the singular lines is neither vertical nor horizontal or, equivalently, if there are some oblique lines at the intersection points (\textit{i.e} if the argument of at least one of the singular gamma functions depends on $z_1$ and $z_2$ simultaneously), one finds the more general expression
\begin{equation}
\label{type2}
\omega=\frac{h(z_1,z_2)}{(z_1+a\,z_2)^p(z_1-b\,z_2)^q\ z_1^r\ z_2^s}\ dz_1\wedge dz_2\,,
\end{equation}
where $a$, $b$ are positive numbers\footnote{Notice that in the case of the Mellin-Barnes integrals met in perturbative quantum field theory these positive numbers are in fact integers. It is important since this is a condition required by Horn's theorem to find the regions of convergence of the series (see Appendix \ref{ConvHorn}).} and $p$, $q$, $r$ and $s$ positive integers (possibly null for $q$ if $p\neq0$ or conversely, \textit{idem} for $r$ and $s$).

To find the expression ot the residue of the form (\ref{Cauchy_form}) one may directly apply the Cauchy formula
\begin{equation}
\label{cauchy}
\Res{\omega}=\frac{1}{(n-1)!(m-1)!}\frac{\partial^{n+m-2}\,h(z_1,z_2)}{\partial z_1^{n-1}\partial z_2^{m-1}}\Bigg\vert_{(0,0)}\doteq\frac{1}{(n-1)!(m-1)!}h^{(n-1,m-1)}(0,0),
\end{equation} 
but one has to care about the fact that an overall sign may contribute, related to the (clockwise or anti-clockwise) integration contours. As we will see, to find this sign it is more convenient to consider (\ref{Cauchy_form}) as a trivial case of the more general situation (\ref{type2}) with\footnote{We stress that this sign is undecidable for spurious singularities of the type (\ref{Cauchy_form}) when considered as a trivial case of (\ref{type2}). This reflects their spuriousity.} $p=q=0$.

On (\ref{type2}) one cannot trivially apply the Cauchy formula, one first has to use the so-called transformation law \cite{griffiths, Aguilar:2008qj, Aguilar}. 
The transformation law applied to (\ref{type2}) reads
\begin{equation}
\label{TL}
\Res{\frac{h(z_1,z_2)}{(z_1+az_2)^p(z_1-bz_2)^q\ z_1^r\ z_2^s}\ dz_1\wedge dz_2}=\Res{\frac{h(z_1,z_2)\det A}{z_1^n\ z_2^m}\ dz_1\wedge dz_2}\;,
\end{equation}
if one can find a $2\times 2$ matrix $A$ whose elements are analytic and which satisfies $A \boldsymbol{f}=\boldsymbol{g}$ where $\boldsymbol{g}=(z_1^i,z_2^j)^T$ ($i$ and $j$ are  positive chosen integers) and $\boldsymbol{f}=(f_1,f_2)^T$, $f_1$ being the product of a \emph{specific} subset\footnote{This point is very important
 since an arbitrary subset gives wrong results.} of the following terms: $(z_1+az_2)^p$, $(z_1-bz_2)^q$, $\ z_1^r$ and $\ z_2^s$, while $f_2$ is composed of the product of the terms which do not belong to $f_1$ (in some cases one of these specific subsets may be reduced to only one term, then there is of course no need for a product). Moreover one has to check that $f^{-1}(0,0)=(0,0)$ and $g^{-1}(0,0)=(0,0)$. Once this has been done, (\ref{cauchy}) may be applied on the right hand side of (\ref{TL}).
 
\begin{figure}[h]
\hspace{1.2cm}
\begin{tabular}{cc}
\includegraphics[width=0.43\textwidth]{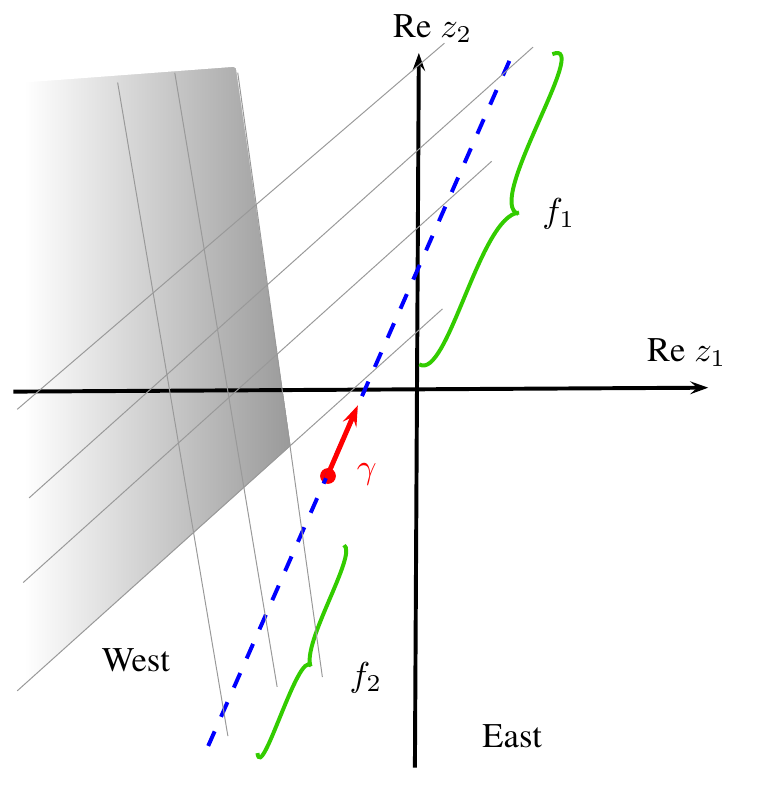} &
\includegraphics[width=0.4\textwidth]{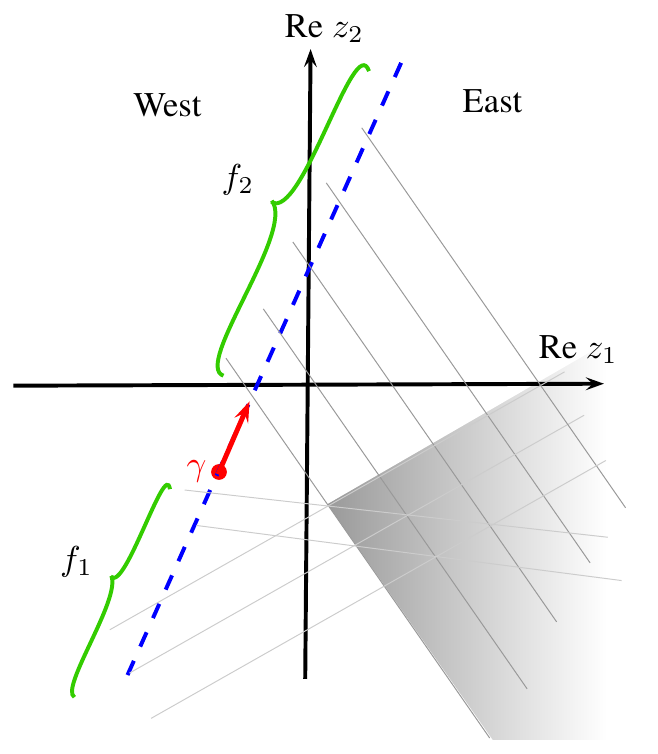} \\
\end{tabular}
\caption{Definition of $\boldsymbol{f}$.\label{orientation}}
\end{figure}

In fact the expression of $f_1$ and $f_2$ are precisely defined in Figure \ref{orientation} which summarizes the two different possibilities. 
One can see that the red direction vector defines the orientation of the $l$-line which allows to consider two half-planes: the "West" and the "East" half-planes (the direction vector gives the "North"). If the cone under study belongs to the East half-plane, all singular lines which cross the $l$-line on the side where the arrow of the direction vector lies contribute to $f_2$. All others singular lines contribute to $f_1$. If the cone considered belongs to the West half-plane, this is the converse situation.

The correct sign of the residue, which is automatically given with this approach in the case of (\ref{type2}), is then also easily obtained for Cauchy type residues (\ref{Cauchy_form}) by the value of $\det A=\pm 1$ in the trivial application of the transformation law
\begin{equation}\label{trivialLaw1}
\begin{pmatrix}
1 & 0  \\
0 & 1 
\end{pmatrix}\;
\begin{pmatrix}
z_1^n\\
z_2^m
\end{pmatrix}=
\begin{pmatrix}
z_1^n\\
z_2^m
\end{pmatrix}
\end{equation}
or
\begin{equation}\label{trivialLaw2}
\begin{pmatrix}
0 & 1  \\
1 & 0 
\end{pmatrix}\;
\begin{pmatrix}
z_2^m\\
z_1^n
\end{pmatrix}=
\begin{pmatrix}
z_1^n\\
z_2^m
\end{pmatrix},
\end{equation}
depending on the vector $f$ that has been found.

This definition of the transformation law explains why one cannot decide for the value of the residues in the case of the spurious singularities that we mentioned above: in this particular case the $f$ vector cannot be obtained since the singular lines cross the $l$-line only on one of its side. Therefore the transformation law cannot be applied, even in its trivial versions (\ref{trivialLaw1}) or (\ref{trivialLaw2}).

In our present case of study, it is clear that for all residues associated to the upper cone of Figure \ref{cones1} (Cone 1), a direct application of Cauchy formula is possible, the sign of the residues being fixed by eq. (\ref{trivialLaw1}). On the contrary, in the lower cone (Cone 5), one will have to apply a "non-trivial" transformation law as an intermediate step in the calculation of each of the different contributions.
\\

$\bullet$ At the end, the partial derivatives of (\ref{cauchy}) or (\ref{TL}) may be easily obtained by hand (see Appendix A for a formulary and some notations used in the calculations) in terms, among others, of polygamma functions $\psi^{(k)}$ of different orders which may be translated as harmonic numbers $H_M=\sum_{i=1}^M\frac{1}{i}$ and their generalizations $H^{(k)}_M=\sum_{i=1}^M\frac{1}{i^k}$ \cite{FlaSal} if desired, using the relations 
\begin{equation}
\label{PsiH}
\psi(N+1)=H(N)-\gamma_E
\end{equation}
and
\begin{equation}
\label{PsiNHN}
\psi^{(k)}(N+1)=(-1)^{k+1}k!\left[\zeta_{k+1}-H^{(k+1)}_{N}\right],
\end{equation}
where $N$ and $k$ are integers.

The infinite double sums of the final results may in turn be expressed in terms of nested harmonic sums if one wants to rewrite them as a combination of multiple polylogarithms with the help of the different algorithms developed in \cite{Moch:2001zr}.

\paragraph{Back to the regions of convergence\label{para_conv}}

Remember that our aim is to find the region of convergence associated to a given cone without a full calculation of the corresponding double series. For this, we need some of the tools described in the last paragraph. 

It is easy to see that from the use of (\ref{reflexion}), after the change of variables performed on $\omega$, the $h$ function in (\ref{Cauchy_form}) or (\ref{type2}) has the form (\ref{h_formulary}) with $x=\frac{1}{u_1}$, $y=\frac{1}{u_2}$, $s=z_1$, $t=z_2$ and where $e_j$ and $f_k$ are linear combinations of $m$ and/or $n$. Moreover, from (\ref{cauchy}) as well as eqs. (\ref{h10}), (\ref{h11}), (\ref{h20}), (\ref{h02}), (\ref{h21}), (\ref{h12}) and their higher-order derivatives, one can infer that any residue may be written as $h(0,0)$ multiplied by a combination of polygamma functions of different orders. The latter depend also on the summation indices $m$ and/or $n$. The case where both $m$ and $n$ are involved in the argument of the gamma functions of $h(0,0)$ is in general the most restrictive concerning the regions of convergence and corresponds of course to the occurence of a double series. We proved in Appendix \ref{ConvHorn} section \ref{AlmostHorn} that the polygamma functions do not affect the region of convergence that would be obtained by simply considering that the general term of the double series is $h(0,0)$.

Therefore, since $h(0,0)$ is a ratio of gamma functions, and since in the quantum field theory perturbative calculations the coefficients of the integration variables in the argument of these gamma functions are integers, one may use Horn's theorem to find the regions of convergence (see Appendix B, section \ref{horn_conv}). We show how to proceed in the following.

\subparagraph{Cone 1}

To Cone 1, as we said above, corresponds only one double series, coming from the residues of the poles at $(1+m,1+n)$. We then have
\begin{equation}
\left. h(0,0)\right\vert_{(1+m,1+n)}=-\frac{u_1^{m+1}u_2^{n+1}\ \Gamma(3+m+n)}{(1+m)(1+n)\Gamma(m+2)\Gamma(n+2)}.
\end{equation}
By eqs. (\ref{fg}) and (\ref{FG}) we find
\begin{equation}\label{FCone1}
\mathrm{F}(m,n)=\frac{m+n}{m}
\end{equation}
and
\begin{equation}\label{FCone2}
\mathrm{G}(m,n)=\frac{m+n}{n}.
\end{equation}
Since we are in the $p=r$ and $q=s$ case (in the notation of Appendix \ref{ConvHorn} section \ref{horn_conv}), the region of convergence of the double series is then given by 
$\mathcal{R}_1=\mathcal{C}\cap\mathcal{D}$
where the domains $\mathcal{C}$ and $\mathcal{D}$ are defined as  (see Appendix \ref{ConvHorn})
\begin{equation}
\mathcal{C} = \left\{ (|u_1|, |u_2|) \; \; \big \vert \;\;  0 < |u_1| < \frac{1}{\left|\mathrm{F}(1,0)\right|} \; \;  \text{and} \; \;  0 < |u_2| < \frac{1}{\left|\mathrm{G}(0,1)\right|} \right\}
\end{equation}
and 
\begin{equation}\label{Dregion}
\mathcal{D} = \left\{ (|u_1|, |u_2|) \; \; \big \vert \;\; \forall (m,n)\in \mathds{R}_+^2: 0 < |u_1| < \frac{1}{\left|\mathrm{F}(m,n)\right|} \; \;  \text{or} \; \;  0 < |u_2| < \frac{1}{\left|\mathrm{G}(m,n)\right|} \right\}.
\end{equation}
Therefore, one has
\begin{equation}
\mathcal{C} = \left\{0 < |u_1| < 1 \; \;  \text{and} \; \;  0 < |u_2| < 1 \right\}
\end{equation}
and
\begin{equation}\label{DregionCone1}
\mathcal{D} = \left\{  |u_1| +|u_2|< 1 \right\}.
\end{equation}
Indeed, (\ref{DregionCone1}) fulfills the disjunction constraints of (\ref{Dregion}): if $0 < |u_1| < \frac{1}{\left|\mathrm{F}(m,n)\right|}$ and $0 < |u_2| < \frac{1}{\left|\mathrm{G}(m,n)\right|}$ it is straightforward to prove that (\ref{FCone1}) and (\ref{FCone2}) lead to $|u_1| +|u_2|< 1$. Now suppose that $0 < |u_1| < \frac{1}{\left|\mathrm{F}(m,n)\right|}$ is false, \textit{i.e} that $|u_1| > \frac{1}{\left|\mathrm{F}(m,n)\right|}$. Then, inserting this in the relation $|u_1| +|u_2|< 1$ one obtains, still by (\ref{FCone1}) and (\ref{FCone2}), that  $0 < |u_2| < \frac{1}{\left|\mathrm{G}(m,n)\right|}$.

The single series corresponding to the poles at $(1+m,0)$ and $(0,1+m)$ do not lead to more restrictive constraints since we have 
\begin{equation}
\left. h(0,0)\right\vert_{(1+m,0)}=-\frac{u_1^{m+1}}{1+m}
\end{equation}
and
\begin{equation}
\left. h(0,0)\right\vert_{(0,1+m)}=-\frac{u_2^{m+1}}{1+m}
\end{equation}
from which, by d'Alembert ratio test, one obtains respectively  $\{ |u_1|< 1\}$ and $\{|u_2|< 1\}$.

\begin{figure}[h]
\begin{center}
\includegraphics[width=0.5\textwidth]{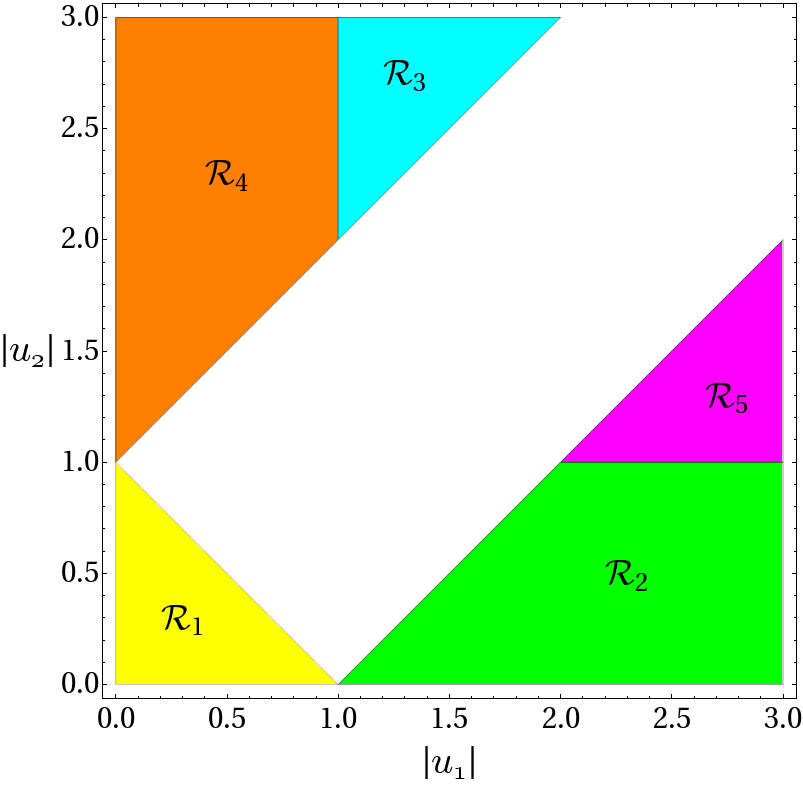}
\end{center}
\caption{Regions of convergence of the series representations of $R_{-1}(u_1,u_2)$.\label{reg_conv_R1}}
\end{figure}

The region of convergence of Cone 1 is then given by $\mathcal{R}_1=\left\{  |u_1| +|u_2|< 1 \right\}$, and it is represented in yellow in Figure \ref{reg_conv_R1}. Let us remind that we did not need the full computation of the general term of the double series, which is given in (\ref{R1type2}), to get its region of convergence.

From the symmetry of the integral $R_{-1}(u_1,u_2)$ that we mentioned above and which is also visible in Figure \ref{conesR1}, it is obvious to deduce that the region of convergence of Cone 2 may be obtained from the one of Cone 4, and that the region of convergence of Cone 3 may be obtained from the one of Cone 5 (or conversely).

\subparagraph{Cone 4 and Cone 2}
In the case of Cone 4 we have three types of intersections lying at $(1+m+n,-1-m)$, $(1+m,-2-m-n)$ and $(0,-1-m)$. Therefore we have two double series and one single series. 

The single series gives the constraint $\{|u_2|> 1\}$. The series associated to the poles at $(1+m+n,-1-m)$ has
\begin{equation}
\left. h(0,0)\right\vert_{(1+m+n,-1-m)}=(-1)^m\left(\frac{u_1}{u_2}\right)^{m+1}u_1^n\ \frac{\Gamma(1+m+n)\Gamma^2(1+m)\Gamma(1+n)}{\Gamma^2(2+m+n)\Gamma(2+m)}.
\end{equation}
Therefore
\begin{equation}
\mathrm{F}(m,n)=-\frac{m}{m+n}
\end{equation}
and
\begin{equation}
\mathrm{G}(m,n)=\frac{n}{m+n}
\end{equation}
and we find
\begin{equation}
\mathcal{C}_1 = \left\{0< |u_1| < |u_2| \; \;  \text{and} \; \;  0 < |u_1| < 1 \right\}
\end{equation}
and
\begin{equation}
\mathcal{D}_1 = \left\{  |u_1| -1< |u_2| \right\}.
\end{equation}
Using the same procedure, one finds for the last series (poles at $(1+m,-2-m-n)$)
\begin{equation}
\mathcal{C}_2 = \left\{0< |u_1| < |u_2| \; \;  \text{and} \; \;   1<|u_2| \right\}
\end{equation}
and
\begin{equation}
\mathcal{D}_2 = \left\{  |u_1| +1< |u_2| \right\}.
\end{equation}
The region of convergence of Cone 4 is then $\mathcal{R}_4=\mathcal{C}_1\cap\mathcal{D}_2$ (the orange region in Figure \ref{reg_conv_R1}).

This immediately gives the region of convergence of Cone 2 (the green region in Figure \ref{reg_conv_R1}) as 
\begin{equation}
\mathcal{R}_2=\left\{0< |u_2| < |u_1| \; \;  \text{and} \; \;  0 < |u_2| < 1 \right\}\cap\left\{  |u_2| +1< |u_1| \right\}.
\end{equation} 

\subparagraph{Cone 3 and Cone 5}
The same procedure applied to Cone 3 gives
\begin{equation}
\mathcal{R}_3=\left\{1< |u_1| \; \;  \text{and} \; \;  1 < |u_2| \right\}\cap\left\{  |u_1| +1< |u_2| \right\}
\end{equation}
(the cyan region in Figure \ref{reg_conv_R1}) from what we deduce that
\begin{equation}
\mathcal{R}_5=\left\{1< |u_1| \; \;  \text{and} \; \;  1 < |u_2| \right\}\cap\left\{  |u_2| +1< |u_1| \right\}
\end{equation}
(the magenta region in Figure \ref{reg_conv_R1}).\\

In conclusion we see that the symmetry of the integral (\ref{Rmoins1}) is of course also reflected in the regions of convergence of Figure \ref{reg_conv_R1}. However one may notice that a semi-infinite white band is not reachable by any of the series representations that our method allows to obtain. In fact we believe that it is possible to extract, from the Mellin-Barnes representation (\ref{Rmoins1}), other series that would probably fill this gap, from a complementary method which is presently under investigation.

We checked that the regions of convergence $\mathcal{R}_1$, ..., $\mathcal{R}_5$ may be obtained from the results of section \ref{Kampe}, since the corresponding series (with general term $h(0,0)$) are in fact Kamp\'e de F\'eriet double series.

Let us now apply the computational strategy described above to obtain the different series representations which converge in each of the regions of Figure \ref{reg_conv_R1}.

\subsubsection{Cone 1: series representation valid in the region $\mathcal{R}_1=\left\{  |u_1| +|u_2|< 1 \right\}$}

We begin with the simplest case, namely Cone 1.
As we said in section \ref{R-1}, there are four types of intersections that have to be considered separately and whose contributions will be subsequently added. From the discussion on the transformation law, we also concluded that there is no need for "non-trivial" applications of the transformation law in this cone and, moreover, that no additional overall sign appears for every residues associated to this cone since Figure \ref{orientation} implies that for Cone 1 the sign is fixed by the determinant of the matrix in (\ref{trivialLaw1}).

\paragraph{Type 1:} the singular point $(z_1,z_2)=(0,0)$.

In this case obviously no change of variable is necessary and one therefore has
\begin{equation}
\omega\big\vert_{Type\;1}^{\textrm{Cone 1}}=\frac{h_1(z_1,z_2)}{z_1^3z_2^3}\,dz_1\wedge dz_2
\end{equation}
where
\begin{equation}
h_1(z_1,z_2)\doteq- u_1^{z_1} u_2^{z_2}\, \Gamma^2\left(1-z_1\right)\, \Gamma\left(1+z_1\right)\Gamma^2\left(1-z_2\right)\, \Gamma \left(1+z_2\right)\Gamma \left(z_1+z_2+1\right)\,.
\end{equation}
Using the Cauchy Formula (\ref{cauchy}) one obtains
\begin{flalign}
&R_{-1}(u_1,u_2)\big\vert_{Type\;1}^{\textrm{Cone 1}}\nonumber \\
&=\frac{1}{4}h_1^{(2,2)}(0,0)\nonumber\\
&=-\left(\frac{1}{4}\ln^2u_1+\zeta_2\right)\left(\ln^2u_2+4\zeta_2\right)-\zeta_2 \ln u_1\ln u_2-\zeta_3(\ln u_1+\ln u_2)-\frac{1}{4}\zeta_2^2-\frac{3}{2}\zeta_4 \;.\label{R1type1}
\end{flalign}

\paragraph{Type 2:}

the singular points $(z_1,z_2)=(1+m,1+n)$ where $m, n\geq 0$ are integers. 

Performing the change of variables $z_1\mapsto z_1+1+m, z_2\mapsto z_2+1+n$ one has
\begin{align}
\omega\big\vert_{Type\;2}^{\textrm{Cone 1}} &= -u_1^{z_1+1+m}\, u_2^{z_2+1+n}\, \Gamma^2 \left(-z_1-1-m\right)\, \Gamma\left(z_1+1+m\right) \nonumber\\
&\hspace{1.5cm}\times \Gamma^2 \left(-z_2-1-n\right) \Gamma \left(z_2+1+n\right) \Gamma \left(z_1+z_2+3+m+n\right)\;dz_1 \wedge dz_2.
\end{align}
After applying (\ref{reflexion}), one obtains
\begin{equation}
 \omega\big\vert_{Type\;2}^{\textrm{Cone 1}} = \frac{h_2(z_1,z_2)}{z_1^2z_2^2} \; dz_1\wedge dz_2
\end{equation}
where
\begin{flalign}
h_2(z_1,z_2)&\doteq-u_1^{z_1+1+m}\, u_2^{z_2+1+n}\, \frac{\Gamma^2\left(1+z_1\right)\Gamma^2\left(1-z_1\right)}{\Gamma^2 \left(z_1+m+2\right)}\, \Gamma\left(z_1+1+m\right)&\nonumber\\
&\hspace*{1.5cm}\times\;\frac{\Gamma^2 \left(1+z_2\right)\Gamma^2 \left(1-z_2\right)}{\Gamma^2\left(z_2+n+2\right)}\Gamma \left(z_2+1+n\right)\Gamma \left(z_1+z_2+3+m+n\right)\,.
\end{flalign}
Using now the Cauchy Formula one has
\begin{equation}
R_{-1}(u_1,u_2)\big\vert_{Type\;2}^{\textrm{Cone 1}} = \sum_{m=0}^{\infty}\sum_{n=0}^{\infty}h_2^{(1,1)}(0,0)
\end{equation}
and from equations (\ref{h11}), (\ref{A}), (\ref{B}) and (\ref{Akl}) or (\ref{Bkl}) a straightforward (mental) calculation gives
\begin{flalign}
&R_{-1}(u_1,u_2)\big\vert_{Type\;2}^{\textrm{Cone 1}}= \nonumber\\
&-\sum_{m=0}^{\infty}\sum_{n=0}^{\infty}\frac{u_1^{m+1}u_2^{n+1}\ \Gamma(3+m+n)}{(1+m)(1+n)\Gamma(m+2)\Gamma(n+2)}\Big\{\Big[\ln u_1+\psi(1+m)+\psi(3+m+n)-2\psi(2+m)\Big] \nonumber\\
&\hspace{6.5cm} \times \Big[\ln u_2+\psi(1+n)+\psi(3+m+n)-2\psi(2+n)\Big]\nonumber\\
&\hspace{6cm} +\psi^{(1)}(3+m+n)\Big\}\label{R1type2}\;.
\end{flalign}
It may be useful to express the polygamma functions in terms of harmonic numbers and their generalizations  using the relations (\ref{PsiH}) and (\ref{PsiNHN}).

One then finds
\begin{flalign}
&R_{-1}(u_1,u_2)\big\vert_{Type\;2}^{\textrm{Cone 1}} \nonumber\\
&=-\sum_{m=0}^{\infty}\sum_{n=0}^{\infty}\frac{u_1^{m+1}u_2^{n+1}\ \Gamma(3+m+n)}{(1+m)(1+n)\Gamma(m+2)\Gamma(n+2)}\bigg\{\bigg[-\frac{2}{m+1}-H_m+H_{2+m+n}+\ln u_1\bigg] \nonumber\\
&\hspace{6.3cm}\times\bigg[-\frac{2}{n+1}-H_n+H_{2+m+n}+\ln u_2\bigg]\nonumber\\
&\hspace{6.2cm}+\zeta_2-H^{(2)}_{2+m+n}\bigg\}\;.\label{R1type2v2}
\end{flalign}

\paragraph{Type 3:} the singular points $(z_1,z_2)=(1+m,0)$ where $m\geq 0$ is an integer. \\

Performing the same change of variable as in Type 2 for $z_1$ and no change of variable for $z_2$ one has
\begin{equation}
\omega\big\vert_{Type\;3}^{\textrm{Cone 1}}=\frac{h_3(z_1,z_2)}{z_1^2z_2^3}\,dz_1\wedge dz_2
\end{equation}
where
\begin{flalign}
h_3(z_1,z_2)&\doteq -u_1^{z_1+1+m}\, u_2^{z_2}\, \frac{\Gamma^2 \left(1-z_1\right) \Gamma^2\left(1+z_1\right)}{\Gamma^2 \left(2+m+z_1\right)}\, \Gamma\left(z_1+1+m\right)\nonumber\\
&\hspace{4cm}\times\, \Gamma^2\left(1-z_2\right)\, \Gamma \left(1+z_2\right)\Gamma \left(z_1+z_2+2+m\right)\,,
\end{flalign}
and, by the Cauchy Formula (\ref{cauchy}) and equation (\ref{h12}), one obtains 
\begin{flalign}
&R_{-1}(u_1,u_2)\big\vert_{Type\;3}^{\textrm{Cone 1}} \nonumber\\ &=\frac{1}{2}\sum_{m=0}^{\infty}h_3^{(1,2)}(0,0)
\nonumber\\
&=-\frac{1}{2}\sum_{m=0}^{\infty}\frac{u_1^{m+1}}{ 1+m }\left\{\left[\ln u_1+\psi(1+m)-\psi(2+m)\right]\left(\left[\ln u_2-\psi(1)+\psi(2+m)\right]^2+3\psi^{(1)}(1)\right.\right. \nonumber \\
&\hspace{3cm}\left.\left.+\psi^{(1)}(2+m)\right)+2\left[\ln u_2-\psi(1)+\psi(2+m)\right]\psi^{(1)}(2+m)+\psi^{(2)}(2+m)\right\}\nonumber\\
&=-\frac{1}{2}\sum_{m=0}^{\infty}\frac{u_1^{m+1}}{ 1+m }\left\{\left[\left(\ln u_1-\frac {1}{1+m}\right)\left(\ln u_2+H_{m+1}\right)+2\left(\zeta_2-H^{(2)}_{m+1}\right)\right]\right.\big[\ln u_2+H_{m+1}\big] \nonumber \\
&\hspace{3cm}\left.+\left(\ln u_1-\frac {1}{1+m}\right)\left(4\zeta_2-H^{(2)}_{m+1}\right)-2\left(\zeta_3-H^{(3)}_{m+1}\right)\right\}\label{R1type3}.
\end{flalign}

\paragraph{Type 4:} the singular points $(z_1,z_2)=(0,1+m)$ where $m\geq 0$ an integer. 

Because of the symmetry of the integral $R_{-1}$ this case may be obtained from Type 3 by exchanging $u_1$ and $u_2$ in (\ref{R1type3}). 
\\

As a numerical check of our results, we directly compute the integral (\ref{Rmoins1}) for $u_1=0.1$ and $u_2=0.2$ and find
\begin{equation}
R_{-1}(0.1,0.2)\simeq-37.976338.
\end{equation}
Now, a truncation of the series expansion of Type 2, 3, 4 at $M=N=10$ added to the contribution of Type 1 for the same values of the arguments gives the following result
\begin{flalign}
&R_{-1}(0.1,0.2)\big\vert_{Type\; 2, M=N=10}^{\textrm{Cone 1}}+ R_{-1}(0.1,0.2)\big\vert_{Type\; 3, M=10}^{\textrm{Cone 1}} \nonumber\\
&\hspace{4cm}+ R_{-1}(0.1,0.2)\big\vert_{Type\; 4, M=10}^{\textrm{Cone 1}} +R_{-1}(0.1,0.2)\big\vert_{Type\; 1}^{\textrm{Cone 1}} \nonumber\\
&\simeq-0.3035221+1.0827389+2.2579423-41.0134972 \\
&\simeq-37.976338. 
\end{flalign}
We also checked that eqs. (C.3), (C.4) and (C.5) in Appendix C of \cite{DelDuca:2010zg} may be obtained from eqs. (\ref{R1type1}), (\ref{R1type2v2}) and (\ref{R1type3}) and its symmetrical expression.

\subsubsection{Cone 3: series representation valid in  the region $\mathcal{R}_3=\left\{1< |u_1| \; \;  \text{and} \; \;  1 < |u_2| \right\} \cap\left\{  |u_1| +1< |u_2| \right\}$}

We now turn to a less trivial cone, in the sense that the intersection types all involve oblique lines and therefore imply a use of the transformation law. This adds to the computation an intermediate step, but apart from this the calculation of Cone 3 follows exactly the same reasoning as for Cone 1 and the complexity is not increased.

\begin{figure}[h]
\begin{center}
\includegraphics[width=0.5\textwidth]{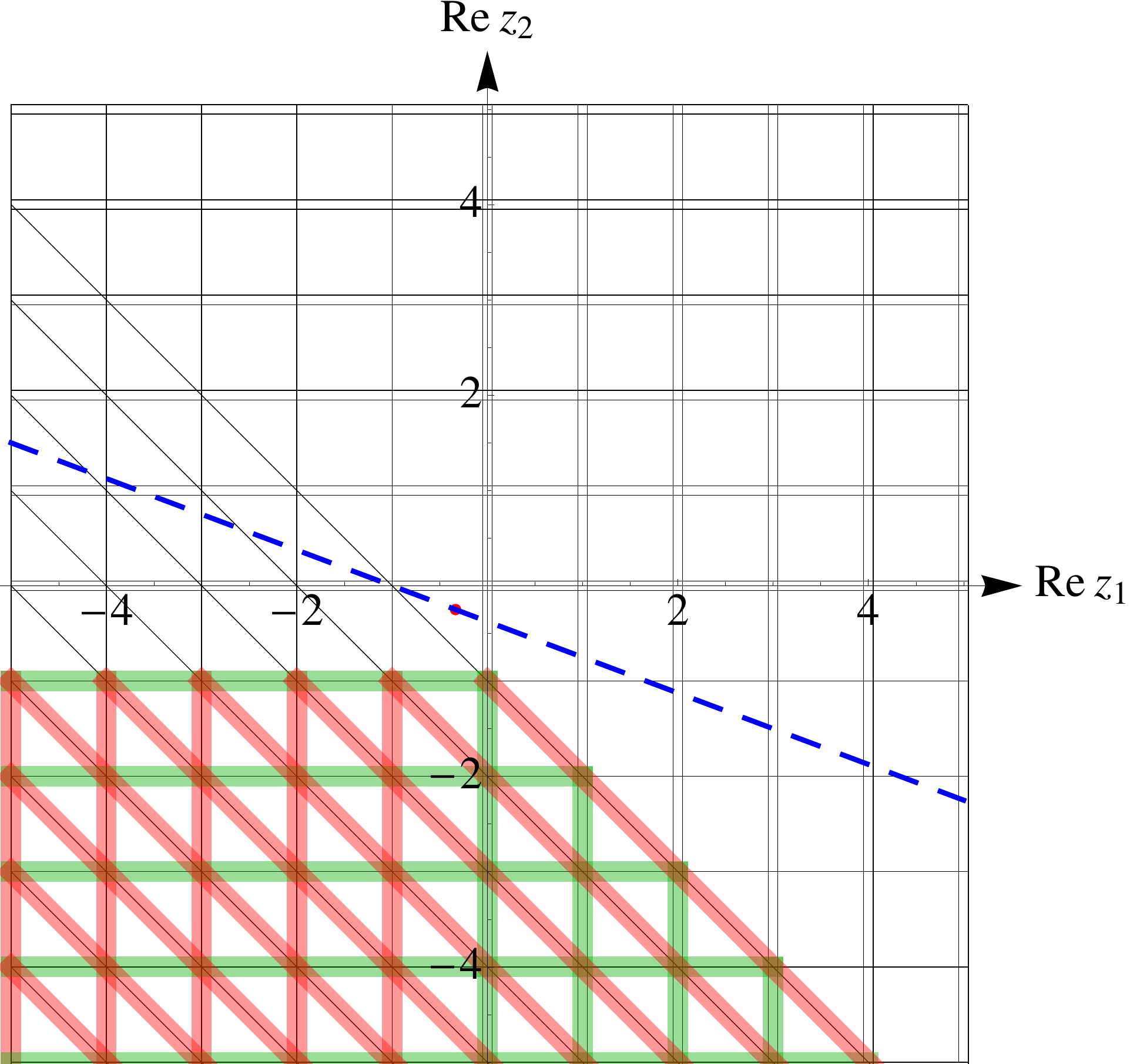}
\end{center}
\caption{Cone 3.\label{conesR12}}
\end{figure}

Let us begin by looking for the different types of intersections in Cone 3. It is clear from Figure \ref{conesR12} that there are 3 different types. 

\paragraph{Type 1:} the singular points $(z_1,z_2)=(-1-m,-1-n)$ where $m, n\geq 0$ are integers. \\

Performing the change of variables $z_1\mapsto z_1-1-m, z_2\mapsto z_2-1-n$ and using (\ref{reflexion}) one has
\begin{equation}
\omega\big\vert_{Type\;1}^{\textrm{Cone 3}} = \frac{h_1(z_1,z_2)}{z_1z_2(z_1+z_2)}\,dz_1\wedge dz_2
\end{equation}
where
\begin{flalign}
h_1(z_1,z_2)&\doteq u_1^{z_1-1-m}\, u_2^{z_2-1-n}\, \Gamma^2\left(1+m-z_1\right) \frac{\Gamma\left(1+z_1\right)\Gamma\left(1-z_1\right)}{\Gamma\left(2+m-z_1\right)}\nonumber \\
&\hspace{1cm}\times\, \Gamma^2 \left(1+n-z_2\right) \frac{\Gamma\left(1+z_2\right)\Gamma\left(1-z_2\right)}{\Gamma\left(2+n-z_2\right)}\frac{\Gamma\left(1+z_1+z_2\right)\Gamma\left(1-z_1-z_2\right)}{\Gamma\left(2+m+n-z_1-z_2\right)} \;.
\end{flalign}
We now fix the 2-vector $\boldsymbol{f}$ by the geometrical rule described when we presented the transformation law and choose the matrix $A$ and 2-vector $\boldsymbol{g}$ as simple as possible. It is clear that the singular gamma functions of $R_{-1}$ in this type of intersection are $\Gamma(z_1)$, $\Gamma(z_2)$ and $\Gamma(1+z_1+z_2)$ and that the singular lines corresponding to the first and last of these three gamma functions intersect one side of the $l$-line, whereas the singular lines corresponding to $\Gamma(z_2)$ intersect the other side (see Figure \ref{conesR12}). Therefore, from the conventions of Figure \ref{orientation} one finds $\boldsymbol{f}=\big(z_1(z_1+z_2), z_2\big)^T$. It is easy to find that the simplest $g$ and $A$ which fulfill all the requirements of the transformation law are $\boldsymbol{g}=(z_1^2,z_2)^T$ and 
\begin{equation}
A=\left( \begin{array}{cc}
1 & -z_1  \\
0 & 1 \end{array} \right)\;.
\end{equation}
Since $\det A=1$ one then finds
\begin{equation}
\Res{\frac{h_1(z_1,z_2)}{z_1z_2(z_1+z_2)}}=\Res{\frac{h_1(z_1,z_2)}{z_1^2\ z_2}}.
\end{equation}
Therefore
\begin{flalign}
&R_{-1}(u_1,u_2)\big\vert_{Type\;1}^{\textrm{Cone 3}}\nonumber\\
&=\sum_{m=0}^{\infty}\sum_{n=0}^{\infty}h_1^{(1,0)}(0,0)\nonumber\\
&=\sum_{m=0}^{\infty}\sum_{n=0}^{\infty}\frac{u_1^{-m-1}u_2^{-n-1}\Gamma(m+1)\Gamma(n+1)}{(m+1)(n+1)\Gamma(2+m+n)}\left[\ln u_1-2\psi(1+m)+\psi(2+m)+\psi(2+m+n)\right]\nonumber\\
&=\sum_{m=0}^{\infty}\sum_{n=0}^{\infty}\frac{u_1^{-m-1}u_2^{-n-1}\Gamma(m+1)\Gamma(n+1)}{(m+1)(n+1)\Gamma(2+m+n)}\left[\frac{1}{m+1}-H_m+H_{1+m+n}+\ln u_1\right].
\label{R1type1cone2}
\end{flalign}

Notice that if the choice of $\boldsymbol{f}$ is unique in the transformation law, this is of course not the case for $A$ and $g$. But with other choices, one obviously finds the same result at the end. Indeed, instead of $\boldsymbol{g}=(z_1^2,z_2)^T$ one could have chosen for instance $\boldsymbol{g}=(z_1^3,z_2)^T$. In this case, one possible matrix of the transformation law would have been 
\begin{equation}
A=\left( \begin{array}{cc}
z_1-z_2 & z_1z_2  \\
z_2 & 1-z_1z_2-z_1^2  \end{array} \right)\;,
\end{equation}
with $\det A=z_1-z_1^3-z_2$.
In this case one finds
\begin{equation}
\Res{\frac{h_1(z_1,z_2)}{z_1z_2(z_1+z_2)}}=\Res{\frac{h_1(z_1,z_2)}{z_1^2\ z_2}} -\Res{\frac{h_1(z_1,z_2)}{z_2}}-\Res{\frac{h_1(z_1,z_2)}{z_1^3}}\;.
\end{equation}
By the Cauchy formula the second and third terms are zero and one obtains the same result as in eq. (\ref{R1type1cone2}).

\paragraph{Type 2:} the singular points $(z_1,z_2)=(0,-1-m)$ where $m\geq 0$ is an integer. \\

Performing the same change of variable for $z_2$ as in the case of Type 1 and no change of variable for $z_1$ one has
\begin{equation}
\omega\big\vert_{Type\;2}^{\textrm{Cone 3}}=\frac{h_2(z_1,z_2)}{z_1^3z_2(z_1+z_2)}\,dz_1\wedge dz_2
\end{equation}
where
\begin{flalign}
h_2(z_1,z_2)& \doteq u_1^{z_1}\, u_2^{z_2-1-m}\, \Gamma^2\left(1-z_1\right) \Gamma\left(1+z_1\right)\Gamma^2\left(1+m-z_2\right)\nonumber\\ 
&\hspace{2cm}\times\, \frac{\Gamma \left(1+z_2\right)\Gamma \left(1-z_2\right)}{\Gamma \left(2+m-z_2\right)}\, \frac{\Gamma \left(1+z_1+z_2\right)\Gamma \left(1-z_1-z_2\right)}{\Gamma \left(1+m-z_1-z_2\right)}\,.
\end{flalign}
Looking at Figure \ref{conesR12} (and Figure \ref{orientation})  one finds $\boldsymbol{f}=(z_1+z_2, z_1^3z_2)^T$. Now, with $\boldsymbol{g}=(z_1^4,z_2^5)^T$ and 
\begin{equation}
A=\left( \begin{array}{cc}
z_1^3 & -1  \\
z_2(z_2-z_1)(z_1^2+z_2^2) & z_1  \end{array} \right)\;,
\end{equation}
whose determinant is $\det A=z_1^4+z_1^2z_2^2+z_2^4-z_1^3z_2-z_1z_2^3$, one finds
\begin{multline}
\Res{\frac{h_2(z_1,z_2)}{z_1^3z_2(z_1+z_2)}}=\frac{1}{2}h_2^{(1,2)}(0,0)-\frac{1}{2}h_2^{(2,1)}(0,0)
+\frac{1}{3!}h_2^{(3,0)}(0,0)-\frac{1}{3!}h_2^{(0,3)}(0,0)\, ,
\end{multline}
so that
\begin{flalign}
&R_{-1}(u_1,u_2)\big\vert_{Type\;2}^{\textrm{Cone 3}} \nonumber\\
&=\frac{1}{6}\sum_{m=0}^{\infty}\frac{u_2^{-1-m}}{1+m}\bigg\{\left[\ln \frac{u_1}{u_2}+H_m-\frac{1}{1+m}\right]^3 \nonumber\\
&\hspace{3cm}+\left[\ln \frac{u_1}{u_2}+H_m-\frac{1}{1+m}\right] \left[18\zeta_2+\frac{3}{(1+m)^2}-3H^{(2)}_m\right]\nonumber\\
&\hspace{3cm}+2H^{(3)}_m-\frac{2}{(1+m)^3}\bigg\}
\end{flalign}
(we do not give here the equivalent expression in terms of polygamma functions).

\paragraph{Type 3:} the singular points $(z_1,z_2)=(1+m,-2-m-n)$ where $m, n\geq 0$ are integers. \\

Performing the change of variables $z_1\mapsto z_1+1+m, z_2\mapsto z_2-2-m-n$, one has
\begin{equation}\label{omegaT3c2}
\omega\big\vert_{Type\;3}^{\textrm{Cone 3}}=\frac{h_3(z_1,z_2)}{z_1^2z_2(z_1+z_2)}\,dz_1\wedge dz_2
\end{equation}
where
\begin{multline}
h_3(z_1,z_2)\doteq(-1)^{m+1}u_1^{z_1+1+m}\, u_2^{z_2-2-m-n}\, \frac{\Gamma^2 \left(1+z_1\right)\Gamma ^2\left(1-z_1\right)}{\Gamma^2 \left(2+m+z_1\right)}\Gamma\left(1+m+z_1\right)\\ 
\times\,\Gamma^2\left(2+m+n-z_2\right) \frac{\Gamma \left(1+z_2\right)\Gamma \left(1-z_2\right)}{\Gamma \left(3+m+n-z_2\right)}\, \frac{\Gamma \left(1+z_1+z_2\right)\Gamma \left(1-z_1-z_2\right)}{\Gamma \left(1+n-z_1-z_2\right)}\label{h3c2}\,.
\end{multline}
Looking at Figure \ref{conesR12} one finds $\boldsymbol{f}=(z_1+z_2, z_1^2z_2)^T$. Now, with $\boldsymbol{g}=(z_1^3,z_2^3)^T$ and 
\begin{equation}
A=\left( \begin{array}{cc}
z_1^2 & -1  \\
z_2(z_2-z_1) & 1  \end{array} \right)\;,
\end{equation}
whose determinant is $\det A=z_1^2+z_2^2-z_1z_2$, one finds
\begin{equation}\label{LawT3Cone3}
\Res{\frac{h_3(z_1,z_2)}{z_1^2z_2(z_1+z_2)}}=\frac{1}{2}h_3^{(2,0)}(0,0)+\frac{1}{2}h_3^{(0,2)}(0,0)-h_3^{(1,1)}(0,0),
\end{equation}
so that
\begin{flalign}
&R_{-1}(u_1,u_2)\big\vert_{Type\;3}^{\textrm{Cone 3}} \nonumber\\
&=\sum_{m=0}^{\infty}\sum_{n=0}^{\infty}(-1)^{m+1} u_1^{1+m}u_2^{-2-m-n} \frac{\Gamma(1+m)\Gamma^2(2+m+n)}{\Gamma^2(2+m)\Gamma(3+m+n)\Gamma(1+n)} \nonumber\\
&\hspace{2cm}\times\left\{\frac{1}{2}\left[\ln \frac{u_1}{u_2}+H_{m}-2H_{1+m}+ 2H_{1+m+n}-H_{2+m+n}\right]^2\right. \nonumber\\
&\hspace{3.5cm}\left.+3\zeta_2-\frac{1}{2}H^{(2)}_{m}+H^{(2)}_{m+1}-H^{(2)}_{1+m+n}+\frac{1}{2}H^{(2)}_{2+m+n}\right\}\; \label{R1type3cone2}
\end{flalign}
(we do not give here the equivalent expression in terms of polygamma functions).

\bigskip

We numerically checked all these results as we did for the case of Cone 1.
Notice that thanks to the symmetry of $R_{-1}$ it is straightforward to deduce the series representation corresponding to Cone 5 from the results of this section.

\subsubsection{Other cones, other regions}

As we saw in Figure \ref{conesR1}, there are two other cones associated to $R_{-1}$: Cone 2 and Cone 4 and, thanks again to the symmetry of (\ref{Rmoins1}) only one of them has to be considered.

This is what we did by computing the corresponding series for Cone 4 and, once more, we checked  numerically the expressions that we obtained by a comparison with the direct evaluation of the integral representation (\ref{Rmoins1}) for values of $u_1$ and $u_2$ lying in the relevant convergence region $\mathcal{R}_4$. Since these results do not really bring important information, we do not list here the expressions of the series in order to shorten the paper. They are however as straightforward to obtain as the others.

Nevertheless, it is perhaps instructive to briefly consider one particular type of contributions, which belong to the overlapping regions of Cone 3 and Cone 4, to insist on one point concerning the derivation of the 2-vector $\boldsymbol{f}$, in the transformation law.
More precisely, we will have a look to the contribution of the singular points $(z_1,z_2)=(1+m,-2-m-n)$, where $m, n\geq 0$ are integers, in Cone 4. 
One can indeed see that we already met this type of singularities in Cone 3 (Type 3). Therefore part of the calculation has already been done and one may directly take the result of eq. (\ref{omegaT3c2}) and eq. (\ref{h3c2}).

The important point is that, after a look at Cone 4 in Figure \ref{conesR1}, one finds $\boldsymbol{f}=\big(z_2(z_1+z_2), z_1^2\big)^T$ which is not the same expression as the one of Cone 3, that we recall to be $\boldsymbol{f}=(z_1+z_2, z_1^2z_2)^T$ (see the text after eq. (\ref{h3c2})). This is of course due to the fact that the geometrical rule which allows to fix $\boldsymbol{f}$ does not give the same results in different cones even if the singular structure under study is the same in these cones. Now, with $\boldsymbol{g}=(z_1^2,z_2^3)^T$ and 
\begin{equation}
A=\left( \begin{array}{cc}
0 & 1  \\
z_2-z_1 & z_2  \end{array} \right)\;,
\end{equation}
whose determinant is $\det A=z_1-z_2$, one finds
\begin{equation}\label{LawCone4}
\Res{\frac{h_3(z_1,z_2)}{z_1^2z_2(z_1+z_2)}}=\frac{1}{2}h_3^{(0,2)}(0,0)-h_3^{(1,1)}(0,0),
\end{equation}
where $h_3$ is given in (\ref{h3c2}). 

Therefore, although in our notations the left hand side of (\ref{LawCone4}) and (\ref{LawT3Cone3}) are formally the same, their right hand side are not, so that obviously the result given in eq. (\ref{R1type3cone2}) do not coincide with the expression of $R_{-1}(u_1,u_2)\big\vert_{Type\ (1+m,-2-m-n)}^{\text{Cone 4}}$.

\subsection{The $R_{-1-z_1}(u_1,u_2,u_3)$ integral }

After our exposition of the method in detail for the simple case of the $R_{-1}$ integral considered in \cite{DelDuca:2010zg}, we want to mention a few facts about the $R_{-1-z_1}(u_1,u_2,u_3)$ integral of the same paper (second line of eq. (3.6) in \cite{DelDuca:2010zg}):
\begin{multline}
R_{-1-z_1}(u_1,u_2,u_3) = \int\limits_{c-i\infty}^{c+i\infty}{d z_1\over 2\pi i} \int\limits_{d-i\infty}^{d+i\infty}{d z_2\over 2\pi i}\,
u_1^{z_1}\, u_2^{z_2}\,u_3^{-z_1}\\
   \times\frac{1}{z_1}\Gamma^2\left(-z_1\right)\, \Gamma^2
   \left(z_1+1\right)\, \Gamma^2\left(-z_2\right)\, \Gamma \left(z_2-z_1\right)\Gamma \left(z_1+z_2+1\right)\;,\label{Rmoins1z1}
\end{multline}
where $c=-\frac{1}{3}$ and  $d=-\frac{1}{4}$. 

\begin{figure}[h]
\begin{center}
\includegraphics[width=0.5\textwidth]{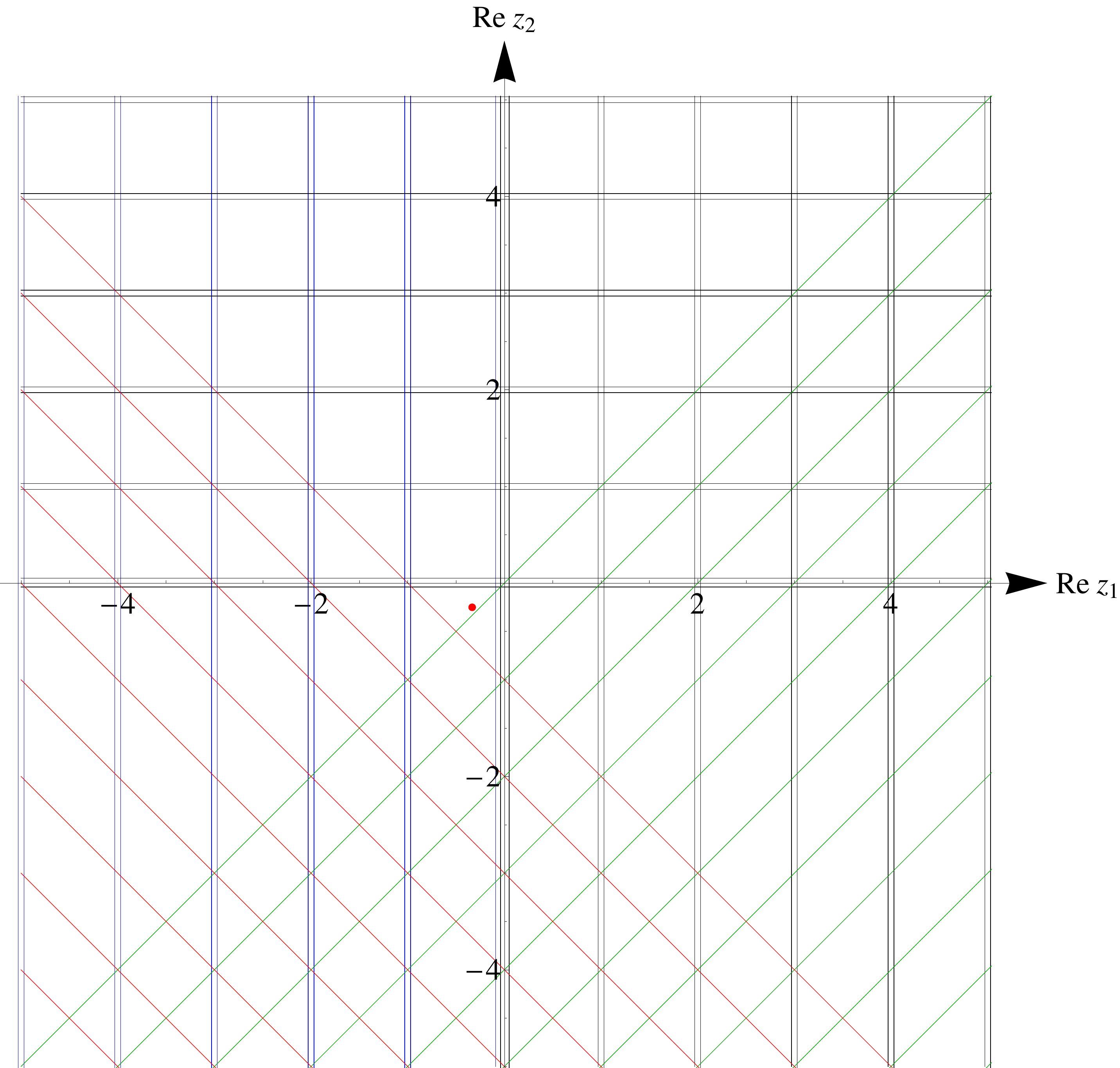}
\end{center}
\caption{Singular structure of the integrand of $R_{-1-z_1}(u_1,u_2,u_3)$. The red dot is the point $(c,d)$. \label{plane2}}
\end{figure}

The singular structure of the integrand of (\ref{Rmoins1z1}) is represented in Figure \ref{plane2}. 

\begin{figure}[h]
\hspace{1.2cm}
\begin{tabular}{cc}
   \includegraphics[width=0.4\textwidth]{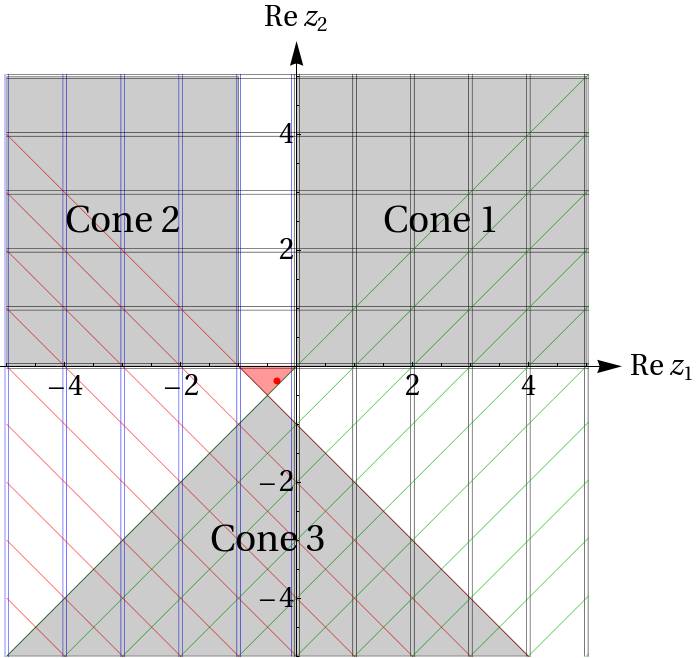} &
   \includegraphics[width=0.4\textwidth]{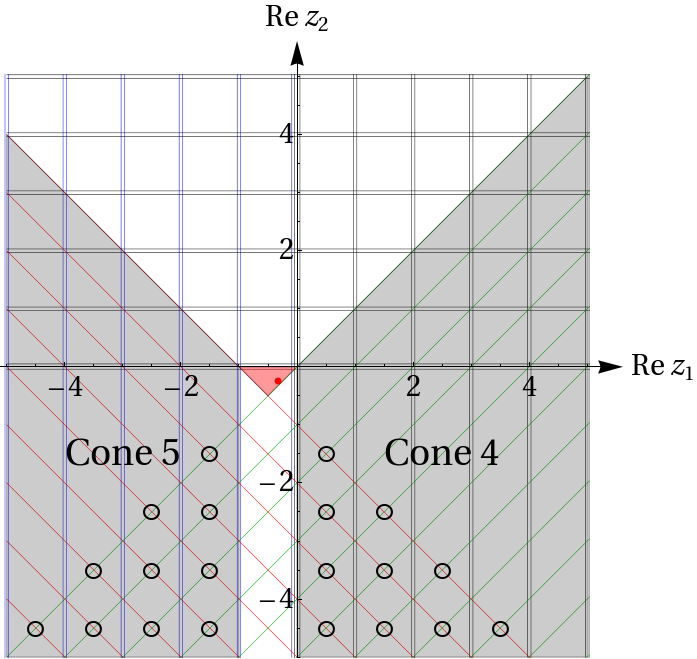} \\
\end{tabular}
\caption{The five different cones  of $R_{-1-z_1}(u_1,u_2,u_3)$. Small black circles distinguish spurious singularities.\label{conesR2}}
\end{figure}

Drawing the straight $l$-line and rotating around $\gamma=(c,d)$ it is easy to conclude that five cones are to be found, that we represented on Figure \ref{conesR2}. 

It is interesting to present (part of) the calculation of the $R_{-1-z_1}(u_1,u_2,u_3)$ integral for several reasons. 

At first sight it is a less simple integral than $R_{-1}(u_1,u_2)$ since one can see that it is not symmetric under an exchange of $\frac{u_1}{u_3}$ and $u_2$. Notice however that apart from its $\frac{1}{z_1}$ term the integrand of (\ref{Rmoins1z1}) is symmetric under the replacement $z_1\rightarrow -z_1-1$ and this is reflected in a partial symmetry of the different cones (see Figure \ref{conesR2}, where one can indeed remark that the symmetry between Cone 1 and Cone 2, as well as between Cone 4 and Cone 5, is broken only by the $z_1=0$ line) and in fact also of their corresponding regions of convergence. 

The calculations of the different cone's contributions are not harder than the computations of previous sections and one may follow the same systematic evaluation method: we do not need to perform any manipulation at the integral level (by inserting Euler representations) as the authors of \cite{DelDuca:2009au,DelDuca:2010zg} did (see Appendix C.2 of \cite{DelDuca:2010zg}).

Moreover, one will see that the regions of convergence of the different series representations of (\ref{Rmoins1z1}) are slightly more complicated than those of $R_{-1}(u_1,u_2)$ (see Figure \ref{Region_Smirnov2}).
This means that the region of convergence of the series representation of the two-loop hexagon Wilson loop (see \cite{DelDuca:2009au,DelDuca:2010zg}) is probably tricky to find (without the help of analytic continuations).

At last, the $R_{-1-z_1}(u_1,u_2,u_3)$ integral presents the feature that (as we already said in section \ref{R-1}), for two of its cones, some subsets of singularities are spurious. This fact, that we did not meet in the case of $R_{-1}(u_1,u_2)$, should however not be thought of a rare occurence in general and we think that it is important to show explicitly on this simple example how spurious singularities may appear.

\subsubsection{Regions of convergence}

Let us now give the region of convergence of each cone (to obtain them we follow the same procedure as in section \ref{reg_conv_section}).

\begin{figure}[h]
\begin{center}
\includegraphics[width=0.5\textwidth]{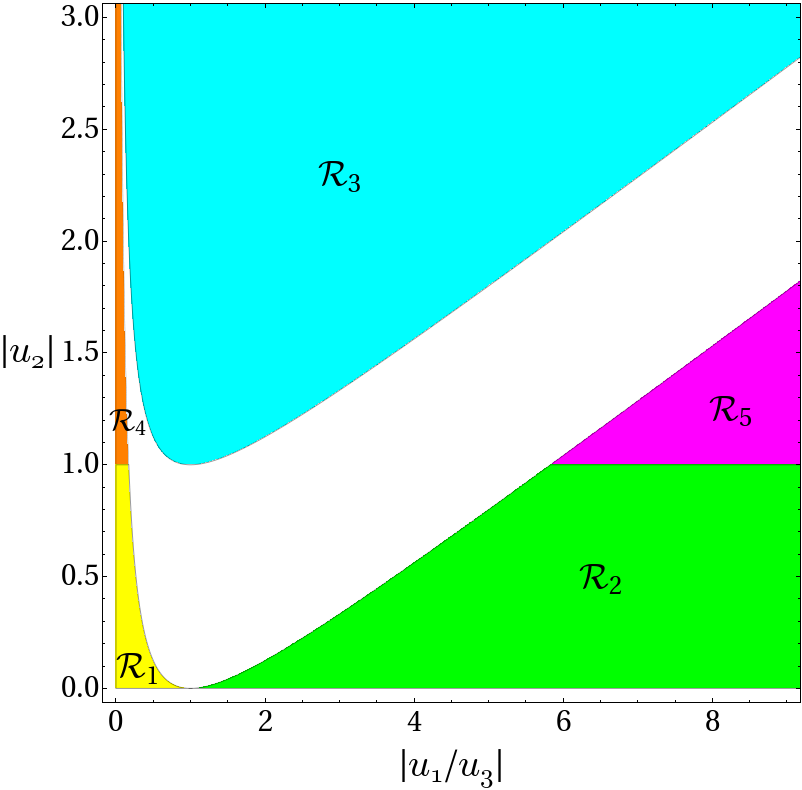}
\end{center}
\caption{Regions of convergence of the series representations of $R_{-1-z_1}(u_1,u_2,u_3)$.\label{Region_Smirnov2}}
\end{figure}

\paragraph{Cone1} The region of convergence is given by $\mathcal{R}_1=\{\left|u_2\right|<1\;\; \text{and}\;\; \left|\frac{u_1}{u_3}\right|+2\sqrt{\left|\frac{u_1u_2}{u_3}\right|}<1\}$ (yellow region in Figure \ref{Region_Smirnov2}).

\paragraph{Cone2} The region of convergence is $\mathcal{R}_2=\{\left|u_2\right|<1\;\; \text{and}\;\; \left|\frac{u_3}{u_1}\right|+2\sqrt{\left|\frac{u_3u_2}{u_1}\right|}<1\}$ (green region).

\paragraph{Cone3} The region of convergence is $\mathcal{R}_3=\{\sqrt{\left|\frac{u_1}{u_2u_3}\right|}<1+\sqrt{1-\left|\frac{1}{u_2}\right|}\;\; \text{and}\;\; \sqrt{\left|\frac{u_3}{u_2u_1}\right|}<1+\sqrt{1-\left|\frac{1}{u_2}\right|}\}$ (cyan region).

\paragraph{Cone4} The region of convergence is $\mathcal{R}_4=\{\sqrt{\left|\frac{u_1}{u_2u_3}\right|}<1+\sqrt{1-\left|\frac{1}{u_2}\right|}\;\; \text{and}\;\; \left|\frac{u_1}{u_3}\right|+2\sqrt{\left|\frac{u_1u_2}{u_3}\right|}<1\}$ (orange region).

\paragraph{Cone5} The region of convergence is $\mathcal{R}_5=\{\sqrt{\left|\frac{u_3}{u_2u_1}\right|}<1+\sqrt{1-\left|\frac{1}{u_2}\right|}\;\; \text{and}\;\; \left|\frac{u_3}{u_1}\right|+2\sqrt{\left|\frac{u_3u_2}{u_1}\right|}<1\}$ (magenta region).\\

The (partial) symmetry mentioned above between the cones is clearly seen here as we perform the change $\frac{u_1}{u_3}\rightarrow\frac{u_3}{u_1}$ and corresponds indeed to the change $z_1\rightarrow -z_1-1$. It is however completely manifest at this level since the $\frac{1}{z_1}$ term of the integrand has no influence on the different regions of convergence.

As for the case of $R_{-1}(u_1,u_2)$ one can see in Figure \ref{Region_Smirnov2} that an infinite white band is not reachable by any of the series representations that our method allows to derive.

Contrary to the case of $R_{-1}(u_1,u_2)$ the regions of convergence $\mathcal{R}_1$, ..., $\mathcal{R}_5$ cannot be obtained from the results of section \ref{Kampe}, since the corresponding series are not Kamp\'e de F\'eriet double series.

We give now the results of the calculation of the series associated to Cone 1.

\subsubsection{Cone 1: region of convergence $\mathcal{R}_1=\{\left|u_2\right|<1\;\; \text{and}\;\; \left|\frac{u_1}{u_3}\right|+2\sqrt{\left|\frac{u_1u_2}{u_3}\right|}<1\}$}

As can be seen from Figure \ref{conesR2} there are 5 different types of intersections in this cone.

\paragraph{Type 1:} the singular points $(z_1,z_2)=(1+m,2+m+n)$ where $m, n\geq 0$ are integers. 

This type of singularities does not require to use the transformation law and one finds
\begin{flalign}
&R_{-1-z_1}(u_1,u_2,u_3)\big\vert_{Type\; 1}^{\textrm{Cone 1}} \nonumber\\
&=\sum_{m=0}^{\infty}\sum_{n=0}^{\infty}\left(\frac{u_1}{u_3}\right)^{1+m}u_2^{2+n+m}\frac{\Gamma(1+n)\Gamma(4+n+2m)}{(m+1)\Gamma^2(m+n+3)} \nonumber\\
&\hspace{1cm}\times\Bigg\{\Big[\ln u_2-2\psi(m+n+3)+\psi(1+n)+\psi(4+n+2m)\Big] \nonumber\\
&\hspace{3cm}\times\Big[\ln\frac{u_1}{u_3}-\psi(m+2)+\psi(m+1)-\psi(1+n)+\psi(4+n+2m)\Big] \nonumber\\
&\hspace{1.5cm}-\psi^{(1)}(1+n)+\psi^{(1)}(4+n+2m)\Bigg\}.
\end{flalign}

\paragraph{Type 2:} the singular points $(z_1,z_2)=(1+m,0)$ where $m\geq 0$ is an integer. \\

In this case the transformation law is necessary and one finds
\begin{flalign}
&R_{-1-z_1}(u_1,u_2,u_3)\big\vert_{Type\; 2}^{\textrm{Cone 1}} \nonumber\\
&=-\sum_{m=0}^{\infty}\left(-\frac{u_1}{u_3}\right)^{1+m}\frac{1}{m+1}\Bigg\{\Big[\ln\frac{u_1}{u_3}+\psi(1+m)-\psi(2+m)\Big] \nonumber\\
&\times\Bigg[\Big(\ln\frac{u_1}{u_3}+\psi(1+m)-\psi(2+m)\Big)\left(\frac{1}{6}\ln\frac{u_1}{u_3}+\frac{1}{6}\psi(1+m)+\frac{1}{2}\ln u_2-\psi(1)+\frac{5}{6}\psi(2+m)\right) \nonumber\\
&\hspace{2cm}+\psi^{(1)}(1)+\frac{1}{2}\psi^{(1)}(1+m)\left.+\frac{3}{2}\psi^{(1)}(2+m)\right] \nonumber\\
&+\Big[\ln u_2-2\psi(1)+2\psi(2+m)\Big]\left[3\psi^{(1)}(1)+\frac{1}{2}\psi^{(1)}(1+m)-\frac{1}{2}\psi^{(1)}(2+m)\right] \nonumber\\
&+\frac{1}{6}\psi^{(2)}(1+m)+\frac{5}{6}\psi^{(2)}(2+m) \Bigg\}\label{R1z_1type2cone1}.
\end{flalign}

\paragraph{Type 3:} the singular points $(z_1,z_2)=(1+m+n,1+n)$ where $m, n\geq 0$ are integers. \\

This type of singularities does not impose to use the transformation law and one finds
\begin{flalign}
&R_{-1-z_1}(u_1,u_2,u_3)\big\vert_{Type\; 3}^{\textrm{Cone 1}}\nonumber\\
&=\sum_{m=0}^{\infty}\sum_{n=0}^{\infty}(-1)^{m+1}\left(\frac{u_1}{u_3}\right)^{1+m+n}u_2^{1+n}\frac{\Gamma(3+m+2n)}{(1+m+n)\Gamma(1+m)\Gamma^2(2+n)}\nonumber\\
&\frac{1}{3!}\left\{\left[\ln\frac{u_1}{u_3}-\psi(2+m+n)+\psi(1+m+n)-\psi(1+m)+\psi(3+2n+m)\right]^2\right.\nonumber\\
& \hspace{1cm}\times \Big[\ln\frac{u_1}{u_3}-\psi(2+m+n)+\psi(1+m+n)+3\ln u_2\nonumber\\
&\hspace{4cm}-6\psi(2+n)+2\psi(1+m)+4\psi(3+m+2n)\Big]\nonumber\\
&+3\left[\ln\frac{u_1}{u_3}-\psi(2+m+n)+\psi(1+m+n)+\ln u_2-2\psi(2+n)+2\psi(3+m+2n)\right]\nonumber\\
&\hspace{0.5cm}\times\left[6\psi^{(1)}(1)-\psi^{(1)}(2+m+n)+\psi^{(1)}(1+m+n)-\psi^{(1)}(1+m)+\psi^{(1)}(3+m+2n)\right]\nonumber\\
&+6\left[\ln\frac{u_1}{u_3}-\psi(2+m+n)+\psi(1+m+n)-\psi(1+m)+\psi(3+2n+m)\right]\nonumber\\
&\hspace{1cm}\times\left[-2\psi^{(1)}(1)+\psi^{(1)}(1+m)+\psi^{(1)}(3+m+2n)\right]\nonumber\\
&-\psi^{(2)}(2+m+n)+\psi^{(2)}(1+m+n)+2\psi^{(2)}(1+m)+4\psi^{(2)}(3+m+2n)\Bigg\}.
\end{flalign}

\paragraph{Type 4:} the singular points $(z_1,z_2)=(0,1+m)$ where $m\geq 0$ is an integer. \\

This type of singularities does not require to use the transformation law and one finds
\begin{flalign}
&R_{-1-z_1}(u_1,u_2,u_3)\big\vert_{Type\; 4}^{\textrm{Cone 1}}\nonumber
\\
&=\frac{1}{2}\sum_{m=0}^{\infty}u_2^{1+m}\frac{1}{m+1}\left\{\Big(\ln u_2-\psi(2+m)+\psi(1+m)\Big)\left[4\psi^{(1)}(1)+\psi^{(1)}(1+m)\right.\right.\\
&\left.+\psi^{(1)}(2+m)+\Big(\ln \frac{u_1}{u_3}-\psi(1+m)+\psi(2+m)\Big)^2\right]-2\left(\psi^{(1)}(1+m)-\psi^{(1)}(2+m)\right)\\
&\left.\times\Big(\ln \frac{u_1}{u_3}-\psi(1+m)+\psi(2+m)\Big)+\psi^{(2)}(1+m)+\psi^{(2)}(2+m)\right\}.
\end{flalign}

\paragraph{Type 5:} the singular point $(z_1,z_2)=(0,0)$. \\

For this last point the transformation law is necessary and one finds
\begin{multline}
R_{-1-z_1}(u_1,u_2,u_3)\big\vert_{Type\; 5}^{\textrm{Cone 1}} \\
= \bigg\{-\psi^{(2)}(1)-\Big[3\psi^{(1)}(1)+\frac{1}{6}\ln^2\frac{u_1}{u_3}\Big]\ln u_2-\Big[\frac{3}{2}\psi^{(1)}+\frac{1}{24}\ln^2\frac{u_1}{u_3}\Big]\ln\frac{u_1}{u_3}\bigg\}\ln\frac{u_1}{u_3} \\
-\frac{9}{2}\big[\psi^{(1)}(1)\big]^2-\frac{1}{4}\psi^{(3)}(1).
\label{R1z_1type5cone1}
\end{multline}

\subsubsection{Other cones}

As for the case of $R_{-1}(u_1, u_2)$, we have computed the series representations of $R_{-1-z_1}(u_1, u_2, u_3)$ coming from all the cones of Figure \ref{conesR2} and checked that the results were in agreement with direct numerical evaluations of the integral representation (\ref{Rmoins1z1}) for some values of $u_1$, $u_2$ and $u_3$ in the corresponding different regions of convergence. 

The contributions of all cones were easy to compute but we will not write the corresponding results to shorten the paper. 

One point has however to be underlined: in each of the cones 4 and 5 there is an infinite subset of singularities which do not have to be included in the calculation of their respective contributions.
For Cone 4 these singularities are located at $(z_1,z_2)=(\frac{1}{2}+n,-\frac{3}{2}-m-n)$ and for cone 5 they are at $(z_1,z_2)=(-\frac{3}{2}-m,-\frac{3}{2}-m-n)$, where as usual $m,n\geq0$ are integers. We recall that the reason why one has to discard these singularities is that in each case all singular lines which cross each other at these singular points are also crossing the same side of the $l$-line corresponding to the cone under consideration. Therefore they do not fulfill the condition presented in section \ref{R-1} for being considered as relevant singularities. Notice that these singularities which are spurious for Cone 4 and Cone 5 are however completely relevant for the calculation of the series associated to Cone 3.

\subsection{The scalar box integral with one external mass}

As a last twofold example, we consider the scalar box integral with one external mass. This divergent integral, computed in dimensional regularisation, will give an example of how the $\epsilon$ singularity may be simply solved graphically with our approach.

This integral is also a nice example since, due to its simplicity, its double series representations may be resummed in terms of simple functions as well as combinations of the Gauss hypergeometric function $_2\rm F_1$. By well-known analytic continuation formulas, this allows to reach in a simple way the regions in the parameters space which are not directly accessible by the series representations.

\begin{figure}[h]
\hspace{1.2cm}
\begin{tabular}{cc}
\includegraphics[width=0.4\textwidth]{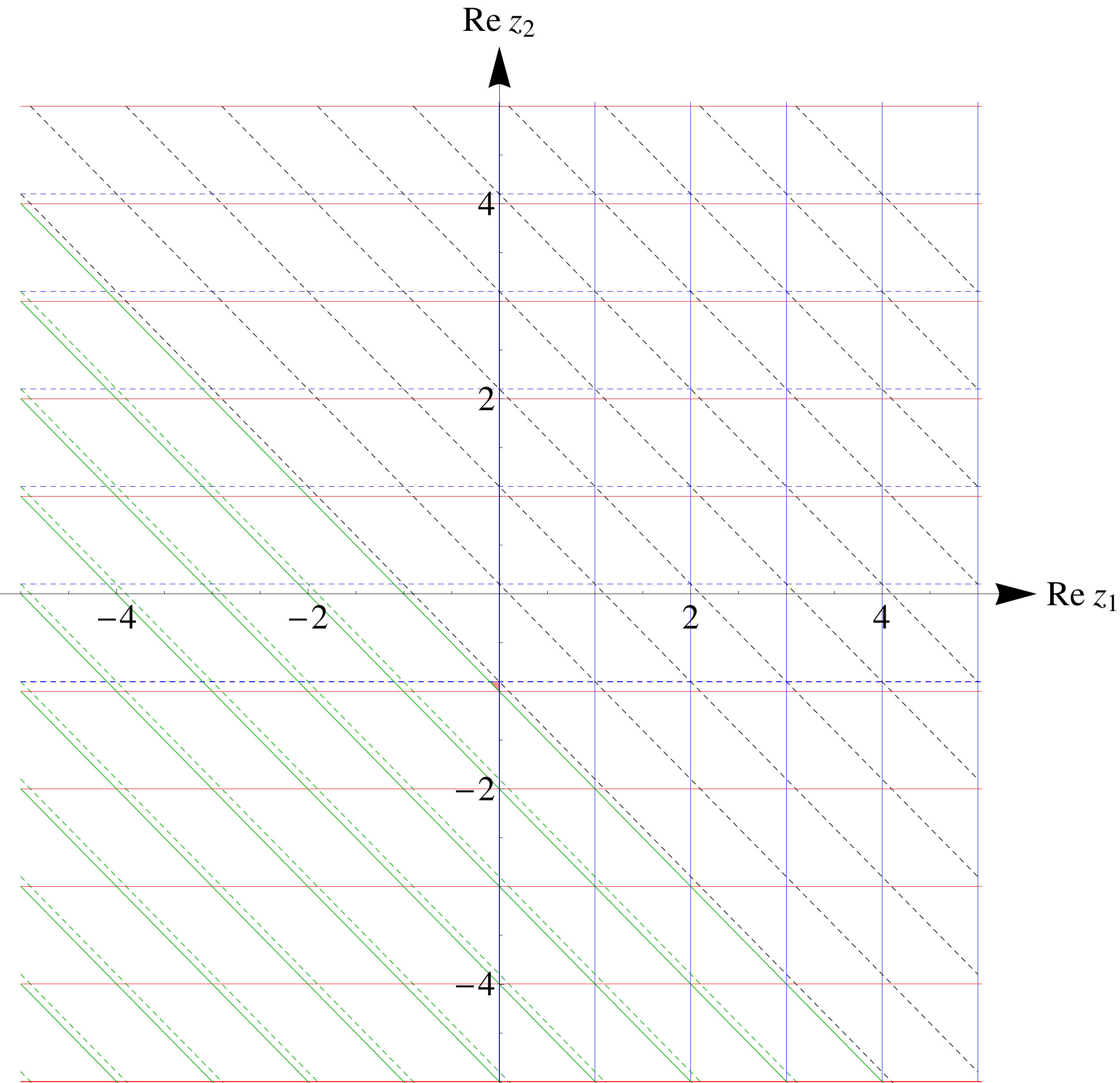} &
\includegraphics[width=0.5\textwidth]{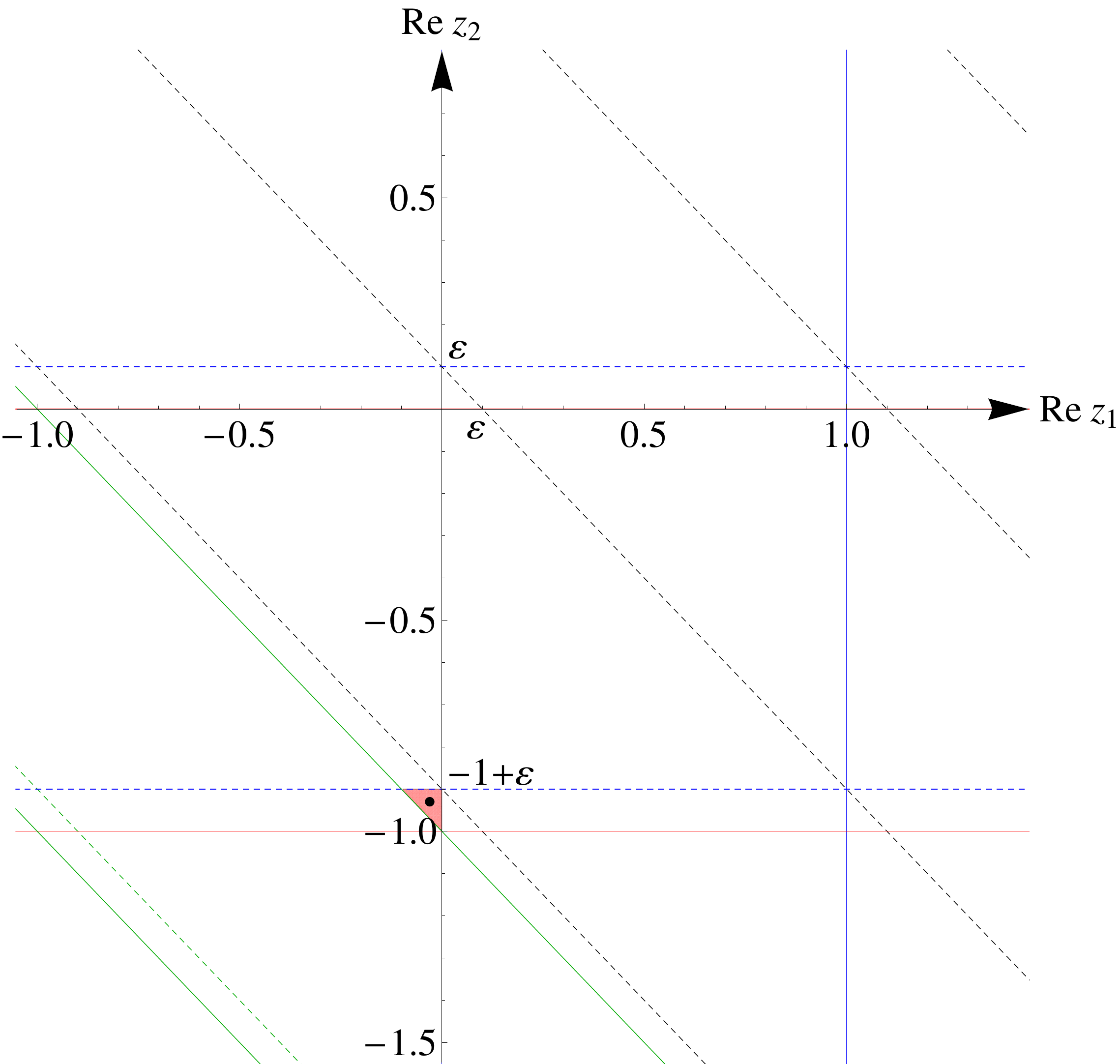} \\
\end{tabular}
\caption{Singular structure of the integrand of the box integral (the right figure is a zoom of the left one, in order to better show the fundamental triangle). The black dot is the point $\gamma=(c,d)$. \label{plane3}}
\end{figure}

Modulo an overall factor, the box integral with one external mass in $D=4+2\epsilon$ dimensions has been expressed in \cite{Valtancoli:2011kr} as
\begin{multline}\label{boxMB}
I(s,t,m^2) = \frac{i}{(4\pi)^{2+\epsilon}}\left(\frac{-t}{\mu^2}\right)^{\epsilon}\frac{(-t)^{-2}}{\Gamma(2\epsilon)}\int\limits_{c-i\infty}^{c+ i\infty}\frac{dz_1}{2i\pi}\int\limits_{d-i\infty}^{d+ i\infty}\frac{dz_2}{2i\pi}\left(\frac{t}{m^2} \right)^{-z_1} \left(\frac{t}{s}\right)^{-z_2}\Gamma(-z_1)\Gamma(-z_2)  \\
\times \Gamma(2-\epsilon + z_1+z_2)\Gamma(\epsilon -1 - z_1 -z_2) \Gamma(1+z_2) \Gamma(\epsilon -1 - z_2) \Gamma(1+z_1+z_2)\;,
\end{multline}
where the conditions $\, c < 0$, $d - \epsilon < -1$ and $c +d > -1 \,$ 
follow from the integration over Feynman parameters and fix the fundamental polyhedron (the red triangle in Figure \ref{plane3}). 

The singular structure of the integrand of (\ref{boxMB}) is represented in Figure \ref{plane3}. 

It is obvious from Figure \ref{plane3} that the $\epsilon$ dependence of the integral implies a splitting of some of the singular lines that would be identical if $\epsilon$ were zero and, therefore, that this will allow to obtain the results in a simpler form since the multiplicities of corresponding the poles are decreased. It is however clear that when $\epsilon\rightarrow0$ the fundamental triangle collapses, giving birth to a pinch singularity located at $(z_1,z_2)=(0,-1)$ which of course reflects the divergence of the integral. 

\begin{figure}[h]
\hspace{1.2cm}
\begin{center}
\begin{tabular}{cc}
\includegraphics[width=0.4\textwidth]{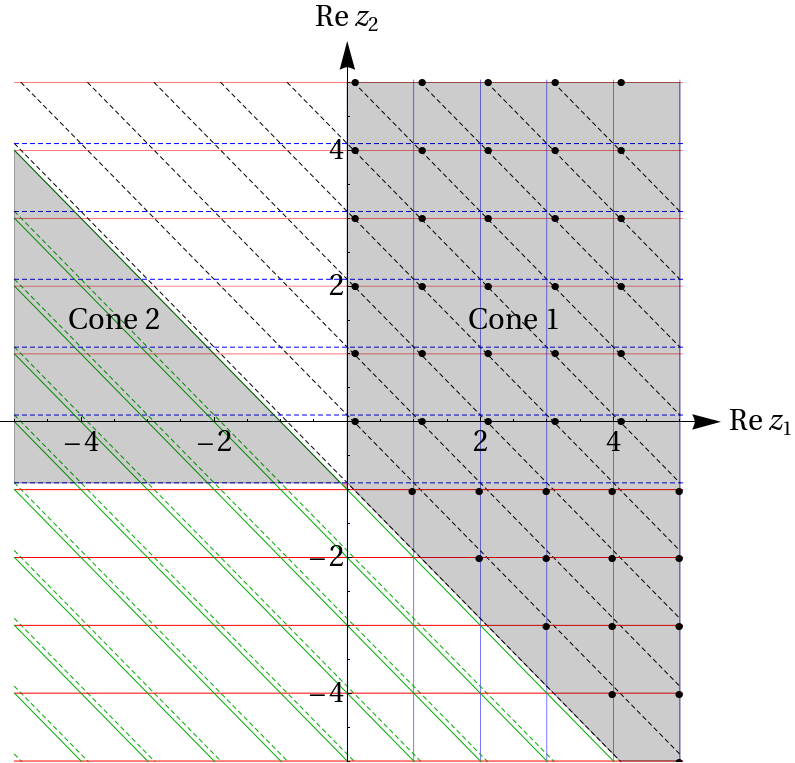} &
\includegraphics[width=0.4\textwidth]{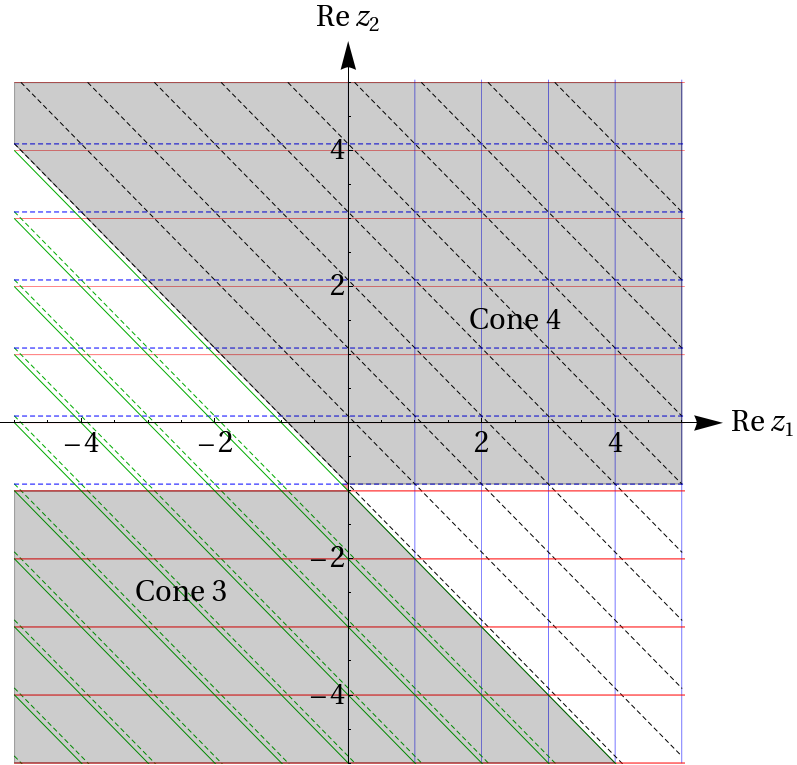} \\
\end{tabular}
\includegraphics[width=0.4\textwidth]{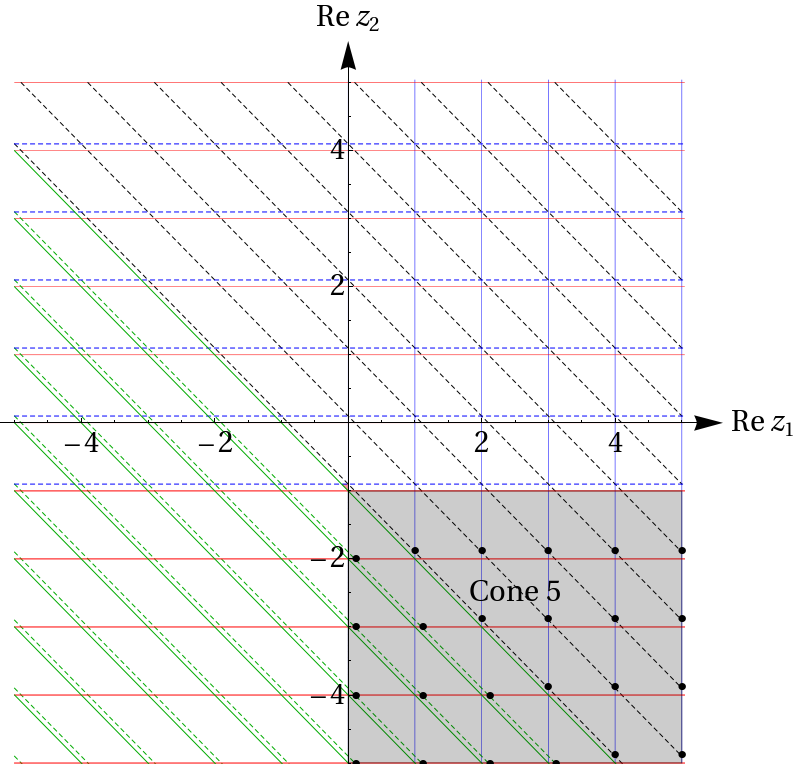}
\caption{The five different cones  of the box integral. The dashed lines are those which have an $\epsilon$ dependence and the small black filled circles distinguish spurious singularities as described in section \ref{reg_conv_section}. \label{conesBox}}
\end{center}
\end{figure}

Drawing the straight $l$-line and performing the rotation around $\gamma=(c,d)$, it is easy to conclude that only five cones may be found, that are represented on Figure \ref{conesBox}. 
Cone 2 corresponds in fact to the calculation performed recently in \cite{Valtancoli:2011kr}, which allows a cross-check of the calculation. Notice also that for each of the three integrals considered until here  in this paper there were five differents cones. However, this is of course not always the case in general.

The series representations of the box integral associated to all the cones but Cone 2 do have polygamma functions in their general term, even if $\epsilon$ is kept at a non-zero value, since they contain subsets of singular points coming from the intersection of three singular lines. The simplest cone is therefore Cone 2 whose singular points are formed only by the intersection of two different singular lines. In fact, its corresponding double series may be resummed in terms of the Gauss hypergeometric function \cite{Valtancoli:2011kr} and this allows, since one knows the $\epsilon$ expansion of this function \cite{Kalmykov:2006hu}, to obtain the $\epsilon$ expansion of the box integral from an exact result. We will compare the first few terms of this expansion with the result that may be obtained with our graphical approach for solving the $\epsilon$ singularities.

Before this we discuss the convergence properties of the series representations of the box integral, and we give the expressions of the series associated to Cone 1 and Cone 2 (the series corresponding to the three other cones have also been obtained easily).

\subsubsection{Regions of convergence}

\begin{figure}[h]
\begin{center}
\includegraphics[width=0.5\textwidth]{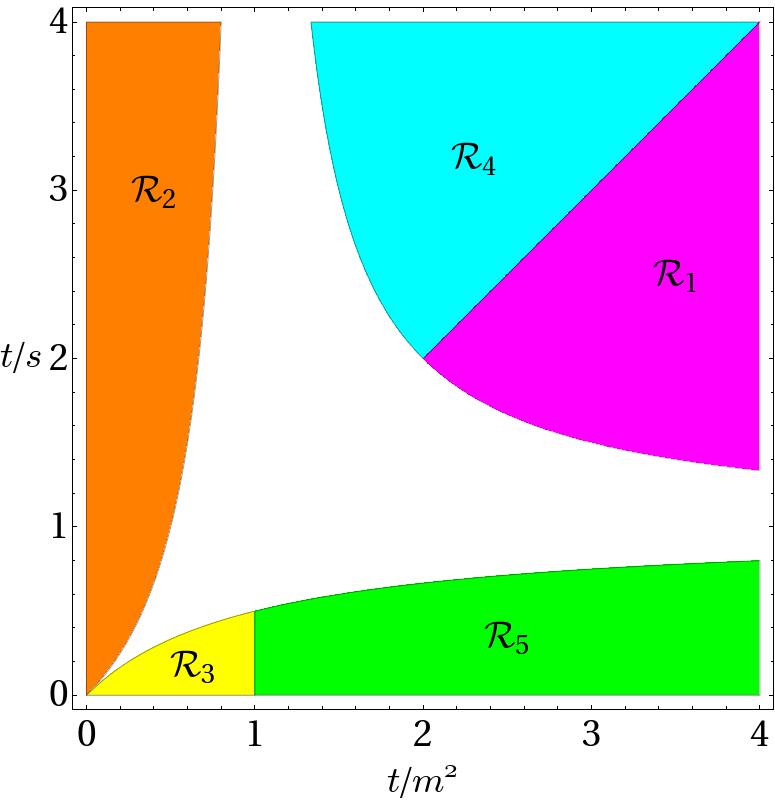} 
\caption{Regions of convergence of the series representations of the box integral obtained from Horn's theorem.} \label{scalarboxregion}
\end{center}
\end{figure}

From Horn's theorem, one gets the following regions of convergence (summarized in Figure \ref{scalarboxregion}):

\paragraph{Cone 1} The region of convergence is given by $\mathcal{R}_1=\{\left|\frac{m^2}{t}\right| + \left|\frac{s}{t}\right| < 1\;\; \text{and}\;\; \left|\frac{m^2}{s}\right| < 1\}$ (magenta region in Figure \ref{scalarboxregion}).

\paragraph{Cone 2} The region of convergence is $\mathcal{R}_2=\{\left|\frac{t}{m^2}\right| + \left|\frac{s}{m^2}\right| < 1\}$ (orange region).

\paragraph{Cone 3} The region of convergence is $\mathcal{R}_3=\{\left|\frac{m^2}{s}\right| + \left|\frac{t}{s}\right| < 1\;\; \text{and}\;\; \left|\frac{t}{m^2}\right|< 1\}$ (yellow region).

\paragraph{Cone 4} The region of convergence is $\mathcal{R}_4=\{\left|\frac{m^2}{t}\right| + \left|\frac{s}{t}\right| < 1\;\; \text{and}\;\; \left|\frac{s}{m^2}\right| < 1\}$ (cyan region).

\paragraph{Cone 5} The region of convergence is $\mathcal{R}_5=\{\left|\frac{m^2}{s}\right| + \left|\frac{t}{s}\right| < 1\;\; \text{and}\;\; \left|\frac{m^2}{t}\right|< 1\}$ (green region).\\

However, as we will see explicitly in sections\footnote{All results given in these sections have been checked by a comparison with the direct numerical computation of (\ref{boxMB}) for particular values of the expansion parameters and $\epsilon$.} \ref{cone1Box} and \ref{cone2Box} for the cases of Cone 1 and Cone 2, 
the double series representations of the box integral may be reduced to combinations of the Gauss hypergeometric function $_2\text{F}_1$ and of some trivial functions. Therefore, well-known analytic continuation formulas  allow to cover the whole $\left(\left|\frac{t}{m^2}\right|,\left|\frac{t}{s}\right|\right)$ plane but a few exceptional singular lines.

To be more precise, we give the following simple concrete example, taken from \cite{Tsikh:1998}, which is also met here in (\ref{type1Cone1Box}) and (\ref{type1Cone2Box}).

Let us consider the series
\begin{equation}\label{TsikhF2_1}
\Phi(t_1,t_2)=\sum_{m=0}^\infty\sum_{m=0}^\infty\frac{(-1)^{m+n}}{m!\ n!}\Gamma(a+m+n)t_1^mt_2^n.
\end{equation}
The region of convergence of this series  is given in \cite{Tsikh:1998} as $\mathcal{R}=|t_1|+|t_2|<1$, which is also what one may conclude from Horn's theorem.

However one may rewrite (\ref{TsikhF2_1}) as
\begin{equation}\label{TsikhF2_2}
\Phi(t_1,t_2)={\rm F_2}(a,\beta,\beta',\beta,\beta';-t_1,-t_2)=\frac{1}{(1+t_1+t_2)^a},
\end{equation}
where $\rm F_2$ is one of the Appell hypergeometric series.
It is clear that the domain of definition of the right hand side of (\ref{TsikhF2_2}) is not restricted to $\mathcal{R}$, and allows an analytic continuation on the whole ($|t_1|,|t_2|$) plane except on the line $1+t_1+t_2=0$.

\subsubsection{Results for Cone 1\label{cone1Box}}

Notice that in this cone there are two different subsets of spurious singularities located at $(z_1,z_2)=(1+k+n,-1-k)$ and $(z_1,z_2)=(\epsilon+k,n)$, where $k, n\geq 0$ are integers. 

Let us now give the results corresponding to the relevant singularities.

\paragraph{Type 1:} the singular points $(z_1,z_2)=(k,n-1+\epsilon)$ where $k, n\geq 0$ are integers.
\begin{align}\label{type1Cone1Box}
I(s,t,m^2)\Big\vert_{Type\, 1}^\text{Cone 1} =
&\frac{i}{(4\pi)^{2+\epsilon}}\left(\frac{-t}{\mu^2}\right)^{\epsilon}\frac{(-t)^{-2}}{\Gamma(2\epsilon)} \left(\frac{t}{s}\right)^{1-\epsilon} \frac{\pi}{\sin(\pi\epsilon)}\nonumber\\
&\times\sum_{k=0}^\infty \sum_{n=0}^\infty \left(\frac{m^2}{t} \right)^{k}\left(-\frac{s}{t} \right)^{n} \frac{\Gamma(n+k+\epsilon)}{\Gamma(1+k)\Gamma(1+n)}\nonumber\\
&\times \left[\ln \frac{t}{s}+ \psi(1+n)+ \psi(1-\epsilon)- \psi(\epsilon)  - \psi(n+k+\epsilon) \right]\nonumber\\
=
&\frac{i}{(4\pi)^{2+\epsilon}}\left(\frac{-t}{\mu^2}\right)^{\epsilon}\frac{(-t)^{-2}}{\Gamma(2\epsilon)} \left(\frac{t}{s}\right)^{1-\epsilon} \frac{\pi}{\sin(\pi\epsilon)}\Gamma(\epsilon)\left(1-\frac{m^2}{t}+\frac{s}{t}\right)^{-\epsilon}\nonumber\\
&\times\left[\ln \frac{t}{s}+  \psi(1-\epsilon)- \psi(\epsilon)+\frac{1}{\epsilon}\left(-1-\frac{t}{s}+\frac{m^2}{s}\right)^{\epsilon}\hyper{2}{1}{\epsilon,\epsilon}{1+\epsilon}{1+\frac{t}{s}-\frac{m^2}{s}}\right.\nonumber\\
&\left.+\ln\left(1-\frac{m^2}{t}\right)+\ln\left(1+\frac{s}{t-m^2}\right)-\ln\left(-1+\frac{m^2}{s}-\frac{t}{s}\right)\right]
\end{align}

\paragraph{Type 2:} the singular points $(z_1,z_2)=(k,n)$ where $k, n\geq 0$ are integers.

\begin{multline}
I(s,t,m^2)\Big\vert_{Type\, 2}^\text{Cone 1} =\frac{i}{(4\pi)^{2+\epsilon}}\left(\frac{-t}{\mu^2}\right)^{\epsilon}\frac{(-t)^{-2}}{\Gamma(2\epsilon)} \\
\times\sum_{k=0}^\infty \sum_{n=0}^\infty \left(-\frac{m^2}{t} \right)^{k}\left(-\frac{s}{t} \right)^{n} \frac{\Gamma(1+k+n)\Gamma(2+k+n-\epsilon)\Gamma(-1-n+\epsilon)\Gamma(-1-k-n+\epsilon)}{\Gamma(1+k)}.
\end{multline}
Using (\ref{reflexion}) this double series may be expressed as a special case of the Appell hypergeometric function $\rm F_2$ which in fact may be reduced to the Gauss hypergeometric function $_2\rm F_1$
\begin{multline}
I(s,t,m^2)\Big\vert_{Type\, 2}^\text{Cone 1} =\frac{i}{(4\pi)^{2+\epsilon}}\left(\frac{-t}{\mu^2}\right)^{\epsilon}\frac{(-t)^{-2}}{\Gamma\left(2\epsilon\right)\Gamma\left(2-\epsilon\right)}\left(\frac{\pi}{\sin(\pi\epsilon)}\right)^2\\
\frac{t}{t-m^2}\hyper{2}{1}{1,1}{2-\epsilon}{\frac{s}{m^2}-1}.
\end{multline}

\paragraph{Type 3:} the singular points $(z_1,z_2)=(\epsilon+k+n,-1-n)$ where $k, n\geq 0$ are integers.
\begin{multline}
I(s,t,m^2)\Big\vert_{Type\, 3}^\text{Cone 1} =\frac{i}{(4\pi)^{2+\epsilon}}\left(\frac{-t}{\mu^2}\right)^{\epsilon}\frac{(-t)^{-2}}{\Gamma(2\epsilon)}  \left(\frac{m^2}{t}\right)^\epsilon\frac{t}{s}\\
\times\sum_{k=0}^\infty \sum_{n=0}^\infty \left(-\frac{m^2}{t} \right)^{k}\left(-\frac{m^2}{s} \right)^{n}
\Gamma(- k-n-\epsilon)\Gamma( k+\epsilon)\Gamma(n+\epsilon).
\end{multline}
Once again this series may be reduced to the $_2\rm F_1$ function with the help of the generalized reflexion formula. It is a special case of the Appell function $\rm F_3$ which can be expressed as a special case of the Appell function $\rm F_1$, and the latter simplifies to the $_2\rm F_1$ function as follows:
\begin{multline}
I(s,t,m^2)\Big\vert_{Type\, 3}^\text{Cone 1} =-\frac{i}{(4\pi)^{2+\epsilon}}\left(\frac{-m^2}{\mu^2}\right)^{\epsilon}\frac{(-t)^{-2}}{\Gamma\left(2\epsilon\right)}\frac{\pi}{\sin(\pi\epsilon)}\frac{\Gamma(\epsilon)}{\epsilon}\frac{t}{s}\\
\hyper{2}{1}{\epsilon,1}{1+\epsilon}{1-\left(1-\frac{m^2}{t}\right)\left(1-\frac{m^2}{s}\right)}.
\end{multline}

\paragraph{Type 4:} the singular points $(z_1,z_2)=(1+k+n,-2-k+\epsilon)$ where $k, n\geq 0$ are integers.
\begin{multline}\label{type4cone2box}
I(s,t,m^2)\Big\vert_{Type\, 4}^\text{Cone 1} = \frac{i}{(4\pi)^{2+\epsilon}}\left(\frac{-t}{\mu^2}\right)^{\epsilon}\frac{(-t)^{-2}}{\Gamma(2\epsilon)} \frac{m^2}{t}\left(\frac{t}{s} \right)^{2-\epsilon}\frac{\pi}{\sin(\pi\epsilon)}\\
\times \sum_{k=0}^\infty \sum_{n=0}^\infty \left(\frac{m^2}{s} \right)^k \left(\frac{m^2}{t} \right)^{n} \frac{ \Gamma (k+1)  \Gamma (n+\epsilon
   )}{\Gamma (k+n+2)}.
\end{multline}
In this case, one first finds that (\ref{type4cone2box}) may be written as a special case of the Appell function $\rm F_3$ which can be expressed as a special case of the Appell function $\rm F_1$, and the latter simplifies to a sum of $_2\rm F_1$ functions as follows:
\begin{align}
I(s,t,m^2)\Big\vert_{Type\, 4}^\text{Cone 1} = \frac{i}{(4\pi)^{2+\epsilon}}\left(\frac{-t}{\mu^2}\right)^{\epsilon}\frac{(-t)^{-2}}{\Gamma(2\epsilon)} \frac{s}{t}\left(\frac{t}{s} \right)^{2-\epsilon}\frac{\pi}{\sin(\pi\epsilon)}\frac{\Gamma(\epsilon)}{\epsilon-1}\frac{t-m^2}{m^2-s-t}\left(\frac{t}{t-m^2}\right)^{\epsilon}\nonumber\\
\times\left[-\hyper{2}{1}{1,1-\epsilon}{2-\epsilon}{1-\frac{s}{s+t-m^2}}+\left(\frac{t}{t-m^2}\right)^{1-\epsilon}\hyper{2}{1}{1,1-\epsilon}{2-\epsilon}{\frac{t}{s+t-m^2}}\right].
\end{align}
\subsubsection{Results for Cone 2\label{cone2Box}}

As we said above, the calculation corresponding to this cone has been considered recently in \cite{Valtancoli:2011kr}. We give anyway the final results since it will be useful in the following. Indeed, this cone allows for the simplest exposition of the way we treat the $\epsilon$ singularities with our computational approach.
 
The double series corresponding to the following four different intersections types are special cases of the Appell $\rm F_2$ function which can each time be simplified. 

\paragraph{Type 1:} the singular points $(z_1,z_2)=(-n-\epsilon-k,n-1+\epsilon)$ where $k, n\geq 0$ are integers. 
\begin{align}\label{type1Cone2Box}
I(s,t,m^2)\Big\vert_{Type\, 1}^\text{Cone 2} =&  \frac{i}{(4\pi)^{2+\epsilon}}\left(-\frac{t}{\mu^2}\right)^{\epsilon}\frac{(-t)^{-2}}{\Gamma(2\epsilon)} \frac{t}{s} \left(\frac{s}{m^2}\right)^{\epsilon}\left(\frac{\pi}{\sin(\pi\epsilon)}\right)^2\nonumber\\
&\times \sum_{k=0}^\infty \sum_{n=0}^\infty \left(\frac{t}{m^2} \right)^k\left(\frac{s}{m^2} \right)^{n} \frac{\Gamma(k+n+\epsilon)}{\Gamma(1+n)\Gamma(1+k)}\nonumber\\
=&  \frac{i}{(4\pi)^{2+\epsilon}}\left(-\frac{t}{\mu^2}\right)^{\epsilon}\frac{1}{\Gamma(2\epsilon)} \frac{1}{st} \left(\frac{\pi}{\sin(\pi\epsilon)}\right)^2\Gamma(\epsilon)\left(\frac{s}{m^2-s-t}\right)^{\epsilon}.
\end{align}

\paragraph{Type 2:} the singular points $(z_1,z_2)=(-1-n-k,n)$ where $k, n\geq 0$ are integers.

\begin{align}
I(s,t,m^2)\Big\vert_{Type\, 2}^\text{Cone 2} =&  \frac{i}{(4\pi)^{2+\epsilon}}\left(-\frac{t}{\mu^2}\right)^{\epsilon}\frac{(-t)^{-2}}{\Gamma(2\epsilon)}\frac{t}{m^2}\left(\frac{\pi}{\sin(\pi\epsilon)}\right)^2\nonumber\\
&\times \sum_{k=0}^\infty \sum_{n=0}^\infty \left(\frac{t}{m^2} \right)^k\left(\frac{s}{m^2} \right)^{n} \frac{\Gamma(k+n+1)}{\Gamma(1+k)\Gamma(2+n-\epsilon)}\nonumber\\
=&  -\frac{i}{(4\pi)^{2+\epsilon}}\left(-\frac{t}{\mu^2}\right)^{\epsilon}\frac{1}{\Gamma(2\epsilon)}\left(\frac{\pi}{\sin(\pi\epsilon)}\right)^2\frac{1}{\Gamma(2-\epsilon)}\nonumber\\
&\times\frac{1}{t(m^2-t)}\hyper{2}{1}{1,1}{2-\epsilon}{\frac{s}{m^2-t}}.
\end{align}

\paragraph{Type 3:} the singular points $(z_1,z_2)=(-2-k-n+\epsilon,n)$ where $k, n\geq 0$ are integers.
\begin{align}
I(s,t,m^2)\Big\vert_{Type\, 3}^\text{Cone 2} =&  \frac{i}{(4\pi)^{2+\epsilon}}\left(-\frac{t}{\mu^2}\right)^{\epsilon}\frac{(-t)^{-2}}{\Gamma(2\epsilon)}\left(\frac{t}{m^2}\right)^{2-\epsilon}\left(\frac{\pi}{\sin(\pi\epsilon)}\right)^2\nonumber\\
&\times \sum_{k=0}^\infty \sum_{n=0}^\infty \left(\frac{t}{m^2} \right)^k\left(\frac{s}{m^2} \right)^{n} \frac{\Gamma(k+n+2-\epsilon)}{\Gamma(k+2-\epsilon)\Gamma(n+2-\epsilon)}\nonumber\\
=&  \frac{i}{(4\pi)^{2+\epsilon}}\left(-\frac{m^2}{\mu^2}\right)^{\epsilon}\frac{1}{\Gamma(2\epsilon)}\left(\frac{\pi}{\sin(\pi\epsilon)}\right)^2\frac{1}{\Gamma(2-\epsilon)}\nonumber\\
&\times\frac{1}{m^2-t}\frac{1}{m^2-s}\hyper{2}{1}{1,1}{2-\epsilon}{\frac{st}{(m^2-t)(m^2-s)}}.
\end{align}

\paragraph{Type 4:} the singular points $(z_1,z_2)=(-1-n-k,n-1+\epsilon)$ where $k, n\geq 0$ are integers.
\begin{align}
I(s,t,m^2)\Big\vert_{Type\, 4}^\text{Cone 2} =& - \frac{i}{(4\pi)^{2+\epsilon}}\left(-\frac{t}{\mu^2}\right)^{\epsilon}\frac{(-t)^{-2}}{\Gamma(2\epsilon)}\frac{t}{m^2}\left(\frac{t}{s}\right)^{1-\epsilon}\left(\frac{\pi}{\sin(\pi\epsilon)}\right)^2\nonumber\\
&\times \sum_{k=0}^\infty \sum_{n=0}^\infty \left(\frac{t}{m^2} \right)^k\left(\frac{s}{m^2} \right)^{n} \frac{\Gamma(k+n+1)}{\Gamma(1+n)\Gamma(2+k-\epsilon)}\nonumber\\
=& - \frac{i}{(4\pi)^{2+\epsilon}}\left(-\frac{s}{\mu^2}\right)^{\epsilon}\frac{1}{\Gamma(2\epsilon)}\left(\frac{\pi}{\sin(\pi\epsilon)}\right)^2\frac{1}{\Gamma(2-\epsilon)}\nonumber\\
&\times\frac{m^2}{m^2-s}\hyper{2}{1}{1,1}{2-\epsilon}{\frac{t}{m^2-s}}.
\end{align}

\subsubsection{Cone 2: $\epsilon$ expansion}
Until here our results have an exact dependence in the $\epsilon$ dimensional regularisation parameter. We will now show how to solve the $\epsilon$ singularities in a simple way from Figure \ref{plane3}, for the case of Cone 2 (obviously the same method may be used for the other cones).

\begin{figure}[h]
\hspace{1.2cm}
\begin{center}
\begin{tabular}{cc}
\includegraphics[width=0.4\textwidth]{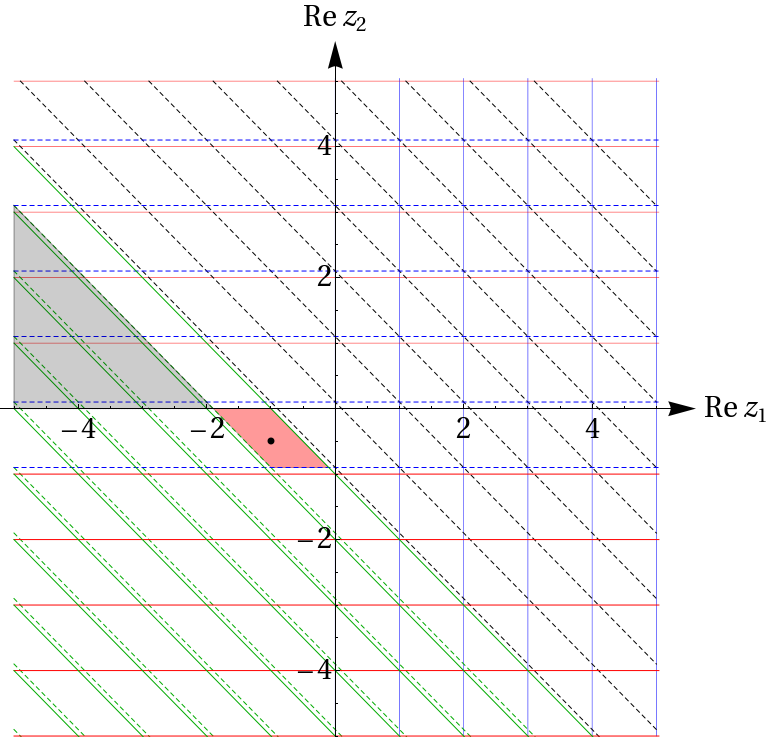} &
\includegraphics[width=0.4\textwidth]{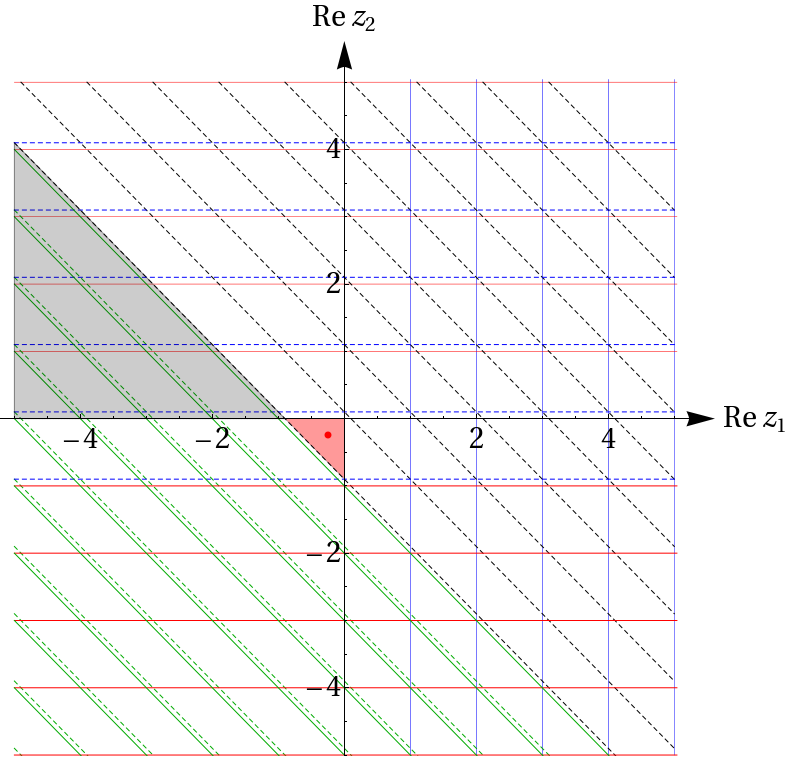} \\
\end{tabular}
\caption{Solving the $\epsilon$ singularities. The black dot is the point $\gamma_1=(c_1,d_1)=(-1,-\frac{1}{2})$, the red one is $\gamma_2=(c_2,d_2)=(-\frac{1}{4},-\frac{1}{4})$. \label{conesBoxshift}}
\end{center}
\end{figure}

Let us first recall that it is easy to compute the first few terms in the $\epsilon$ expansion of the exact results obtained in section \ref{cone2Box}, by using the $\epsilon$ expansion of the Gauss hypergeometric function \cite{Kalmykov:2006hu}
\begin{equation}
\hyper{2}{1}{1,1\pm\epsilon}{2\pm\epsilon}{z} \underset{\epsilon \rightarrow
0}{=} \frac{1\pm\epsilon}{z}\left[-\ln (1-z) \mp \epsilon\; \text{Li}_2\,(z) +\epsilon^2\; \text{Li}_3\,(z)+
\mathcal{O}(\epsilon^3)  \right]\;,
\end{equation}
and one finds that the divergent part of the result associated to Cone 2 is of the form $\frac{A_{-2}}{\epsilon^2}+\frac{A_{-1}}{\epsilon}$ where
\begin{equation}
A_{-2}=\frac{i}{8\pi^2st}
\end{equation}
and
\begin{equation}
A_{-1}=\frac{i}{8\pi^2st}\left[\gamma_E-\ln(4\pi)+\ln\left(-\frac{st}{m^2\mu^2}\right)\right].
\end{equation}
In fact, without using the exact result as a starting point, it is very simple to get these terms (\textit{i.e} to solve the $\epsilon$ singularities), just by having a look at the picture representing the singular structure of the integral, and the one where Cone 2 is shown.
Indeed since the fundamental triangle collapses when $\epsilon$ goes to zero (see Figure \ref{plane3}), to avoid the pinch singularity it is necessary to "move" the $\gamma$ point outside the fundamental triangle in order to put it in a safe region. There are many ways to do this and, as an example, we show two different possibilities in Figure \ref{conesBoxshift}, where one can see that the new $\gamma$ points ($\gamma_1$ and $\gamma_2$) will not be pinched if $\epsilon$ is taken equal to zero. 

Now, if one looks for the cones associated to the integral $I(s,t,m^2)$ with $\gamma$ replaced by $\gamma_1$ (resp. $\gamma_2$), one finds for the relevant cone\footnote{By relevant cone we mean the cone which has the largest overlap with Cone 2.} the region $C_2$ (resp $C_2'$) (see Figure \ref{conesBoxshift}).
It is then obvious that
\begin{equation}
I(s,t,m^2)\vert_{\gamma}^{\text{Cone 2}} = I(s,t,m^2)\vert_{\gamma_1}^{C_2}+Res._1
\end{equation}
or 
\begin{equation}
I(s,t,m^2)\vert_{\gamma}^{\text{Cone 2}} = I(s,t,m^2)\vert_{\gamma_2}^{C_2'}+Res._2,
\end{equation}
where
\begin{equation}
Res._1\doteq \sum_{k=0}^{\infty}\left[Res.\vert_{(-\epsilon-k,-1+\epsilon)}+Res.\vert_{(-1-k,-1+\epsilon)}+Res.\vert_{(-1-k,k)}+Res.\vert_{(-1-\epsilon-k,k+\epsilon)}\right]
\end{equation}
and
\begin{equation}
Res._2\doteq \sum_{k=0}^{\infty}\left[Res.\vert_{(-\epsilon-k,-1+\epsilon)}+Res.\vert_{(-1-k,-1+\epsilon)}-Res.\vert_{(-1-k,k+\epsilon)}-Res.\vert_{(-1-\epsilon-k,k)}\right]
\end{equation}
and that the $\frac{1}{\epsilon^2}$ and $\frac{1}{\epsilon}$ terms will of course be entirely given by $Res._1$ or by $Res._2$ (which we checked explicitly in both cases).

\section{Mellin-Barnes integrals of higher dimension\label{MBofHD}}

In this section we give a foretaste of the extension of the method to Mellin-Barnes integrals of higher dimension. We only consider here a simple threefold integral, as a toy model, and one of its $n$-dimensional extension, since the general procedure is presently under study a will be treated in a forthcoming publication. 

\subsection{A toy integral}

Let us consider the integral
\begin{multline}
\label{toy}
I(u_1,u_2,u_3)= \int\limits_{c-i\infty}^{c+i\infty} {d s\over 2\pi i}\int\limits_{d-i\infty}^{d+i\infty}{d t\over 2\pi i} \int\limits_{e-i\infty}^{e+i\infty}{d u\over 2\pi i}\\\\
\times u_1^{s}\; u_2^{t}\; u_3^{u}\; \Gamma^2 \left(-s\right)\, \Gamma^2 \left(-t\right)\, \Gamma^2 \left(-u\right)\, \Gamma^2 \left(1+s+t+u\right)\,,
\end{multline}
where $c=-\frac{1}{8}$, $d=-\frac{1}{9}$ and $e=-\frac{1}{10}$.

As before, to obtain the triple series representations of (\ref{toy}) we have to find the different cones associated to this integral, in order to know the different convergence regions as well as over which residues the corresponding triple sums will be performed in each case. However, the singular structure of the integrand in (\ref{toy}) can not be represented in a picture as readable as, for instance, Figure \ref{plane2} and, moreover, we do not have a simple geometrical procedure to obtain the cones as the one used in the twofold case. Nevertheless it is possible to obtain them from an exhaustive search of the possible simultaneous constraints coming from the different triplets of gamma functions, in the same spirit as what was done in the beginning of section \ref{det_cone}. Indeed, only the following four different regions of the $(\textrm{Re}\ s, \textrm{Re}\ t, \textrm{Re}\ u)$ three dimensional space have to be considered:
\vspace{0.5cm}

$\bullet$ Cone 1: $\textrm{Re}\ s> 0$, $\textrm{Re}\ t> 0$ and $\textrm{Re}\ u> 0$.

$\bullet$ Cone 2: $\textrm{Re}\ s> 0$, $\textrm{Re}\ t> 0$ and $\textrm{Re}\ (1+s+t+u)< 0$.

$\bullet$ Cone 3: $\textrm{Re}\ s> 0$, $\textrm{Re}\ u> 0$ and $\textrm{Re}\ (1+s+t+u)< 0$.

$\bullet$ Cone 4: $\textrm{Re}\ t> 0$, $\textrm{Re}\ u> 0$ and $\textrm{Re}\ (1+s+t+u)< 0$.
\vspace{0.5cm}
 
\begin{figure}[h]
\begin{center}
\includegraphics[width=0.5\textwidth]{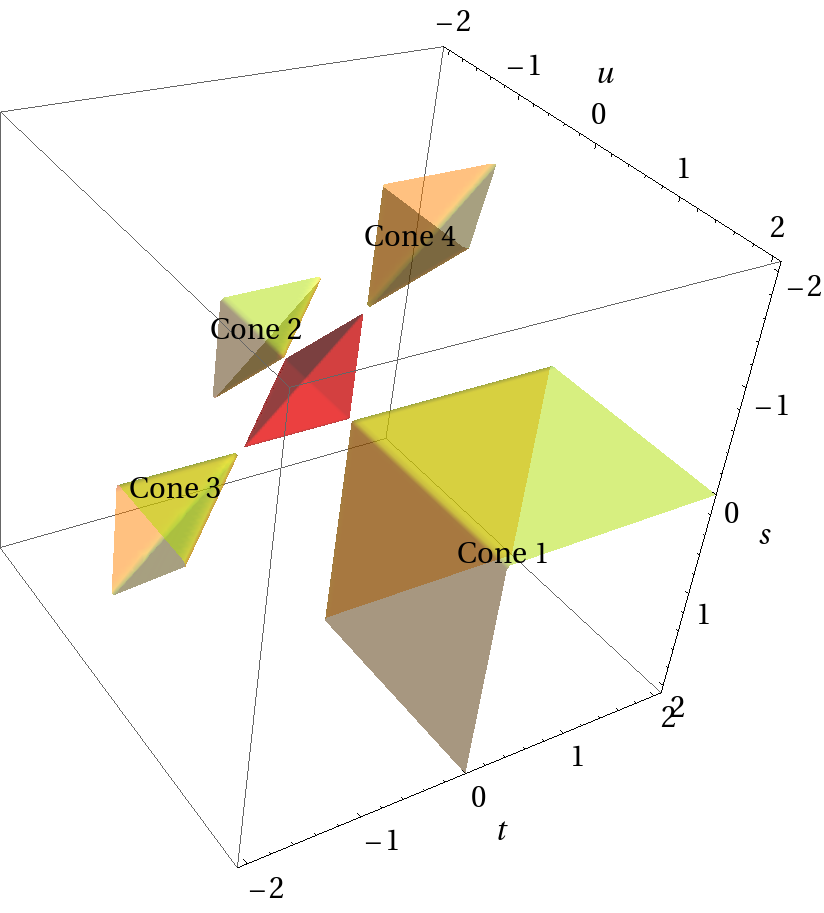}
\end{center}
\caption{The fundamental tetrahedron and different cones of $I(u_1,u_2,u_3)$.\label{toy3D}}
\end{figure}

As can be seen on Figure \ref{toy3D} the cones are in fact infinite pyramids, the fundamental polyhedron of the integral, given by the constraint $\textrm{Re}\ s< 0$, $\textrm{Re}\ t< 0$, $\textrm{Re}\ u< 0$ and $\textrm{Re}\ (1+s+t+u)> 0$, being a tetrahedron.

It is easy to see that the singularities in Cone 1 are only of Cauchy type, therefore the corresponding residues will be trivial to compute. The three other cones, however, will necessitate a three dimensional extension of the transformation law. In fact, due to the symmetry of the integral,  one will see that in principle only one cone has to be computed, the three others being deduced from the latter by simple changes of variables. This will give us the opportunity to check that our implementation of the transformation law at the 3-dimensional level is correct. 

\subsubsection{Cone 1}

Let us begin by the calculation of the triple series corresponding to Cone 1. 
In this simple example there is only one type of intersection in the whole cone: the points $(s,t,u)=(m,n,p)$ where $m, n, p\geq 0$ are integers.

Performing the change of variables $s\mapsto s+m, t\mapsto t+n, u\mapsto u+p$ and using (\ref{reflexion}) one has
\begin{equation}
  \omega=\frac{h(s,t,u)}{s^2t^2u^2}\,ds\wedge dt\wedge du
\end{equation}
where
\begin{multline}
h(s,t,u)\doteq u_1^{s+m}\, u_2^{t+n}\, u_3^{u+p}\, \frac{\Gamma^2 \left(1+s\right)\Gamma^2 \left(1-s\right)}{\Gamma^2 \left(s+m+1\right)}\, \frac{\Gamma^2\left(1+t\right)\Gamma^2 \left(1-t\right)}{\Gamma^2\left(t+n+1\right)}\\
\times\frac{\Gamma^2 \left(1+u\right)\Gamma^2\left(1-u\right)}{\Gamma^2\left(u+p+1\right)}\Gamma^2\left(s+t+u+m+n+p+1\right)\,.
\end{multline}
Using now the three dimensional extension of Cauchy Formula (\ref{cauchy}) one has
\begin{equation}
I(u_1,u_2,u_3)\big\vert_{\textrm{Cone 1}}=-\sum_{m=0}^{\infty}\sum_{n=0}^{\infty}\sum_{p=0}^{\infty}\left.\frac{\partial^3h(s,t,u)}{\partial s\ \partial t\ \partial u}\right\vert_{(0,0,0)},
\end{equation}
the overall sign coming from the $(-1)^3$ due to the closing of the integration contours.

It is then straightforward to show that
\begin{flalign}
&I(u_1,u_2,u_3)\big\vert_{\textrm{Cone 1}}\nonumber\\
&=-\sum_{m=0}^{\infty}\sum_{n=0}^{\infty}\sum_{p=0}^{\infty} \;u_1^mu_2^nu_3^p\; \frac{\Gamma^2(1+m+n+p)}{\Gamma^2(1+m)\Gamma^2(1+n)\Gamma^2(1+p)} \nonumber\\
&\times \Big\{f(u_1,m+n+p,m)f(u_2,m+n+p,n)f(u_3,m+n+p,p) \nonumber\\
&+2\psi^{(1)}(1+m+n+p)\big[f(u_1,m+n+p,m)+f(u_2,m+n+p,n)+f(u_3,m+n+p,p)\big] \nonumber\\
&+2\psi^{(2)}(1+m+n+p)\Big\}. \label{cauchy3D}
\end{flalign}
where
\begin{equation}
\label{f}
f(x,y,z)\doteq\ln x+2\psi(1+y)-2\psi(1+z).
\end{equation}

From the results in sections \ref{AlmostHorn} and \ref{Horn3D} this triple sum converges in the region $\sqrt{|u_1|}+\sqrt{|u_2|}+\sqrt{|u_3|}<1$ (notice that this region of convergence may also be obtained from the statements of section \ref{Kampe}, since the series with general term $h(0,0,0)$ is a triple Kampé de Fériet series).

For particular numerical values of the parameters $u_1$, $u_2$ and $u_3$ in the region of convergence we checked that one recovers the value obtained from a direct numerical evaluation of the integral (\ref{toy}), with the same values of the parameters, to a very good accuracy.

\subsubsection{Cone 3}

Once more, in the simple example of (\ref{toy}) there is only one type of intersection in the whole Cone 3: the points $(s,t,u)=(m,-1-m-n-p,n)$ where $m, n, p\geq 0$ are integers.

Performing the change of variables $s\mapsto s+m, t\mapsto t-1-m-n-p, u\mapsto u+n$ and using (\ref{reflexion}) one has
\begin{equation}\nonumber
  \omega=\frac{h(s,t,u)}{s^2u^2(s+t+u)^2}\,ds\wedge dt\wedge du
\end{equation}
where
\begin{multline}
h(s,t,u)\doteq u_1^{s+m}\, u_2^{t-1-m-n-p}\, u_3^{u+n}\, \frac{\Gamma^2 \left(1+s\right)\Gamma^2 \left(1-s\right)}{\Gamma^2 \left(s+m+1\right)}\,\Gamma^2 \left(1+m+n+p-t\right) \\
\times\, \frac{\Gamma^2\left(1+u\right)\Gamma^2 \left(1-u\right)}{\Gamma^2\left(u+n+1\right)}\, \Gamma^2 \left(s+t+u+m+n+p+1\right)\,.
\end{multline}
We now have to apply the transformation law, therefore we have to fix the 3-vector $\boldsymbol f$ and choose a $3\times3$ matrix $A$ and a 3-vector $\boldsymbol g$ as simple as possible, in order to use the natural three dimensional extension of (\ref{TL}). In our present case of study the 3-vector $\boldsymbol f$ may be obtained without a general procedure as the one presented for the twofold case 
since there are only three different sources of divergent behaviour: $\frac{1}{s^2}$, $\frac{1}{u^2}$ and $\frac{1}{(s+t+u)^2}$. Therefore, one finds $\boldsymbol{f}=\big(s^2,u^2,(s+t+u)^2\big)^T$. Although not completely straightforward, it is also not too hard to find a relevant matrix $A$ and its corresponding vector $g$. For instance $\boldsymbol{g}=(s^2,t^4,u^2)^T$ and the matrix elements
\begin{flalign}
&A_{11}=A_{32}=1\nonumber\\
&A_{12}=A_{13}=A_{31}=A_{33}=0\nonumber\\
&A_{21}=2st(u+4)+s^2(3+t)+2t^2(3+s+u)-6u^2+t^3,\nonumber\\
&A_{22}=-2tu(s-4)+2t^2(3-s-u)-u^2(t-3)-t^3,\nonumber\\
&A_{23}=-2s(t-3u)-2tu-s^2(3+t)+t^2+u^2(t-3).
\end{flalign}

fullfill all the requirements of the transformation law.

Since $\det A=3 s^2 + 2 s t + s^2 t - t^2 - 6 s u + 2 t u + 3 u^2 - t u^2$, one then finds, keeping only the non-zero contributions
\begin{multline}
\Res{\frac{h(s,t,u)}{s^2u^2(s+t+u)^2}}=-\Res{\frac{6h(s,t,u)}{st^4u}}+\Res{\frac{2h(s,t,u)}{st^3u^2}}\\
+\Res{\frac{2h(s,t,u)}{s^2t^3u}}-\Res{\frac{h(s,t,u)}{s^2t^2u^2}}\;,
\end{multline}
so that
\begin{multline}
I(u_1,u_2,u_3)\big\vert_{\textrm{Cone 3}}=\sum_{m=0}^{\infty}\sum_{n=0}^{\infty}\sum_{p=0}^{\infty}\Bigg\{\left.-\frac{\partial^3h(s,t,u)}{\partial t^3}\right\vert_{(0,0,0)}+\left.\frac{\partial^3h(s,t,u)}{\partial t^2\ \partial u}\right\vert_{(0,0,0)}\\
+\left.\frac{\partial^3h(s,t,u)}{\partial s\ \partial t^2}\right\vert_{(0,0,0)}
-\left.\frac{\partial^3h(s,t,u)}{\partial s\ \partial t\ \partial u}\right\vert_{(0,0,0)}\Bigg\}.
\end{multline}

Using the same way to write the derivatives than in section 2, one has 
\begin{flalign}
&I(u_1,u_2,u_3)\big\vert_{\textrm{Cone 3}}\nonumber\\
&=-\sum_{m=0}^{\infty}\sum_{n=0}^{\infty}\sum_{p=0}^{\infty}\; u_1^m u_2^{-1-m-n-p} u_3^n\; \frac{\Gamma^2(1+m+n+p)}{\Gamma^2(1+m)\Gamma^2(1+n)\Gamma^2(1+p)}\nonumber\\
&\times\left\{f\left(\frac{u_1}{u_2},m+n+p,m\right)f\left(\frac{1}{u_2},m+n+p,n\right)f\left(\frac{u_3}{u_2},m+n+p,p\right)\right.\nonumber\\
&\hspace{1cm}+2\psi^{(1)}(1+m+n+p)\left[f\left(\frac{u_1}{u_2},m+n+p,m\right)+f\left(\frac{1}{u_2},m+n+p,n\right)\right.\nonumber\\
&\hspace{5.5cm}\left.\left.+f\left(\frac{u_3}{u_2},m+n+p,p\right)\right]+2\psi^{(2)}(1+m+n+p)\right\},\label{law3D}
\end{flalign}
where $f(x,y,z)$ is given in (\ref{f}) and which is nothing but (\ref{cauchy3D}) multiplied by $\frac{1}{u_2}$, with $u_1\rightarrow\frac{u_1}{u_2}$, $u_2\rightarrow\frac{1}{u_2}$ and $u_3\rightarrow\frac{u_3}{u_2}$.

The triple sum (\ref{law3D}) converges in the region $\sqrt{\left|\frac{u_1}{u_2}\right|}+\sqrt{\left|\frac{1}{u_2}\right|}+\sqrt{\left|\frac{u_3}{u_2}\right|}<1$.

Once more, we mention that all results have been checked numerically.

Moreover, we see that in the simple case of (\ref{toy}), the triple series representation corresponding to Cone 3 could have been derived from the one obtained from Cone 1. This is of course due to the fact that under the change of variable $s\rightarrow-1-s-t-u$ we have
\begin{equation}
I(u_1,u_2,u_3) = \frac{1}{u_2}I\left(\frac{u_1}{u_2},\frac{1}{u_2},\frac{u_3}{u_2}\right)
\end{equation}
and it provides a check that our implementation of the 3-dimensional transformation law is correct.

Thanks to this nice symmetry of the integral, which may also be observed at the $n$-dimensional level in (\ref{toyND}), it is then possible de get the $n$-dimensional version of (\ref{law3D})  directly from eq. (\ref{cauchyND}), by mutiplying it by $\frac{1}{u_2}$ and by doing a replacement of the arguments similar as the one we did above.

\subsection{Extension to $n$ dimensions}
Notice that, if not totally obvious, it is however possible to get the $n$-dimensional extension of (\ref{cauchy3D}) corresponding to the integral
\begin{multline}
\label{toyND}
I(u_1,...,u_n) = \int\limits_{c_1-i\infty}^{c_1+i\infty} {d s_1\over 2\pi i}\cdots\int\limits_{c_n-i\infty}^{c_n+i\infty}\;{d s_n\over 2\pi i}\\\\
\times u_1^{s_1}\cdot\cdot\cdot u_n^{s_n}\; \Gamma^p(-s_1)\cdots\Gamma^p(-s_n)\ \Gamma^p\left(1+s_1+ \cdots + s_n\right)
\end{multline}
where the $c_j\ (j=1,...,n)$ belong to the $n$-dimensional fundamental hyperpolyhedron (which reduces to the red tetrahedron of Figure \ref{toy3D} when $n=3$) and $p$ is an arbitrary positive integer.

Indeed, the residues to compute in this case, in the $n$-dimensional equivalent of Cone 1 defined by $\textrm{Re}\ s_1>0, ..., \textrm{Re}\ s_n> 0$, are those of the singular points  $(s_1,...,s_n)=(M_1,...,M_n)$ where $M_1,..., M_n\geq 0$ are integers.
Performing the change of variables $s_j\mapsto s_j+M_j\ (j=1,...,n)$ and using (\ref{reflexion}) one has
\begin{equation}
\omega=\frac{h(s_1,...,s_n)}{s_1^p\cdot\cdot\cdot s_n^p}\,ds_1\wedge \cdots \wedge ds_n
\end{equation}
where
\begin{multline}
h(s_1,...,s_n)\doteq u_1^{s_1+M_1}\cdots u_n^{s_n+M_n}\; \frac{\Gamma^p\left(1+s_1\right)\Gamma^p\left(1-s_1\right)}{\Gamma^p \left(s_1+M_1+1\right)}\times\cdots\times\frac{\Gamma^p\left(1+s_n\right)\Gamma^p\left(1-s_n\right)}{\Gamma^p \left(s_n+M_n+1\right)}\\
\times\; \Gamma^p \left(1+M_1+\cdots +M_n+s_1+ \cdots + s_n\right)\,.
\end{multline}
The $n$-dimensional Cauchy formula then allows to write
\begin{equation}\label{partialND}
I(u_1,...,u_n)\big\vert_{\textrm{Cone 1}}=(-1)^n\sum_{M_1=0}^{\infty}\cdot\cdot\cdot\sum_{M_n=0}^{\infty}\frac{1}{(p-1)!^n}\left.\frac{\partial^{n(p-1)}h(s_1,...,s_n)}{\partial s_1^{p-1}\cdot\cdot\cdot\partial s_n^{p-1}}\right\vert_{(0,...,0)}.
\end{equation}

To compute the multiple partial derivative, one needs the following relation \cite{Hardy}
\begin{equation}
{\partial^{k_1+\cdots+k_n} \over \partial x_1^{k_1}\,\cdots\,
\partial x_n^{k_n}} (uv) 
=\sum_{\ell_1=0}^{k_1}\cdots\sum_{\ell_n=0}^{k_n}
{k_1 \choose \ell_1}\cdots {k_n \choose \ell_n}
{\partial^{\ell_1+\cdots+\ell_n} u
\over \partial x_1^{\ell_1}\,\cdots\,\partial x_n^{\ell_n}}
\cdot {\partial^{k_1-\ell_1+\cdots+k_n-\ell_n} v
\over \partial x_1^{k_1-\ell_1}\,\cdots
\,\partial x_n^{k_n-\ell_n}}
\end{equation}
as well as the Fa\`a di Bruno formula
\begin{equation}\label{Faa.di.Bruno}
{d^k \over dx^k} f(y)=\sum
{k! \over 1!^{m_1}\cdots k!^{m_k} m_1!\cdots m_k!}
f^{(m_1+\cdots+m_k)}(y)\prod_{j=1}^{k}
{\left(d^{j} y \over dx^{j}\right)^{m_j}}
\end{equation}
where the sum runs over all $k$-tuples $(m_1,\dots,m_k)$
of non-negative integers satisfying the constraint
$m_1+2m_2+3m_3+\cdots+km_k=k$, and the Leibniz rule
\begin{equation}\label{Leibniz}
\frac{d^k}{dx^k}(uv)=\sum_{\ell=0}^k{k \choose \ell}
\frac{d^\ell u}{dx^\ell}\cdot\frac{d^{k-\ell} v}{dx^{k-\ell}}.
\end{equation}
With the help of these relations, the calculation of (\ref{partialND}), if a bit heavy, is however easy to perform and one finally finds
\begin{equation}\nonumber
I(u_1,...,u_n)\big\vert_{\textrm{Cone 1}}=\frac{(-1)^n}{(p-1)!^n}\sum_{M_1=0}^{\infty}\cdots\sum_{M_n=0}^{\infty}\sum_{k_1=0}^{p-1}\cdots\sum_{k_n=0}^{p-1}
{p-1 \choose k_1}\cdots {p-1 \choose k_n}
\end{equation}
\begin{equation}
\nonumber
\prod_{j=1}^n\left\{u_j^{M_j}\frac{1}{\Gamma(M_j+1)^2}\sum_{l=0}^{k_j}{k_j \choose l}\ln^l u_j\sum_{q=0}^{k_j-l}{k_j-l \choose q}\left[\sum\frac{q!}{1!^{m_1}\cdots q!^{m_q}m_1!\cdots m_q!}\right.\right.
\end{equation}
\begin{equation}
\nonumber
\left.\prod_{i=1}^q\left(-2\psi^{(i-1)}(M_j+1)\right)^{n_i}\right]\sum_{r=0}^{k_j-l-q}{k_j-l-q \choose r}\left[\sum\frac{r!}{1!^{m_1}\cdots r!^{m_r}m_1!\cdots m_r!}\right.
\end{equation}
\begin{equation}
\nonumber
\left.\prod_{i=1}^r\left(2\psi^{(i-1)}(1)\right)^{m_i}\right]\left[\sum\frac{(k_j-l-q-r)!}{1!^{m_1}\cdots (k_j-l-q-r)!^{m_{k_j-l-q-r}}m_1!\cdots m_{k_j-l-q-r}!}\right.
\end{equation}
\begin{equation}
\nonumber
\left.\left.\prod_{i=1}^{k_j-l-q-r}\left(-2\psi^{(i-1)}(1)\right)^{m_i}\right]\right\}\Gamma\left(1+\sum_{j=1}^{n}M_j\right)^2\left[\sum\frac{A!}{1!^{m_1}\cdots A!^{m_A}m_1!\cdots m_A!}\right.
\end{equation}
\begin{equation}
\label{cauchyND}
\left.\prod_{i=1}^{A}\left(2\psi^{(i-1)}\left(1+\sum_{j=1}^{n}M_j\right)\right)^{m_i}\right],
\end{equation}
where $A=n(p-1)-\sum_{j=1}^{n}k_j$ and where the sums without indices are understood as for (\ref{Faa.di.Bruno}).

Although the expression (\ref{cauchyND}) looks a bit complicated, it is straightforward to extract from it simple formulas like (\ref{cauchy3D}) for particular values of $p$ and $n$.

More generally, the results corresponding to the two other cones of Figure \ref{toy3D}, as well as those of the $n-2$ other cones of the $n$-dimensional integral (\ref{toyND}), may be straightforwardly obtained from (\ref{cauchy3D}) and its $n$-dimensional version (\ref{cauchyND}) by simply exchanging the arguments as follows: $u_1\rightarrow\frac{u_1}{u_j},..., u_j\rightarrow\frac{1}{u_j},..., u_n\rightarrow\frac{u_n}{u_j}$. Of course one also has to multiply by an overall $\frac{1}{u_j}$ factor.

\section{Conclusions}

In this paper, we have shown with a general method how to derive, from a given twofold MB integral, several convergent series representations as well as their associated regions of convergence, the latter being obtainable before a full computation of the general term of the series. MB integrals of higher dimension have also been considered.

In the twofold case, the method may be based on a very simple graphical approach: once the singular structure of the integrand has been drawn, it is easy to perform the extraction of different sets of poles whose residues give the mathematical expressions of the different series. The extension of the method to MB integrals of higher dimension is possible in principle although, obviously, one cannot keep a graphical view, at least when the dimension is greater than 3. 

In each of the examples treated for illustration, we began by looking for a parametrization of the relevant sets of singular points and deduced straightforwardly the corresponding regions of convergence of the associated series. Then we computed the residues using simple results of multidimensional complex analysis. 

The different series are analytic continuations of one another and an interesting point of the method is that, as we have shown, one does not need to perform the resummation of the series corresponding to a given region of convergence and to use analytic continuation properties of the result, in order to find the expression of the series associated to another region.

\paragraph{Acknowledgements} We would like to dedicate this work to Philippe Flajolet, who passed away recently. Part of his work has been influential in driving us to the study of Mellin-Barnes integrals. This work has been supported by MICINN (grant FPA2009-09638) and
DGIID-DGA (grant 2009-E24/2) and by the Spanish Consolider-Ingenio 2010 Program
CPAN (CSD2007-00042).  

\appendix 

\section{\label{Formulary}Formulary}

In this appendix, we give some of the conventions and a few simple formulas that have been used to organize the calculations in this paper.

Let us consider the function
\begin{equation}\label{h_formulary}
h(s,t)= x^{-s}y^{-t}\frac{\prod_{j=1}^m\Gamma^{q_j}(a_js+b_jt+e_j)}{\prod_{k=1}^p\Gamma^{r_k}(c_ks+d_kt+f_k)}\;,
\end{equation}

where $a_j, b_j, e_j, c_k, d_k$ and $f_k$ are real numbers\footnote{In particle physics perturbative calculations, $a_j, b_j, c_k$ and $d_k$ are integers.} and $q_j$ and $r_k$ are positive integers, and let us define 
\begin{equation}
h^{(1,0)}(s,t)\doteq \frac{\partial}{\partial s}h(s,t)\hspace{1cm} \text{and} \hspace{1cm}h^{(0,1)}(s,t)\doteq \frac{\partial}{\partial t}h(s,t).
\end{equation}
Then
\begin{equation}\label{h10}
h^{(1,0)}(s,t)= h(s,t)\mathsf{A}(s,t)\hspace{1cm} \text{and} \hspace{1cm}h^{(0,1)}(s,t)=h(s,t)\mathsf{B}(s,t),
\end{equation}
where
\begin{equation}\label{A}
\mathsf{A}(s,t)=-\ln x+\sum_{j=1}^mq_ja_j\psi(a_js+b_jt+e_j)-\sum_{k=1}^pr_kc_k\psi(c_ks+d_kt+f_k)
\end{equation}
and
\begin{equation}\label{B}
\mathsf{B}(s,t)=-\ln y+\sum_{j=1}^mq_jb_j\psi(a_js+b_jt+e_j)-\sum_{k=1}^pr_kd_k\psi(c_ks+d_kt+f_k).
\end{equation}
Obviously 
\begin{equation}\label{Akl}
\mathsf{A}^{(k,l)}(s,t)=\sum_{j=1}^mq_ja_j^{k+1}b_j^{l}\psi^{(k+l)}(a_js+b_jt+e_j)-\sum_{k=1}^pr_kc_k^{k+1}d_k^{l}\psi^{(k+l)}(c_ks+d_kt+f_k)
\end{equation}
and
\begin{equation}\label{Bkl}
\mathsf{B}^{(k,l)}(s,t)=\sum_{i=1}^mq_ja_j^{k}b_j^{l+1}\psi^{(k+l)}(a_js+b_jt+e_j)-\sum_{k=1}^pr_kc_k^kd_k^{l+1}\psi^{(k+l)}(c_ks+d_kt+f_k).
\end{equation}
 
Now (we avoid to write the $s$ and $t$ dependence in the following),
\begin{equation}\label{h11}
h^{(1,1)}= h\left[\mathsf{A}\mathsf{B}+\mathsf{A}^{(0,1)}\right]=h\left[\mathsf{A}\mathsf{B}+\mathsf{B}^{(1,0)}\right],
\end{equation}
since of course $\mathsf{A}^{(0,1)}=\mathsf{B}^{(1,0)}$,
\begin{equation}\label{h20}
h^{(2,0)}= h\left[\mathsf{A}^2+\mathsf{A}^{(1,0)}\right],
\end{equation}
\begin{equation}\label{h02}
h^{(0,2)}= h\left[\mathsf{B}^2+\mathsf{B}^{(0,1)}\right],
\end{equation}
\begin{equation}\label{h21}
h^{(2,1)}= h\left[\mathsf{B}\left(\mathsf{A}^2+\mathsf{A}^{(1,0)}\right)+2\mathsf{A}\mathsf{A}^{(0,1)}+\mathsf{A}^{(1,1)}\right]= h\left[\mathsf{B}\left(\mathsf{A}^2+\mathsf{A}^{(1,0)}\right)+2\mathsf{A}\mathsf{B}^{(1,0)}+\mathsf{B}^{(2,0)}\right],
\end{equation}
\begin{equation}\label{h12}
h^{(1,2)}= h\left[\mathsf{A}\left(\mathsf{B}^2+\mathsf{B}^{(0,1)}\right)+2\mathsf{B}\mathsf{B}^{(1,0)}+\mathsf{B}^{(1,1)}\right]= h\left[\mathsf{A}\left(\mathsf{B}^2+\mathsf{B}^{(0,1)}\right)+2\mathsf{B}\mathsf{A}^{(0,1)}+\mathsf{A}^{(0,2)}\right],
\end{equation}
\begin{equation}\nonumber
etc.
\end{equation}

\section{\label{ConvHorn}Regions of convergence}

To ease the reading of this paper, we recall in this appendix some results on the theory of convergence of multiple (hypergeometric) series used in the text. See for instance \cite{Kampe} and \cite{Sri} for more details.
We also give the proof in section \ref{AlmostHorn} that the polygamma functions $\psi^{(k)}$ involved in the multiple series obtained from multiple Mellin-Barnes (in this paper and, we think, more generally in perturbative quantum field theory) do not affect the regions of convergence that may be obtained, when ignoring these polygamma functions, from Horn's theorem (the latter being described in section \ref{horn_conv}).

\subsection{Single series}

In this simple case, one makes use of d'Alembert and Raabe-Duhamel's ratio tests.

\subsection{Double series}

\subsubsection{Horn series\label{horn_conv}}

A general statement as the d'Alembert's ratio test does not exist for arbitrary double series. Nevertheless, for a certain type of double series (those of Horn's type), it is possible to know the region of (absolute) convergence from a kind of extension of d'Alembert's ratio test due to Horn \cite{Kampe, Sri}. 

A double series  $\displaystyle \sum_{m=0}^{\infty}\sum_{n=0}^{\infty} a_{m,n} x^m y^n$ is a Horn's one if the two functions 
\begin{equation}\label{fg}
\mathrm{f}(m,n) \doteq \frac{a_{m+1,n}}{a_{m,n}} \doteq \frac{\mathrm{P}(m,n)}{\mathrm{R}(m,n)} \hspace{1cm} \text{and}  \hspace{1cm} \mathrm{g}(m,n) \doteq \frac{a_{m,n+1}}{a_{m,n}} \doteq \frac{\mathrm{Q}(m,n)}{\mathrm{S}(m,n)}\;
\end{equation}
are rational functions. This means that $\mathrm{P, Q, R}$ and $\mathrm{S}$ are polynomial of degree $p, q, r$ and $s$. 

Defining the functions 
\begin{equation}\label{FG}
\mathrm{F}(m,n) = \lim_{\eta\rightarrow+\infty}\;\mathrm{f}(\eta\, m, \eta\, n) \hspace{1cm} \text{and}  \hspace{1cm} \mathrm{G}(m,n) = \lim_{\eta\rightarrow+\infty}\;\mathrm{g}(\eta\, m, \eta\, n)\;.
\end{equation}
one may distinguish five different cases:

\begin{center}
\renewcommand{\arraystretch}{2}
\begin{tabular}{|c|c|}
\hline
Conditions & Region of convergence\\
\hline
$p>r$ or $q>s$ & $|x| = 0$ and $|y|= 0$\\
\hline
$p<r$ and $q<s$ & $(|x|,|y|) \in \mathds{R}_+^2$\\
\hline
$p<r$ and $q=s$ & $|x| \in \mathds{R}_+$ and $|y| < \frac{1}{\left|\mathrm{G}(0,1)\right|}$\\
\hline
$p=r$ and $q<s$ & $|x|< \frac{1}{\left|\mathrm{F}(1,0)\right|} $ and $|y| \in \mathds{R}_+ $\\
\hline
$p=r$ and $q=s$ & $(|x|,|y|) \in \mathcal{C}\cap\mathcal{D}$\\
\hline
\end{tabular}
\end{center}
where the domains $\mathcal{C}$ and $\mathcal{D}$ are two subsets of $\mathds{R}_+^2$ defined as 
\begin{equation}
\mathcal{C} = \left\{ (|x|, |y|) \; \; \big \vert \;\;  0 < |x| < \frac{1}{\left|\mathrm{F}(1,0)\right|} \; \;  \text{and} \; \;  0 < |y| < \frac{1}{\left|\mathrm{G}(0,1)\right|} \right\}
\end{equation}
and 
\begin{equation}\label{D}
\mathcal{D} = \left\{ (|x|, |y|) \; \; \big \vert \;\; \forall (m,n)\in \mathds{R}_+^2: 0 < |x| < \frac{1}{\left|\mathrm{F}(m,n)\right|} \; \;  \text{or} \; \;  0 < |y| < \frac{1}{\left|\mathrm{G}(m,n)\right|} \right\}.
\end{equation}
We recall that the logical disjunction 'or' in (\ref{D}) implies, among others, that if $0 < |x| < \frac{1}{\left|\mathrm{F}(m,n)\right|}$ is true \textit{and} $0 < |y| < \frac{1}{\left|\mathrm{G}(m,n)\right|}$ is also true, then $\left(0 < |x| < \frac{1}{\left|\mathrm{F}(m,n)\right|} \; \;  \text{or} \; \;  0 < |y| < \frac{1}{\left|\mathrm{G}(m,n)\right|}\right)$ is true.

The case where $p=r$ and $q=s$ is the so-called Horn's theorem \cite{Sri}.

Let us conclude this section by saying that if $a_{m,n}$ is a ratio of gamma functions of the type $\Gamma(am+bn+c)$ where $a, b$ are integers and $c$ is a constant (depending possibly on the dimensional regularisation parameter $\epsilon$), then obviously the corresponding double series will be of Horn's type. This is what we observe in the perturbative calculations of particle physics, modulo polygamma functions (see section \ref{AlmostHorn}).  

\subsection{Three dimensional case\label{Horn3D}}

Horn's theorem may be generalized to the case of triple series or even higher order series \cite{Sri} (but it is not as trivial as what could be thought at first sight, since it seems that wrong results have been given on this subject in a certain number of papers of the specialized mathematics litterature \cite{Karlsson}).
 
Let us consider the series $\displaystyle \sum_{m,n,\ell=0}^\infty a_{m,n,\ell}\, x^m y^n z^\ell$ where 
\begin{equation}
\mathrm{f}(m,n,\ell) \doteq \frac{a_{m+1,n,\ell}}{a_{m,n,\ell}}\;\;, \hspace{1cm} \mathrm{g}(m,n,\ell) \doteq \frac{a_{m,n+1,\ell}}{a_{m,n,\ell}}\;\;,\hspace{1cm} \mathrm{h}(m,n,\ell) \doteq \frac{a_{m,n,\ell+1}}{a_{m,n,\ell}}\;
\end{equation}
are rational functions.
We focus on the particular situation where the degree of each numerator is equal to the degree of its corresponding denominator. 

From the definitions
\begin{equation*}
\mathrm{F}(m,n,\ell) = \lim_{\eta\rightarrow+\infty}\;\mathrm{f}(\eta\, m, \eta\, n,\eta\,\ell) \;\;,  \hspace{1cm} \mathrm{G}(m,n,\ell) = \lim_{\eta\rightarrow+\infty}\;\mathrm{g}(\eta\, m, \eta\, n,\eta\,\ell)\;\;
\end{equation*}
and
\begin{equation}
\mathrm{H}(m,n,\ell) = \lim_{\eta\rightarrow+\infty}\;\mathrm{h}(\eta\, m, \eta\, n,\eta\,\ell)\;,
\end{equation}
the region of convergence of the triple series is given by 
\begin{equation}
(|x|,|y|,|z|) \in \mathcal{C}\cap\mathcal{D}_1\cap\mathcal{D}_2\cap\mathcal{D}_3\cap\mathcal{D}_4\;\;,
\end{equation}
where
\begin{align}
\mathcal{C} &= \Bigg\{ (|x|, |y|,|z|) \; \; \bigg \vert \;\;  0 \leqslant |x| < \frac{1}{\left|\mathrm{F}(1,0,0)\right|} \; \;  \text{and} \; \;  0 \leqslant |y| < \frac{1}{\left|\mathrm{G}(0,1,0)\right|} \nonumber\\
&\hspace{4cm}\text{and} \; \;0 \leqslant |z| < \frac{1}{\left|\mathrm{H}(0,0,1)\right|}\Bigg\},\\
\mathcal{D}_1&=\Bigg\{ (|x|, |y|,|z|) \; \; \bigg \vert \;\; \forall(m,n,\ell)\in\mathds{R}^3_+, |x| < \frac{1}{|\mathrm{F}(m,n,\ell)|} \; \;\text{or}\; \; |y| < \frac{1}{|\mathrm{G}(m,n,\ell)|}\, \nonumber\\
& \hspace{4cm}\; \; \text{or} \; \;|z| < \frac{1}{|\mathrm{H}(m,n,\ell)|} \Bigg \}, \\
\mathcal{D}_2&=\Bigg\{ (|x|, |y|,|z|) \; \; \bigg \vert \;\; \forall(n,\ell)\in\mathds{R}^2_+,|y| < \frac{1}{|\mathrm{G}(0,n,\ell)|}\; \;\text{or}\; \; |z| < \frac{1}{|\mathrm{H}(0,n,\ell)|}\Bigg\}, \\
\mathcal{D}_3&=\Bigg\{ (|x|, |y|,|z|) \; \; \bigg \vert \;\; \forall(m,\ell)\in\mathds{R}^2_+,|x| < \frac{1}{|\mathrm{F}(m,0,\ell)|}\; \;\text{or}\; \; |y| < \frac{1}{|\mathrm{G}(m,0,\ell)|}\Bigg\}, \\
\text{and}\\
\mathcal{D}_4&=\Bigg\{ (|x|, |y|,|z|) \; \; \bigg \vert \;\; \forall(m,n)\in\mathds{R}^2_+,|x| < \frac{1}{|\mathrm{F}(m,n,0)|}\; \;\text{or}\; \; |z| < \frac{1}{|\mathrm{H}(m,n,0)|}\Bigg\}\;.
\end{align}

\subsection{"Almost" Horn series\label{AlmostHorn}}

It is clear that although the integrands of the Mellin-Barnes integrals considered in this paper are products of gamma functions, the coefficients which appear in their (multiple) series representations are not purely composed of gamma functions in general. They contain also polygamma functions $\psi^{(k)}$ whose arguments contain both summation indices $m$ and $n$ (as well as $\epsilon$, the dimensional regularisation parameter, in the case of the box integral), implying that the double series are not strictly of Horn's type. 

In fact, in full generality (see Appendix \ref{Formulary}), we observe that the coefficients of the series obtained from a twofold Mellin-Barnes integral whose integrand is a ratio of gamma functions, cannot be anything else than a ratio of gamma functions multiplied by polygamma functions of different orders.

Let us call $a_{m,n}$ this ratio of gamma functions (with arguments of the type $am+bn+c$ where $a$ and $b$ are integers and $c$ is not a negative integer), then
\begin{equation}
F(x,y) = \sum_{m,n=0}^\infty a_{m,n} x^m y^n
\end{equation}
is a Horn series, converging (absolutely) in a region $\mathcal{R}$. 

Let $\mu$ be a linear combination of $m$ and $n$ such that $\mu \in \mathds{N}$, and let us define
\begin{equation}
\tilde F(x,y) = \sum_{m,n=0}^\infty a_{m,n}\; \psi^{(p)}(\alpha+\mu) \; x^m y^n\;,
\end{equation}
where $p$ is a positiver integer and $\alpha$ is not a negative integer.
We want to show that $\tilde F(x,y)$ converges absolutely in the same region $\mathcal{R}$ as $F(x,y)$. For this we follow the same reasoning than the one used in Chapter 4 of \cite{Sri} to prove that the region of convergence of a hypergeometric series is independent of the parameters.

It is straightforward to prove that 
\begin{equation}
\psi^{(p)}(\alpha+\mu) \underset{\mu \rightarrow +\infty}{=} \mathcal{O}(\mu)\;.
\end{equation}
 
Then, one can find positive numbers\footnote{For the case $p=0$, one has to check that $\alpha+\mu$ is such that $\psi(\alpha+\mu)\neq0$.} $C_0, C_1$ and $C_2$ such that, for $\varepsilon>0$ and $\mu\in \mathds{N}$,
\begin{equation}
C_0 < \left| \psi^{(p)}(\alpha+\mu) \right| < C_1 \; \mu < C_2 (1+\varepsilon)^{m+n}\;.
\end{equation}

Thus one has
\begin{equation}
C_0 F_{abs.}(x,y) < \tilde F_{abs.}(x,y) < C_2 F_{abs.}\left(x(1+\varepsilon),y(1+\varepsilon)\right)\;,
\end{equation}
where the index $abs.$ is meant for the series with absolute values.

Now, for $(x,y)\in \mathcal{R}$, one can choose $\varepsilon$ such that 
\begin{equation}
(x(1+\varepsilon),y(1+\varepsilon))\in \mathcal{R}.
\end{equation}
This ends the proof that $\tilde F$ is absolutely convergent on $\mathcal{R}$. 

This result is straightforwardly extended to multiple series of the same type with an arbitrary number of variables. It may also be used iteratively to prove that a multiple series where the general term has an arbitrary product of polygamma functions of any order do have the same region of convergence $\mathcal{R}$ than the series without these polygamma functions.

Therefore one may conclude that in our cases of study, the region of convergence of a given multiple series may be obtained by ignoring polygamma functions in its general term, and by considering only the term $a_{m,n}$ which is made of gamma functions. 

\subsection{Generalized Kamp\'e de F\'eriet series\label{Kampe}}

A useful class of multiple hypergeometric series, called generalized Kamp\'e de F\'eriet series, is defined as \cite{Sri} 
\begin{equation}
{\rm F}^{p:q_1;\dots;q_n}_{\ell:m_1;\dots;m_n} \left(
x_1, \dots, x_n\right) 
\doteq \sum_{k_1=0}^\infty \cdots \sum_{k_n=0}^\infty a_{k_1,\dots, k_n}
\frac{x_1^{k_1}}{k_1!}\cdots\frac{x_n^{k_n}}{k_n!}
\end{equation}
where 
\begin{equation}
a_{k_1,\dots,k_n} = \frac{\displaystyle \prod_{j=1}^p(a_j)_{k_1+\cdots
k_n}\prod_{i=1}^{n}\prod_{j=1}^{q_i}(b_{i,j})_{k_i}}{\displaystyle
\prod_{j=1}^\ell(\alpha_j)_{k_1+\cdots
k_n}\prod_{i=1}^{n}\prod_{j=1}^{m_i}(\beta_{i,j})_{k_i}}\ .
\end{equation}
For convergence of these series one has to check that
\begin{equation}
1+\ell+ m_i - p - q_i \geq 0, \;\; i= 1,\dots, n.
\end{equation}
Moreover, the region of convergence is given either by 
\begin{equation}
\sum_{j=1}^n |x_j|^{\frac{1}{p-\ell}} <1\;\;\;\; \text{if}\;\;\;\; p>\ell 
\end{equation}
or by
\begin{equation}
\text{max}\{|x_1|,\dots,|x_n|\} <1\;\;\;\; \text{if}\;\;\;\; p \leq \ell.
\end{equation}


\begin{thebibliography}{99}

\bibitem{Friot:2009fw}
  S.~Friot and D.~Greynat,
  \textit{``Non-perturbative asymptotic improvement of perturbation theory and
  Mellin-Barnes representation''},
  SIGMA {\bf 6} (2010) 079
  [arXiv:0907.5593 [hep-th]].


\bibitem{Friot:2005cu}
  S.~Friot, D.~Greynat and E.~De Rafael,
  \textit{``Asymptotics of Feynman diagrams and the Mellin-Barnes representation''},
  Phys.\ Lett.\  B {\bf 628} (2005) 73
  [arXiv:hep-ph/0505038].


\bibitem{Aguilar:2008qj}
  J.~P.~Aguilar, D.~Greynat and E.~De Rafael,
  \textit{``Muon anomaly from lepton vacuum polarization and the Mellin-Barnes
  representation''},
  Phys.\ Rev.\  D {\bf 77} (2008) 093010
  [arXiv:0802.2618 [hep-ph]].

\bibitem{Paris}
R.~B.~Paris and D.~Kaminski, 
\textit{``Asymptotics and Mellin-Barnes integrals''},
Encyclopedia of Mathematics and its applications (2001), Cambridge University Press. 




\bibitem{Tsikh:1994}
M.~Passare, A.~K.~Tsikh and O.~N.~Zhdanov,
\textit{``A multidimensional Jordan residue lemma with an application to Mellin-Barnes integrals''}, 
Contributions to Complex Analysis and Analytic Geometry, Aspects of Mathematics, vol. E{\bf 26}, Vieweg Verlag, Wiesbaden (1994) 233.
 
\bibitem{Passare:1996db}
  M.~Passare, A.~K.~Tsikh and A.~A.~Cheshel,
  \textit{``Multiple Mellin-Barnes integrals as periods of Calabi-Yau manifolds  with
  several moduli''},
  Theor.\ Math.\ Phys.\  {\bf 109} (1997) 1544
  [Teor.\ Mat.\ Fiz.\  {\bf 109N3} (1996) 381]
  [arXiv:hep-th/9609215].



\bibitem{DelDuca:2009au}
  V.~Del Duca, C.~Duhr and V.~A.~Smirnov,
   \textit{``An Analytic Result for the Two-Loop Hexagon Wilson Loop in N = 4 SYM''},
  JHEP {\bf 1003} (2010) 099
  [arXiv:0911.5332 [hep-ph]].



\bibitem{DelDuca:2010zg}
  V.~Del Duca, C.~Duhr and V.~A.~Smirnov,
  \textit{``The two-loop hexagon Wilson loop in $\mathcal{N}=4$ SYM''},
  JHEP {\bf 1005} (2010) 084
  [arXiv:1003.1702 [hep-th]].


\bibitem{Tsikh:1998}
  O.~N.~Zhdanov and A.~K.~Tsikh,
  \textit{``Studying the multiple Mellin-Barnes integrals by means of multidimensional residues''},
  Sib.\ Math.\ J.\  {\bf 39} (1998) 245.






\bibitem{Dorokhov:2008cd}
  A.~E.~Dorokhov and M.~A.~Ivanov,
  \textit{``On mass corrections to the decays $P \to l^+l^-$''},
  JETP Lett.\  {\bf 87} (2008) 531
 [arXiv:0803.4493 [hep-ph]].

\bibitem{Aguilar}
  J.~P.~Aguilar, 
  \textit{``Représentation de Mellin-Barnes et Anomalie Magnétique du Muon''},
  Ph.D. thesis (2008), in french. 


\bibitem{griffiths}
  P.~Griffiths and J.~Harris,
  \textit{``Principles of algebraic geometry''},
 (Wiley, 1978).

\bibitem{FlaSal}
P.~Flajolet and B.~Salvy, 
 \textit{``Euler sums and contour integral representations''},
Experiment. Math. {\bf 7} (1998), 15-35.


\bibitem{Moch:2001zr}
  S.~Moch, P.~Uwer and S.~Weinzierl,
  \textit{``Nested sums, expansion of transcendental functions and multi-scale
  multi-loop integrals''},
  J.\ Math.\ Phys.\  {\bf 43} (2002) 3363
  [arXiv:hep-ph/0110083].
 


\bibitem{Valtancoli:2011kr}
  P.~Valtancoli, \textit{"The scalar box integral and the Mellin - Barnes
representation"},
Int.\ J.\ Mod.\ Phys.\  A {\bf 26} (2011) 2557
  [arXiv:1104.2661 [math-ph]].

\bibitem{Kalmykov:2006hu}
  M.~Y.~Kalmykov, B.~F.~L.~Ward and S.~Yost,
  \textit{``All order epsilon-expansion of Gauss hypergeometric functions with integer
  and half/integer values of parameters''},
  JHEP {\bf 0702} (2007) 040
  [arXiv:hep-th/0612240].


\bibitem{Hardy}
  M.~Hardy,
  \textit{``Combinatorics of partial derivatives''},
 Electron.\ J.\ Combin.\ {\bf 13} (2006), R1
	[arXiv:math/0601149v1].

\bibitem{Kampe}
  P.~Appell and J.~Kamp\'e de F\'eriet, 
  \textit{``Fonctions hyperg\'eométriques et hypersph\'eriques - Polyn\^omes d'Hermite''},
 Gautiers-Villars et $\text{C}^{\text{ie}}$, 1926.


\bibitem{Sri}
H.~M.~Srivastava and P.~W.~Karlsson \textit{"Multiple gaussian hypergeometric serie"}, Ellis Horwood Series in Mathematics and Its Applications, 1985.

\bibitem{Karlsson}
P.~W.~Karlsson,
\textit{``Regions of convergence for hypergeometric series in three variables''}, Math. Scand. {\bf 34} (1974), 241-248








\end{thebibliography}
\end{document}